\def\tilde{\widetilde}
\def\hat{\widehat}

\def\cech{${\rm C}^{\kern-6pt \vbox{\hbox{$\scriptscriptstyle\vee$}\kern2.5pt}}${\rm ech}}
\def\Cech{${\sl C}^{\kern-6pt \vbox{\hbox{$\scriptscriptstyle\vee$}\kern2.5pt}}${\sl ech}}

\def\a{{\alpha}}

\def\b{{\beta}}

\def\d{{\delta}}

\def\G{{\Gamma}}
\def\e{{\epsilon}}

\def\m{{\mu}}

\def\s{{\sigma}}

\def\p{{\phi}}

\def\CD{{\cal D}}

\def\CG{{\cal G}}
\def\CH{{\cal H}}

\def\CK{{\cal K}}
\def\CL{{\cal L}}

\def\CN{{\cal N}}
\def\CO{{\cal O}}
\def\CP{{\cal P}}
\def\CQ{{\cal Q}}

\def\CS{{\cal S}}

\def\BZ{{\bf Z}}

\def\MC{{\mathbb{C}}}

\def\MR{{\mathbb{R}}}

\def\MZ{{\mathbb{Z}}}

 \def\p{\partial}
 
 \def\s{  s }

     \def\e{\varepsilon}
     
     \def\IR{ {\mathbb R}}
     \def\CO{{\cal O}}
          \def\CP{{\cal P}}

\def\b{\beta}

\def\d{\partial}

\def\inv{^{-1}}
\def\STr{{\rm STr}}

 \def\frac#1#2{{\textstyle{#1\over#2}}}
\def\inv{^{\raise.15ex\hbox{${\scriptscriptstyle -}$}\kern-.05em 1}}
 \def\hf{\frac{1}{2}}

\def\({\left(}
\def\){\right)}
\def\<{\left\langle\,}
\def\>{\, \right\rangle}
\def\[{\left[}
\def\]{\right]}

\newcommand{\no}{\nonumber}

\def\la{\label}

\def\sgn{\;{\rm sign}\:}

\def\sn{{\rm sn\,} }

  \def\cn{{\rm cn\,} }
  \def\dn{{\rm dn\,} }

\def\K{ {\rm K}}

\def\up{\uparrow}
\def\don{\downarrow}
\def\updn{\updownarrow}
\def\hole{\circ}

\documentclass[12pt,a4paper]{article}
\input epsf
\usepackage{amsmath}
\usepackage{amssymb}
\usepackage{graphicx}

\newcommand{\pint}{\makebox[0pt][l]{\hspace{3.1pt}$-$}\int}

\newcommand{\atopfrac}[2]{\genfrac{}{}{0pt}{}{#1}{#2}}

\newcommand{\Tr}{{\rm Tr \,}}

\newcommand{\gYM}{g_{\rm YM}}

\def\al{{\alpha}}
\def\dal{{\dot \al}}
\def\be{{\beta}}
\def\dbe{{\dot\be}}
\def\de{{\delta}}
\def\ga{{\gamma}}
\def\dga{{\dot \gamma}}
\def\G{{\Gamma}}
\def\e{{\epsilon}}

\def\m{{\mu}}

\def\s{{\sigma}}

\def\vev#1{\langle #1 \rangle}
\def\ket#1{ | #1 \rangle}
\def\bra#1{ \langle #1 |}
\def\[{\left[}
\def\]{\right]}

\numberwithin{equation}{section}

\begin{document}

\thispagestyle{empty}
\begin{flushright}
{ }
{IPhT-T10/042}\\
\end{flushright}
\vskip 4cm
\vspace{1cm}
\setcounter{footnote}{0}
\vskip 9mm
\begin{center}
{\Large{
\bf   Integrability and the AdS/CFT Correspondence

}}

\vspace{20mm}  Didina Serban 
 \\[7mm] {\it Institut de Physique Th\'eorique, CNRS-URA
2306
\\
C.E.A.-Saclay \\
F-91191 Gif-sur-Yvette, France}\\[1mm]
 \end{center}
 

%

%
%
%






\centerline{\bf \large Abstract}
\vskip 1cm

\noindent{The description of gauge theories at strong coupling is one of the long-standing problems in theoretical physics. The idea of a relation between strongly coupled gauge theories and string theory was pioneered by 't Hooft, Wilson and Polyakov. A decade ago, Maldacena made this relation explicit by conjecturing the exact equivalence of a conformally invariant theory in four dimensions, the maximally supersymmetric Yang-Mills theory, with string theory in the $AdS_5 \times S^5$ background.
Other examples of correspondence between a conformally invariant theory and string theory in an AdS background were discovered recently.
The comparison of the two sides of the correspondence requires the use of non-perturbative methods. The discovery of integrable structures in gauge theory
 and string theory 
led to the conjecture that the two theories are integrable for any value of the coupling constant and that they share the same integrable structure defined non-perturbatively. The last eight years brought remarkable progress in identifying this solvable model and in explicitly solving the problem of computing the spectrum of conformal dimensions of the theory. The progress came from the identification of the dilatation operator with an integrable spin chain and from the study of the string sigma model.  
In this thesis, I present the evolution of the concept of integrability in the framework of the AdS/CFT correspondence and the
the main results obtained using this approach.}

\vfill\eject

\tableofcontents

\vfill\eject

\section{ Introduction: the gauge-string correspondence}

The relation between gauge theories and string theory is a long-standing and unsolved problem in theoretical physics.  Understanding this relation will help us to handle gauge theories at strong coupling and consistently include gravity into the picture of quantum field theory.

String theory itself was discovered in the attempt to describe the strong interactions, once the dual models \cite{Veneziano:1968yb} were proposed to incorporate the so-called Regge trajectories observed in experiments. The advent of the description of strong interaction in terms of Yang-Mills theory reduced the importance of string theory as a model of strong interactions, and for a while the interest in string theory stemmed mainly from the fact that it promised to give an unified description of all the forces in nature, including gravity.

The idea of duality between strings and gauge theories originates in the work of 't Hooft \cite{Hooft:1973jz}, who understood that perturbative expansion of $U(N)$  gauge theories in the large $N$ limit can be reorganized in terms of a genus expansion of the surface spanned by the Feynman diagrams, expansion which is reminiscent of the perturbation expansion in string theory. 
The duality was also suggested by the strong coupling expansion of the lattice gauge theory of Wilson \cite{Wilson:1974sk} and Polyakov \cite{Polyakov:2008zx} in terms of surfaces spanned by the color-electric fluxes, which may end on quarks or make a propagating closed string.

A simpler version of the gauge-string duality was extensively studied via the correspondence  \cite{Kazakov:1985ds} \cite{David:1985nj} \cite{Kazakov:1985ea} between matrix models and non-critical strings, as well as through the study of the 2d quantum gravity. This subject is reviewed in \cite{DiFrancesco:1993nw} \cite{Ginsparg:1993is} \cite{Klebanov:1991qa}.  Many of the ideas and techniques which were developed in this context are now transposed to the study of the modern, full-fledged duality.

The discovery of the D-branes as extended objects on which the strings can start or end \cite{Dai:1989ua} \cite{Polchinski:1995mt} gave a new impetus to the study of non-perturbative effects and string dualities and helped to understand the relation with black hole geometries and gauge theory. 
It was understood  by Witten  \cite{Witten:1995gx} that a stack of $N$ coincident D-branes
is associated to a $U(N)$ symmetric gauge theory. Shortly after, Klebanov \cite{Klebanov:1997kc}
computed the absorption cross section of a stack of $N$ D-branes, both from the point of view of a gauge theory and from the point of view of supergravity. For the case of D3-brane, the two computations gave exactly the same answer.

Building on these ideas, Maldacena \cite{Maldacena:1997re} conjectured in 1997 that the maximally supersymmetric gauge theory in four dimensions is exactly dual
to type IIB strings in the $AdS_5\times S^5$ background. He considered strings in a geometry 
created by a stack of $N$ coincident D3-branes, such that the theory on the branes is the four dimensional ${\cal N}=4$ super Yang-Mills (SYM) theory. The geometry near the branes is that of an anti-de Sitter space times a five-sphere. The four-dimensional space can be recovered as the boundary of the $AdS_5$ space, while the five-sphere is associated with the internal symmetry of 
the gauge fields. 

The conjecture by Maldacena confirmed the idea of Polyakov \cite{Polyakov:1997tj} \cite{Polyakov:1998ju} that the non-critical string theory describing
gauge fields in four dimensions should be completed with an extra, Liouville-like direction, which gives rise to a curved five-dimensional space. In order to insure the full reparametrization invariance of the Wilson loops, the four-dimensional gauge theory should be situated on the boundary of this curved five-dimensional space. The idea that a five-dimensional theory is determined by a four-dimensional theory on the boundary rejoins an earlier proposal by 't Hooft, made in the context of the black hole physics 
and subsequently known under the name of holgraphy \cite{Hooft:1993gx}. A similar property is associated with the topological field theories as the Chern-Simons theory, used to describe the quantum Hall effect.

The gauge theory is characterized by two parameters: the coupling constant $g_{YM}$ and and the
number of colors $N$. In the large $N$ limit, or planar limit, they can be combined into a single parameter, the 't Hooft coupling constant $\lambda=g_{YM}^2N$. These parameters are related to the string parameters by  \cite{Maldacena:1997re} 
\begin{equation}
\label{paramcorr}
4\pi g_s=g_{YM}^2\equiv \lambda/N\;, \qquad R^4/\alpha'^2=\lambda\;,
\end{equation} 
with $g_s$ being the string coupling constant, $\alpha'$ the string tension and $R$ the radius both of $AdS_5$ and of the five-sphere. The identification of the parameters can be done by inspecting the supergravity solution  \cite{Horowitz:1991cd} associated to $N$ coincident D3-branes.

The quantum ${\cal N}=4$ SYM theory is invariant under dilatations, so that the Poincar\'e symmetry is extended to the conformal symmetry in four dimensions, with symmetry group $SO(4,2)\simeq SU(2,2)$. The symmetry group acting on the four copies of supersymmetry generators, the so-called R-symmetry,  is given by $SU(4)\simeq SO(6)$. Together with the supersymmetry generators, the symmetry group of the ${\cal N}=4$ SYM theory is extended to the graded Lie group
$SU(2,2|4)$, the ${\cal N}=4$ superconformal group in four dimensions. The bosonic subgroup $SO(4,2)\otimes SO(6)$ is the isometry group of the $AdS_5\times S^5$ string background. Metsaev and Tseytlin showed \cite{Metsaev:1998it} that the type IIB string theory on the $AdS_5\times S^5$ background can be formulated as a non-linear sigma model on the coset superspace $SU(2,2|4)/SO(4,1)\otimes SO(5)$, thus proving that the two theories share the same global symmetry.

A special role in the correspondence is played by the dilatation operator, which is associated to the Hamiltonian in the radial quantization. The eigenvalues of the dilatation operator are the quantum dimensions of the local operators in the conformally invariant theory. The gauge-invariant local operators in the gauge theory are nothing but the traces over the color indices of products of the fundamental fields.
Under the correspondence, they are associated to the states of the string, and their conformal dimensions are associated to the string energy. The dilatation operator corresponds to a non-compact generator of the global symmetry group, which can get quantum corrections and depends on the coupling constant. The other charges are discretized and correspond to angular momenta for the rotation of the string either on the $AdS_5$ space or on $S^5$.

The conjecture \cite{Maldacena:1997re} is supposed, in its strongest form, to hold for any number of colors $N$. In practice, it is easier to study the planar limit, when the number of colors goes to infinity, while the 't Hooft coupling constant $\lambda$ is kept fixed. In this case, according to eq. (\ref{paramcorr}), the string coupling constant vanishes and we are dealing with the weak formulation of the 
AdS/CFT correspondence which associates the planar limit of the ${\cal N}=4$ SYM theory to non-interacting strings in the $AdS_5\times S^5$ background. In this thesis, we will concentrate exclusively on the planar limit.

As it can be guessed from equation (\ref{paramcorr}), the Maldacena duality is a strong/weak coupling duality, associating the perturbative regime $\lambda\to 0$ in the ${\cal N}=4$ SYM theory to the strongly coupled regime of the Metsaev-Tseytlin sigma model. This is the reason the duality is not easy to prove, since the regimes accessible by perturbative methods  in the two theories do not match.
However, due to supersymmetry, some of the states do not change with the coupling constant. They are the so-called BPS states and they were the main subject of the early checks of the correspondence.

In 2002 the situation changed drastically when a new limit  considered by Berenstein, Maldacena and Nastase (BMN)\cite{Berenstein:2002jq} allowed to explore a regime away from the BPS states. These authors have studied the excitations of around a point-like string configuration which moves with a large angular momentum $J$ along a great circle on $S_5$. The geometry seen by this fast moving string is the so-called plane wave geometry \cite{Penrose},
 and string theory in this background is exactly solvable, the excitations being massive bosons/fermions. The solution depends on the effective parameter $\lambda'=\lambda/J^2$, the BMN coupling constant, which can take an arbitrary value.

The same year, the first integrable structures where discovered, almost simultaneously, in perturbative gauge and string theory. The first mention in the literature concerning integrability in the context of the planar ${\cal N}=4$ SYM theory goes back to Lipatov \cite{Lipatov:1997vu}. However, it is the work of Minahan and Zarembo \cite{Minahan:2002ve}, who mapped the one-loop dilatation operator to an integrable spin chain,  which opened a whole new field of activity. Shortly after this discovery, Bena, Polchinski and Roiban \cite{Bena:2003wd} proved that the Metsaev-Tseylin sigma model is classically integrable. 
When it was understood that integrability extends to all operators at one loop \cite{Beisert:2003jj} and that it persists at two and three loop order in gauge theory \cite{Beisert:2003tq}, it was tempting to conjecture that integrability is an all-loop feature.
The integrable structures discovered in perturbation on the two sides of the correspondence would be then manifestations of the same object, which can be defined non-perturbatively. Given the power of the constraints usually associated with integrability, it is conceivable that integrability may allow to find a complete solution both to the problem of quantizing free strings in a particular curved background and that of solving a non-trivial quantum field theory in four dimensions. 

Other examples of the AdS/CFT correspondence were discovered recently, relating the $\CN=6$ super-conformal Chern-Simons theory in $2+1$ dimensions and string theory in $AdS_4\times CP^3$ \cite{Aharony:2008ug}, as well as the $AdS_3/CFT_2$ correspondence relating the type IIB strings in $AdS_3\times S_3\times M_4$ backgrounds, where $M_4=T_4$ or $S_3 \times S_1$, to some two-dimensional conformal field theory \cite{Babichenko:2009dk}. These theories are probably also integrable in the large $N$ limit \cite{Minahan:2008hf}. Other candidates to integrable correspondences were identified in \cite{Zarembo:2010sg}.

The scope of this thesis is to present the concepts developed during the last eight years
in connection with integrability in the context of the AdS/CFT correspondence. 
The relation between spin chains and sigma models is not new in physics.
It was known for more than thirty years that the low energy physics of a Heisenberg antiferromagnet is captured by a  two dimensional sigma model with a topological term \cite{Haldane:1983ru}. However, the situation in the AdS/CFT correspondence is different, since there should be an exact, one to one  mapping between the states of a spin chain and that of a sigma model. This is presumably made possible by the huge amount of symmetry, in particular the supersymmetry, involved in this particular
theory. The integrable model behind the AdS/CFT correspondence includes as particular cases many known integrable models, but it have features that go beyond any particular model which was already studied. It combines features of super-symmetric, non-compact spin chains with those of non-relativistic integrable sigma models, and it may serve to generalize and unify these different approaches to integrable models.
The development of the subject itself came from the interplay of several approaches and benefitted from experts coming from various backgrounds and with different points of view. Crucial information was obtained at weak coupling in the gauge theory and  in string theory from state-of-the-art perturbative computations. Notions and techniques from integrable spin chains, including super-spin chains, two dimensional sigma models and matrix models were combined together and helped to build a coherent picture of the model. 

 Given the fact that there is no information available for finite coupling constant neither from the gauge theory
nor from string theory, the form of the equations describing the spectrum of conformal dimension was obtained, at least in the beginning, by trial and errors and by conjectures and checks rather than by constructive proofs. 
The formal proof of integrability is still lacking, but the fact that the same set of equation is able to reproduce both
weak and strong coupling results is remarkable and cannot be an accident. 

After the conjecture of Beisert, Kristjansen and Staudacher \cite{Beisert:2003tq} that the dilatation operator of $\CN=4$ SYM is integrable at all loop, the search of the all-loop equations determining the spectrum has started. The dilatation operator at higher loops corresponds to a long range Hamiltonian, with the range of the interaction growing with the loop order. The closest known example of this type is the Inozemtsev Hamiltonian \cite{Inozemtsev:2002vb}. This system can be diagonalized using Bethe ansatz-like equations only for infinite length, with corrections which are exponential in the system size. These equations will be called in the following {\it asymptotic} Bethe ansatz equations.
Serban and Staudacher showed \cite{Serban:2004jf} that the  dilatation operator can be approximated, up to three loops and for a particular class of operators, by a combination of the integrals of motion of the Inozemtsev model and therefore could be diagonalized by the Inozemtsev asymptotic Bethe ansatz. The first corrections in the BMN coupling constant $\lambda'$ coincided with those obtained from string theory \cite{Frolov:2003qc,Frolov:2003xy,Arutyunov:2003uj}, while the third loop order disagree with the string result due to order-of-limit problem.
Staudacher \cite{Staudacher:2004tk} emphasized the link between the asymptotic Bethe ansatz and the concept of scattering matrix for magnons, concept which subsequently proved to be very fruitful. In particular, using the S matrix allowed to circumvent the difficulties related to the changing length \cite{Beisert:2003ys} which renders the usual algebraic Bethe ansatz approach to spin chains inoperable in the present context. 

At the same time, Kazakov, Marshakov, Minahan and Zarembo \cite{Kazakov:2004qf} studied the so-called finite gap classical solutions for the $su(2)$ principal chiral model and expressed them in 
the elegant language of algebraic curves. 
Since the continuum limit of the Bethe ansatz equations leads also to algebraic curves, it became possible to compare directly the sigma model predictions with the spin chain predictions. The same method was used for the $sl(2)$ sector  \cite{Kazakov:2004nh} and later for the whole $psu(2,2|4)$ algebra \cite{Beisert:2005bm}.

The finite size corrections in the language of the spin chains correspond to loop corrections for the string sigma model. These were extensively studied \cite{Frolov:2004bh,Park:2005ji,Hernandez:2005nf,Beisert:2005mq,SchaferNameki:2005tn,Beisert:2005cw,Minahan:2005qj}, and shown to coincide, at least for the first few orders in the BMN coupling constant, where the order-of-limit problem does no show up yet.

The direct comparison between the algebraic curves of the sigma model and the Bethe equations made possible to conjecture the all-loop Bethe equations. First, Beisert, Dippel and Staudacher (BDS) \cite{Beisert:2004hm}
were able to conjecture the dispersion relation and a set of Bethe ansatz equations which reproduce succesfully some of the all-loop features. However, it became soon clear \cite{Arutyunov:2004vx} that a part was missing from the BDS conjecture, the so-called dressing factor. The BDS equations were soon extended to all sectors by Beisert and Staudacher \cite{Beisert:2005fw}, and the Beisert-Staudacher equations were rederived by Beisert \cite{Beisert:2005tm} by the assumption that the excitations of the spin chain are subject to a centrally extended $su(2|2)\times su(2|2)$ symmetry. Subsequently it was found \cite{Rej:2005qt} \cite{Beisert:2006qh} that these equations bear deep links, not yet entirely clarified, with the one-dimension  Hubbard model. The BDS dispersion relation was confirmed, from the string side, by Hofman and Maldacena \cite{Hofman:2006xt}, (see also \cite{Arutyunov:2006gs}), who coined the term {\it giant magnon} for the magnon excitations with momentum of the order of unity. The giant magnons correspond
to string excitations were the effect of the curvature of the sphere is manifest.  Arutyunov, Frolov and Zamaklar \cite{Arutyunov:2006yd} showed that the Bethe ansatz equations for the strings should be essentially the same as that for the spin chain, since the symmetry of the excitations in the string sigma model is the same as that of the magnons in the spin chain \cite{Arutyunov:2006ak}.

The dressing phase was first determined at leading order at strong coupling by Arutyunov, Frolov and Staudacher \cite{Arutyunov:2004vx}. A partial answer for the one-loop string corrections were obtained by Beisert and Tseytlin
\cite{Beisert:2005cw}, building on previous work \cite{Park:2005ji}, and the full one-loop  answer was obtained by Hernandez and Lopez \cite{Hernandez:2006tk}. Janik \cite{Janik:2006dc} used the $su(2|1)$ symmetry and the quantum group structure to derive the equivalent of the crossing equation obeyed by the dressing factor. A solution to the crossing equation was found by Beisert, Hernandez and Lopez (BHL) \cite{Beisert:2006ib}; this solution was shown by Beisert, Eden and Staudacher (BES) \cite{Beisert:2006ez}  to obey the right structure at weak coupling. The BHL/BES solution of the crossing equation allowed to check the fourth loop order in the cusp anomalous dimension, obtained by Bern {\it et al.} \cite{Bern:2006ew} and passed a number of other consistency checks. Dorey, Hofman and Maldacena \cite{Dorey:2007xn} checked that the BHL/BES solution has the expected set of singularities and provided an useful integral representation. Very recently, Arutyunov and Frolov \cite{Arutyunov:2009kf} verified that
the BHL/BES solution satisfies the crossing equation and Volin \cite{Volin:2009uv} showed, conversely, that the solution of the crossing equation with the minimal number of singularities in the physical strip {\it is} the BHL/BES solution. Finding the minimal solution of the crossing equation 
provided the interpolation between the weak coupling and strong coupling, at least for spin chains of large length. 

A considerable effort was spend to analyze the Bethe ansatz equations at strong coupling, especially concerning the
cusp anomalous dimension\footnote{The cusp anomalous dimension governs the renormalization of Wilson loops with cusps  \cite{Polyakov:1980ca} and controls the infrared behavior of the scattering amplitudes in gauge theories \cite{Korchemsky:1986fj}. They are ubiquitous observables in gauge theories, appearing among others in the logarithmic scaling of the anomalous dimensions of  operators with high Lorenz spin.}. An integral equation, the BES equation, giving the leading of the cusp anomalous dimension part at large magnon number (or Lorentz spin) was written down by Eden and Staudacher \cite{Eden:2006rx} and corrected with the dressing factor by Beisert, Eden and Staudacher \cite{Beisert:2006ez}. This equation was analyzed numerically by Benna {\it et al.} \cite{Benna:2006nd}, who confirmed the first two orders obtained previously with string theory techniques \cite{Gubser:2002tv} \cite{Frolov:2002av} and gave a prediction for the third order which was found later \cite{Roiban:2007jf,Roiban:2007dq}. Analytical results were much more difficult to obtain. The first order was reproduced by a series of works \cite{Kotikov:2006ts,Alday:2007qf,Kostov:2007kx,Beccaria:2007tk}, the next term was obtained by Casteill and Kristjansen \cite{Casteill:2007ct} and Belitsky  \cite{Belitsky:2007kf}, and the third term and a recursive procedure for the next terms 
by Basso, Korchemsky and Kotanski \cite{Basso:2007wd} and in \cite{Kostov:2008ax}. 
Another object extensively studied at strong coupling, from the point of view of the Bethe ansatz and the string theory, is the generalized scaling function which also appears in the study of the logarithmic scaling in the $sl(2)$ sector of the theory. An equation giving the generalized scaling function was written down by Freyhult, Rej and Staudacher (FRS) \cite{Freyhult:2007pz}, and rederived and the series expansion in the extra parameter $j$ was studied in \cite{Fioravanti:2008bh}. Basso and Korchemsky \cite{Basso:2008tx} showed that at small values of the parameter $j$ the FRS equation reduces to the equation for the energy of the $O(6)$ sigma model, confirming the relation with the $O(6)$ model identified previously by Alday and Maldacena \cite{Alday:2007mf}. At large values of the parameter $j$, of the order of the coupling constant $g$, the behavior of the generalized scaling function was computed from string theory by Roiban and Tseytlin  \cite{Roiban:2007ju}. The same quantity was computed form the Bethe ansatz, using different approaches,
by Gromov  \cite{Gromov:2008en}, Bajnok {\it et al.}  \cite{Bajnok:2008it} and Volin \cite{Volin:2008kd}, who found a mismatch in one of the coefficients with the string computation  \cite{Roiban:2007ju}, revisited and corrected in \cite{Giombi:2010fa}. 

Among the most recent developments are the study of finite size corrections and the relation between gluon amplitudes and integrability. The finite size corrections in AdS/CFT were considered first by  
Ambjorn, Janik and Kristjansen, \cite{Ambjorn:2005wa}, from the point of view of an integrable field theory. Janik and \L ukowsky \cite{Janik:2007wt} proposed to use the method initiated by L\"uscher \cite{Luscher:1983rk, Luscher:1986pf} to compute the finite size corrections.  This approach was successfully carried out for Konishi operator at four loops  \cite{Bajnok:2008bm} and at five loops \cite{Bajnok:2009vm}. The four loop result coincides coincides with the direct diagrammatic computations \cite{Fiamberti:2007rj,Velizhanin:2008jd}, while there is no five loop direct computation in the gauge theory. 

Another approach for computing finite size corrections is connected with the Thermodynamic Bethe Ansatz, and the underlying idea put forward by Alexei Zamolodchikov \cite{Zamolodchikov:1991et} is to exchange the role of the finite temperature with that of a finite length by a transformation similar to the modular transformation in 2d CFTs. In the context of AdS/CFT, this approach was suggested already in \cite{Ambjorn:2005wa} and a systematic implementation was initiated by Arutyunov and Frolov \cite{Arutyunov:2007tc}. 
Gromov, Kazakov and Vieira \cite{Gromov:2008gj} showed for
the case of the $su(2)$ principal chiral model that these equations, named the Y-system, can be used to determine the finite size corrections. When applied to AdS/CFT, this method allowed to reproduce the four loop \cite{Gromov:2009tv} and five loop \cite{Arutyunov:2010gb} \cite{Balog:2010xa} corrections for the Konishi operator previously obtained by the L\"uscher method. 

While most of the present developments related to integrability concern the determination of the conformal dimensions and the study of the dilatation operator as an integrable spin chain, there are signs that integrability
has deeper consequences on the structure of the supersymmetric $\CN=4$ gauge theory. 
It was discovered \cite{Anastasiou:2003kj}  that the four-gluon amplitude has a particular structure, in which the one-loop contribution exponentiates \cite{Bern:2005iz}. This property, checked at strong coupling by Alday and Maldacena \cite{Alday:2007hr}, does not hold for six or more gluons \cite{Bern:2008ap,Drummond:2008aq}. 
Sokatchev and his collaborators \cite{Drummond:2006rz} discovered that the integrals
which enter the expression of the amplitudes possess a {\it dual} conformal invariance. Shorthly after, it was discovered that the existence of this dual conformal invariance is the consequence of a duality between
the amplitudes and the Wilson loops constructed from light-like segments with cusps. This explains the appearance of the cusp anomalous dimension in the expression of the amplitudes given in \cite{Bern:2005iz}.
 At strong coupling, the relation between amplitudes and the Wilson loops with cusps appear via a T-duality \cite{Alday:2007hr}, which was extended to a fermionic T-duality in \cite{Berkovits:2008ic}. The action of this duality on the integrable structure of the $AdS_5\times S^5$ superstring was analyzed in \cite{Beisert:2008iq} and reviewed in \cite{Beisert:2009cs}. 

Drummond, Henn and Plefka \cite{Drummond:2009fd} proposed that the generators of the super-confor- mal symmetry, together with those of the dual super-conformal theory, constitute the generators of a Yangian.
This would imply the existence of an integrable structure of the amplitudes.  An integrable model for the high energy limit of the amplitudes was proposed by Lipatov \cite{Lipatov:2009nt}. 

Non-planar corrections to the dilatation operator, corresponding to interactions splitting and joining the spin chains, were investigated in \cite{Bellucci:2004ru} \cite{Casteill:2007td}.

The material is structured as follows: in section 2, the basic facts about the $\CN=4$ super Yang-Mills theory are presented, with an emphasis on the symmetry and a simple realization of the symmetry  in terms of oscillators. Section 2 contains also a presentation of the dilatation operator at one loop, together with the exact solution for its spectrum in terms as the spectrum of an integrable spin chain. Section 3 contains the definition of perturbative integrability for the dilatation operator and  the comparison with the Inozemtsev model. Section 4 is devoted to strings on the $AdS_5\times S^5$, its definition as a coset sigma model and the classical integrable structure, including the finite gap solution in terms of algebraic curves. In section 5 are presented the early results concerning the comparison between spin chain solutions and the string solutions  for strings on plane-wave background, as well as the comparison between the 
Bethe ansatz equations for spin chains in the $su(2)$ sector  and the finite gap solutions for the $su(2)$ principal chiral model. 
In section 6, the main results concerning the asymptotic all-loop Bethe equations are described, together with the solution for the dressing phase. Section 7 is devoted to the analysis of the Bethe ansatz equation at strong coupling, concerning the cusp anomalous dimension and the generalized scaling function. In section 8 we present the TBA approach for the  $su(2)$ principal chiral model
and then for AdS/CFT, including the derivation of the wrapping correction for the Konishi operator at four loops.
In section 9 we briefly discuss the recent developments concerning the finite-size corrections and the remarkable properties of the amplitudes.
 
In order to keep the text short  we have skipped some of the details. For a more in-depth presentation of different aspects, one can 
consult the original works or the available reviews, {\it e.g.} the reviews by Plefka \cite{Plefka:2003nb} \cite{Plefka:2005bk}, the PhD thesis of Beisert \cite{Beisert:2004ry} or the recent review on superstrings in $AdS_5\times S^5$  by Arutyunov and Frolov \cite{Arutyunov:2009ga}, as well as the special volume dedicated to the subject by the journal J. Phys A \cite{Tseytlin:2009zz}.

\section{The ${\cal N}=4$ super Yang-Mills theory}

\subsection{The action}
The action for the ${\cal N}=4$ gauge theory in four dimensions can be obtained by dimensional reduction from the ten
dimensional ${\cal N}=1$ gauge theory
\begin{equation}
S=\int d^{10}x\(\frac{1}{4}\Tr F_{MN}F^{MN}+\frac{1}{2}\, \Tr  \psi\, \Gamma^M \CD_M\,\psi\)
\end{equation}
where the covariant derivative is defined as
\begin{equation}
\CD_M=\p_M-i\gYM\, [A_M,\ ]
\end{equation}
and $\psi$ is a sixteen-component Majorana-Weyl spinor in ten dimensions. 
Upon dimensional reduction, six of ten components  of the gauge field become scalars,
while the sixteen-dimensional spinor decomposes into four copies of left and 
right two-component spinors in four dimensions
\begin{eqnarray}
A_M\;, \quad M=1,...,4 \quad &\to& \quad A_\mu \;, \quad \mu =0,...,3\ \\ \nonumber
A_M\;, \quad M=5,...,10 \quad &\to& \quad \Phi_i\;, \quad i=1,...,6\;\\ \nonumber
\psi_A\;,\quad A=1,...,16 \quad& \to& \quad \bar \psi^a_{\dot \al},\ \psi_{a,\al}\;, 
\quad a=1,...,4,\quad \al, \dot \al=1,2
\end{eqnarray}
The action for the four dimensional gauge theory becomes
\begin{eqnarray} 
\label{neq4}
S=\int d^{4}x\Big(\frac{1}{4}\Tr F_{\mu\nu}F^{\mu\nu}+\frac{1}{2} \Tr  \CD_\mu \Phi_i\, \CD^\mu \Phi^i
-\frac{\gYM^2}{4} \Tr [\Phi_i,\Phi_j][\Phi^i,\Phi^j]\\+
 \Tr \bar \psi^a\, \sigma^\mu \CD_\mu\,\psi_a-\frac{i\gYM}{2}\Tr \sigma_i^{ab}\psi_a [\Phi^i,\psi_b] -\frac{i\gYM}{2} \Tr \sigma^i_{ab}\bar \psi^a [\Phi_i,\bar \psi^b]\Big) \nonumber
\end{eqnarray}
where $\sigma^\mu$ and $\sigma^i$ are the chiral projections of the gamma matrices in four and six dimensions respectively\footnote{For a list of properties of the chiral projections of the gamma matrices, see \cite{Beisert:2004yq}.}.
The fields $A_\mu$,  $\Phi_i$ and $\psi_a$ form a supermultiplet.

\subsection{Symmetries}

The action (\ref{neq4}) is Poincar\'e invariant and scale invariant. The classical dimensions of the fields are
\begin{equation}
[A_\mu]=[\Phi_i]=1\quad [\psi_a]=\frac{3}{2}\;.
\end{equation}
The theory remains scale invariant upon quantization, the beta function being zero,
and the theory is conformally invariant at quantum level. The conformal group in four dimensions
is $SO(4,2)\simeq SU(2,2)$.
This group contains two $SU(2)$ components, with generators  $L^\al_\be$ and $\bar L^\dal_\dbe$, the dilatation operator, $D$,
the translations, $P_\mu$ and the special conformal transformations, $K_\mu$. The internal symmetry which
rotates the six scalars into one another, or the $R$ symmetry, is given by $SO(6)\simeq SU(4)$.
Taking into account the super-translations $Q^a_\al,\ \bar Q_{\dal a}$ and the super-version of the special
conformal transformations, $S^\al_a,\ \bar S^{\dal a}$, we obtain the total symmetry group as $PSU(2,2|4)$.
The structure of the generators can be schematically represented as
\begin{equation}
\left(\begin{array}{cc}L,\ \bar L, \  P, \ K,\  D & \ \ \  Q,\ \bar S \vspace{5pt} \\   \bar Q,\ S &\  R\end{array}\right)
\end{equation}
where the generators in the diagonal blocks are bosonic and the ones in the anti-diagonal blocks are fermionic.
The generators have a definite dimension, which is not modified by the radiative corrections
\begin{equation}
\label{classdim}
[D]=[L]=[\bar L]=[R]=0\;, \quad [P]=1\;, \ [K]=-1\;, \quad [Q]=1/2\;,\  [S]=-1/2\;.
\end{equation}

It will be useful in the following to represent the the fields in the spinorial notation, using the properties of the the chiral projections of the gamma matrices
\begin{equation}
{\cal D}_\mu=\sigma_\mu^{\dal \be}{\cal D}_{\dal \be}\;, \qquad \Phi_i =\sigma_i^{ab}\Phi_{ab}\;, \quad
{\rm etc}
\end{equation}

{\it Conformal dimensions and the dilatation operator}
\vskip .2cm

 We are going to consider gauge invariant, local operators, which are traces over the gauge group of products of operators, {\it e.g.}
 \begin{equation}
{\cal O}_{i_1\mu i_2...\al i_n}(x)= \Tr [\Phi_{i_1}(x)\CD_\mu \Phi_{i_2}(x)...\psi_\al (x) \Phi_{i_n}(x)]\;,
 \end{equation}
 Multiple trace operators occur as well, but in the planar limit we can restrain ourselves to the single
 trace operators.
 
 The operators organize in super-multiplets of the $PSU(2,2|4)$ symmetry.
 The operator with the lowest dimension in the multiplet is called a superconformal primary operator.
 
 The unitary representations of the superconformal algebra are labeled by the quantum numbers of the 
 bosonic subgroup
 \begin{eqnarray}
 SO(3,1)\times SO(1,1)\times SU(4)\\ \nonumber
 (s_1, s_2)\qquad\ \Delta(g)\qquad\  [r_1,r_2,r_3]
 \end{eqnarray}
 with $s_1$ and $s_2$ being half-integer, $\Delta(g)$ is the positive conformal dimension
 and $[r_1,r_2,r_3]$ are the (integer) Dynkin labels of $SU(4)$.
 
 The unitary representations of the superconformal group were classified by Dobrev and Petkova \cite{Dobrev:1985qv} \cite{Dobrev:1985vh}.
 There are three discrete series of representations, for which at least one of the generators $Q$ commutes with the superconformal primary operator. These are the BPS or short multiplets, and their dimension
 does not vary with the coupling constant. They correspond to {\it protected operators}. A fourth series
 of representations is continuous, and their dimension vary with the coupling constant. These are the 
 {\it non-BPS} operators, and computing $\Delta (g)$ is the main concern of the activity reported here.
 $\Delta (0)$ is the classical dimension of the field, which is the sum of classical dimensions of the components of a composite field.
 
\subsection{Oscillator representation of $PSU(2,2|4)$}
\label{symmetries}

It is sometimes useful to employ an oscillator realization of $PSU(2,2|4)$ \cite{Bars:1982ep}, which is valid for the free theory, $\gYM=0$. Consider two sets of bosonic oscillators, $(a^\al, a_\al^\dagger)$ and  $(b^\dal, b_\dal^\dagger)$, with $\al,\dal=1,2$ and a set of 
fermionic oscillators, $(c^a, c_a^\dagger)$ with $a=1,2,3,4$, with standard commutation relations
\begin{equation}
[a^\al, a_\beta^\dagger]=\delta^\al_\beta\;, \qquad [b^\dal, b_\dbe^\dagger]=\delta^\dal_\dbe
\;, \qquad \{c^a, c_b^\dagger\}=\delta^a_b\;.
\end{equation}
The fundamental representation  of $SU(2,2|4)$ and its conjugate can be realized in the space spanned by the eight oscillators $(A^A)=(a,b^\dagger,c)$
and $(A^\dagger_A)=(a^\dagger, -b,c^\dagger)$ respectively. The particle-hole transformation on the
 $b$ oscillators is related to the non-compact nature of the group. The generators of the algebra are,
 in this representation, given by the bilinears $A^\dagger_A A^B$ and they act on the states by commutator.  The $su(2)\times su(2)\times su(4)$
 part is realized simply as 
 \begin{eqnarray}
L_\al^\be&=& a^\dagger_\al a^\be-\frac{1}{2}a^\dagger_\ga a^\ga\, \de_\al^\be\;, \\
\bar L_\dal^\dbe&=& b^\dagger_\dal b^\dbe-\frac{1}{2}b^\dagger_\dga b^\dga \,\de_\dal^\dbe\;,\\
R_a^b&=&c^\dagger_a c^b-\frac{1}{4}c^\dagger_d c^d\, \de_a^b\;.
 \end{eqnarray}
 The dilatation operator is given by
 \begin{equation}
 \label{dilzero}
 D_0=1+\frac{1}{2}(n_a+n_b)\;,
 \end{equation}
 while the central charge, which should be zero for $PSU(2,2|4)$, is given by
 \begin{equation}
 C=1-\frac{1}{2}(n_a-n_b+n_c)\;,
 \end{equation}
 which means that for the representation we consider, we have
 $n_c+n_a-n_b=2\;.$
 The off-diagonal generators are 
\begin{eqnarray}
P_{\al \dbe}&=& a^\dagger_\al b^\dagger_\dbe\;, \qquad K^{\al \dbe}=a^\al b^\dbe\;,\\
Q^a_\al&=&a^\dagger_\al c^a\;,\qquad S_a^\al=c^\dagger_a a^\al\;,\\
\bar Q_{\dal a}&=&b^\dagger_\dal c^\dagger_a\;,\qquad \bar S^{\dal a}=b^\dal c^a \;.
 \end{eqnarray}
 On this representation, it is easy to check the commutation relations of the $psu(2,2|4)$ algebra
\begin{eqnarray}
&\ &[S_a^\al, P_{\ga \dbe}]=\de^\al_\ga \;{\bar Q}_{\dbe a} \qquad [K^{\al \dbe}, {\bar Q}_{\dga a}] =
\de_\dga^\dbe\; S^\al_a \\
&\ &[{\bar S}^{\dal a}, P_{\ga \dbe}] =\de^\dal_\dbe\;  Q_\ga^a  \qquad [K^{\al \dbe},  Q_\ga^a] =
\de^\al_\ga\; {\bar S}^{\dbe a}\\
&\ &\{ S^{\be}_b, {\bar S}^{\dal a} \} =\de^a_b \;  K^{\be \dal}  \qquad \{ Q_{\be}^b, {\bar Q}_{\dal a}\} =\de^b_a \;  P_{\be \dal} 
\end{eqnarray}
and
\begin{eqnarray}
&\ &[K^{\al \dbe}, P_{\be \dal}]=\de^\al_\be \;{\bar L}^\dbe_\dal +\de^\dbe_\dal \;L^\al_\be+
\de^\al_\be\, \de^\dbe_\dal \; D \\
&\ &\{ S^{\al}_a, Q_\be^b \} =\de^a_b\; L^\al_\be+ \de^\al_\be \; R^b_a+\frac{1}{2}\de^a_b\,\de^\al_\be\;  D
\\
&\ &\{ \bar S^{\dal a}, \bar Q_{\dbe b} \} =\de^a_b\; \bar L^\dal_\dbe- \de^\dal_\dbe \; R^b_a+\frac{1}{2}\de^a_b\,\de^\dal_\dbe\;  D
\end{eqnarray}

The different ${\cal N}=4$ operators can be realized as states in an infinite dimensional $PSU(2,2|4)$ representation as follows:

a) the bosons $Z=\Phi_1+i\Phi_2$,   $X=\Phi_3+i\Phi_4$ and $Y=\Phi_5+i\Phi_6$
by states with two fermions
$$ Z\sim c_3^\dagger c_4^\dagger |0\rangle\;, \quad  X\sim c_3^\dagger c_2^\dagger |0\rangle\;, \quad
 Y\sim c_3^\dagger c_1^\dagger |0\rangle\;.$$

b) the fermions
$$\psi_{a \al}\sim a^\dagger_\al c^\dagger_a |0\rangle\;, \quad  \bar \psi^a_{ \dal} \sim \e^{abcd}b^\dagger_\dal c^\dagger_b c^\dagger_c c^\dagger_d |0\rangle $$

c) the covariant derivatives
$${\cal D}_{\al\dbe} \sim a^\dagger_\al b^\dagger_\dbe $$

One can make a particle-hole transformation on the fermions $c^3$ and $c^4$
\begin{equation}
d^\dagger_1=c^4\;, \quad d^\dagger_2=c^3\;,
\end{equation}
such that the highest weight state, associated to the operator $Z$, is the ``empty'' state, annihilated by all the annihilation operators $a^i$, $b^i$, $c^i$, $d^i$ with $i=1,2$. 
The reference state $Z$ is manifestly invariant under  $su(2)\times su(2)$, and in fact it is invariant
under the action of $su(2|2)\times su(2|2)$ generated by the oscillators $(a,c)$ and $(b,d)$ respectively.

\vskip .2cm

{\it Dynkin diagrams and the oscillator representation}

\vskip .2cm

There is a close relation between the $sl(N|M)$ root system, as described in Appendix \ref{appa} and the oscillator representation.

For each $\e$ ($\delta$) direction defined in Appendix A one can choose a bosonic (fermionic) oscillator. The bosonic roots will be then represented by bilinears in bosons or of fermions, while the fermionic root will contain one boson and one fermion.
Depending on the real form one wants to choose for the bosonic part,  one can make a particle-hole transformation for a set of the bosons. For $su(2,2|4)$ one recovers the construction from the previous section.

\begin{figure}
  \centering
    \reflectbox{%
    \includegraphics[width=0.5\textwidth]{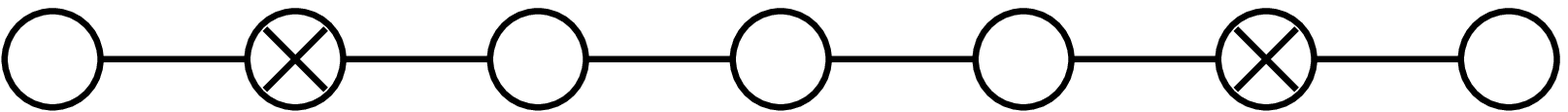}}
    \vskip 1cm 
     \reflectbox{%
        \includegraphics[width=0.5\textwidth]{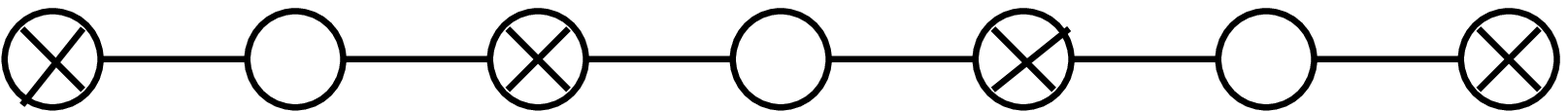}}
  \caption{Two Dynkin diagrams for $psu(2,2|4)$.}
  \label{twdynkin}
\end{figure}

Among the different  Dynkin diagrams for $su(2,2|4)$, the two shown in  the figure \ref{twdynkin} will be especially useful for writing the Bethe ansatz equation. They show the explicit decomposition with respect to $su(2|2)\otimes su(2|2)$, algebra which can be obtained upon deleting the middle node of the two diagrams.

\subsection{The dilatation operator as the Hamiltonian of an integrable spin chain}

The basis of local operators can be written as a collection of words of arbitrary length made by
the letters corresponding to basic operators
\begin{equation}
\Tr (ZZX{\cal D}_3Y\psi_1\psi_2\ldots{\cal  D}_0XY)(x)\;.
\end{equation}
taken at a particular point $x$ in the four-dimensional space-time.
Each basic operator can be interpreted as spin state associated to a site of a periodic spin chain. 
The dilatation operator $D$
maps these spin states into one another. The eigenvalues of the dilatation operator
are the conformal dimensions $\Delta^A(g)$ of the operators $\CO^A(x)$, {\it i.e.}
\begin{equation}
\label{vev}
\vev{\bar \CO^A (x)\CO^A(y)}\sim \frac{1}{|x-y|^{2\Delta^A(g)}}
\end{equation}
The appearance of non-trivial dimensions is a consequence of renormalization. Although the beta function of the theory is zero, there are singularities at higher loops, which need to be removed by
renormalization. For example, the correlation function (\ref{vev}) can be computed in perturbation theory
as 
\begin{equation}
\vev{\bar \CO^A (x)\CO^A(y)}\sim
 \frac{1}{|x-y|^{2\Delta_0^A}}\(1-2g^2\Delta_2^A\ln(\Lambda|x-y|)+\ldots\)
\end{equation}
such that the perturbative expansion of the conformal dimension is given by
\begin{equation}
\Delta^A(g)=\Delta_0^A+g^2\Delta_2^A+\ldots=\Delta_0^A+\delta\Delta^A(g)\;.
\end{equation}
Here we had to introduce the scale $\Lambda$ which plays the role of the UV cutoff.
Alternatively, the anomalous part of dilatation operator $\delta D(g)$ can be read off the 
wave function renormalization for the operators
\begin{equation}
\CO^A_{{\rm ren}}(x)=Z^A_{\ B}\;\CO^B(x)=(e^{\delta D(g) \ln \Lambda})^A_{\ B}\;\CO^B(x)
\end{equation}

The dilatation operator can be therefore computed in perturbation and 
it can be interpreted as a spin Hamiltonian which maps a spin state into another spin state.
Because of the cyclicity of the trace, only the states which are invariant by translation should be considered.
The problem of computing the (anomalous) dimensions is then reduced to diagonalizing the Hamiltonian of the spin chain.

 Let us remind that the dimensions of the symmetry currents $J(g)$ should  not be modified in perturbation theory, so that
 \begin{equation}
[D(g),J(g)]=[D_0,J(g)]={\rm dim}(J) J(g)\;,
 \end{equation}
where $D_0$ is the classical part of the dilatation operator and the dimensions ${\rm dim}(J)$ are given by equation (\ref{classdim}). This means in particular that
the anomalous part of the dilatation operator $D(g)-D_0$ commutes with all the generators of $psu(2,2|4)$
 \begin{equation}
 \label{consham}
[D(g)-D_0,J(g)]=0\;
 \end{equation}
and therefore it can be considered as a $psu(2,2|4)$ symmetric Hamiltonian. In the same time, $D(g)$
is an element of the $psu(2,2|4)$ algebra, a feature that distinguishes it from the usual spin chains. 

Another consequence of equation (\ref{consham}) is that the one-loop part of the dilatation operator is 
invariant under the classical part of the superconformal symmetry
 \begin{equation}
[D_2,J_0]=0\;.
 \end{equation}
 
We notice that a change of sign $g\to -g$ can be absorbed in the ${\cal N}=4$ action by a sign change of the scalar fields $\Phi_i\to -\Phi_i$. Such a change should not modify the conformal dimensions $\Delta(g)$ , so we conclude that they depend only on $g^2$.
  \begin{equation}
\Delta(g)=\sum_{k\geq 0}g^{2k}\Delta_{2k}\;.
 \end{equation}
On the other hand, the dilatation operator itself may contain odd powers of $g$.

The mixing matrix at one loop was first computed in \cite{Berenstein:2002jq}, and it was noticed by Minahan and Zarembo \cite{Minahan:2002ve} that, for the $so(6)$ sector spanned by the $6$ scalar
bosons, the corresponding Hamiltonian is the integrable $so(6)$ spin chain, solved by Reshetikhin \cite{Reshetikhin:1983vw} \cite{Reshetikhin:1986vd}. The full one-loop Hamiltonian was found by 
Beisert \cite{Beisert:2003jj} and the full one-loop Bethe ansatz was written in \cite{Beisert:2003yb}.

\vskip .2cm

{\it The mixing matrix and the sectors}

\vskip .2cm

The spin chain can be represented in terms of oscillators, with a copy of oscillators at each site of the spin chain. While at $g=0$ there is no interaction between sites and the dilatation operator is just the sum classical dimensions at each site (\ref{dilzero}), when the interaction is turned on, the dilatation operator has a non-trivial expression in terms of the creation/annihilation operators. 
Some restrictions may be imposed on the total oscillator content of the states, and some of these restrictions are conserved by the full dilatation operator. The dilatation operator is a Cartan generator for the $psu(2,2|4)$ algebra, and therefore its action cannot change the $psu(2,2|4)$ charges of the states on which it acts, charges that are related to the oscillator content of the states. The restrictions give rise to sectors which are closed in perturbation theory.
Let us remind that any state in the spin chain has to satisfy the constraint
\begin{equation}
\label{pscons}
n_a+n_c-n_b=2L\;.
\end{equation}
\begin{itemize}
\item
The BPS states: in the particular case
\begin{equation}
n_{c_3}=n_{c_4}=L\;, \quad n_{a_i}=n_{b_i}=0
\end{equation}
the constraint (\ref{pscons}) is saturated and the action of the anomalous part of the dilatation operator is zero. They correspond to states $\ket{ZZZZ...ZZZ}$. These are the BPS states protected by supersymmetry, together with any state obtained from it by a$SO(6)$ rotation. 
\item 
The $su(2)$ sector: 
\begin{equation}
n_{a_i}=n_{b_i}=0\quad {\rm and} \quad  n_{c_3}=n_{c_4}+n_{c_2}=L
\end{equation}
These states are of the type $\ket{ZXXZ...ZZZ}$.
 \item 
The $sl(2)$ sector: 
\begin{equation}
n_{a_1}=n_{b_1}=M, \quad n_{a_2}=n_{b_2}=0,\quad \quad {\rm and} \quad  n_{c_3}=n_{c_4}=L;,
\end{equation}
with states are of the type $\ket{\CD^M ZZ...ZZZ}$. 
\item 
The $su(1|1)$ sector: 
\begin{equation}
n_{a_2}=n_{b_i}=0 \quad {\rm and} \quad  n_{c_3}=n_{a_1}+n_{c_4}=L.
\end{equation}
The states are of the type $\ket{Z\psi_1\psi_1 Z...ZZZ}$ with $\psi_1\equiv \psi_{31}$.
\item 
The $su(2|1)$ sector: 
\begin{equation}
n_{a_2}=n_{b_i}=0 \quad {\rm and} \quad  n_{c_3}=n_{a_1}+n_{c_4}+n_{c_2}=L
\end{equation}
spanned by states  of the type $\ket{Z\psi_1\psi_1 Z...ZXZ}$.
\item 
The $su(3|2)$ sector: 
\begin{equation}
n_{b_i}=0 \quad {\rm and} \quad  n_{c_3}=n_{a_1}+n_{a_2}+n_{c_4}+n_{c_2}+n_{c_1}=L
\end{equation}
spanned by states  of the type $\ket{Z\psi_1\psi_2 Z...ZXYZ}$ with $\psi_1\equiv \psi_{31}$ and $\psi_2\equiv \psi_{32}$.
The transformation $\ket{...XYZ...}\to\ket{...\psi_1\psi_2...}$ allowed by diagrammatics corresponds to $c_1^\dagger c_2^\dagger c_3^\dagger c_4^\dagger\to a_1^\dagger a_2^\dagger$. This transformation conserves the $psu(2,2|4)$ charges but reduces the length by one unit, $L\to L-1$. This sector is the smallest sector with variable length.
\item 
The $psu(1,1|2)$ sector: 
\begin{equation}
n_{a_2}=n_{b_2}=0 \quad {\rm and} \quad  n_{c_3}=n_{a_1}-n_{b_1}+n_{c_4}+n_{c_2}=L\;.
\end{equation}
The states  are of the type $\ket{\CD^M Z\psi_1\bar \psi_1 Z...ZXZ}$ with $\bar \psi_1\equiv \bar \psi^1_1$.
\end{itemize}

\subsection{Bethe ansatz solution for the Heisenberg model}
\label{BAHeis}

When reduced to the operators containing only
the two complex combinations of the scalar bosons, for example $Z$ and $X$, the one-loop dilatation operator reduces to the Heisenberg Hamiltonian
\begin{equation}
\label{hamheis}
H_2^{su(2)}=2g^2\sum_{l=1}^{L}(1-P_{l,l+1})\;,
\end{equation}
where $1$ is the identity operator and $P_{l,l+1}$ permutes the states at sites $l$ and $l+1$.

The solution for the Hamiltonian (\ref{hamheis}) can be found by the so-called coordinate Bethe ansatz, which was originally employed by Bethe \cite{Bethe:1931hc}. One starts with a reference state, where all the spins are in the "up"state $\ket{\uparrow}$. In terms of the ${\cal N}=4$ SYM, one starts with a trace composed uniquely from $Z$ operators.
 \begin{equation}
\ket{\Omega}=\ket{\uparrow \uparrow \uparrow \uparrow \ldots \uparrow }
\end{equation}
Starting with this state, one can generate the whole  Hilbert space by reversing an arbitrary number 
of spins. A reversed spin 
$$\ket{\downarrow}=\sigma^-\ket{\uparrow}$$
will be called a {\it magnon}. Since the Hamiltonian (\ref{hamheis}) conserves the number of reversed spins, it can be diagonalized  on a space with fixed number of magnons.
Consider first the one-magnon states
 \begin{equation}
\Psi(p)=\sum_{k=1}^L e^{ipk}\sigma^-_k \ket{\Omega}
\end{equation}
which are eigenstates of (\ref{hamheis}) if $p=2\pi n/L$, $n\in \MZ$, with eigenvalue
 \begin{equation}
E(p)=2g^2(2-e^{ip}-e^{-ip})=8g^2\sin^2p/2\;.
\end{equation}
For a state with two magnons, the eigenfunctions are found with the following {\it ansatz}
 \begin{eqnarray}
 \label{Banz}
\Psi(p_1,p_2)&=&\sum_{1\leq k_1<k_2\leq L} \left[ A(p_1,p_2)e^{i(p_1k_1+p_2k_2)} +A(p_2,p_1)e^{i(p_2k_1+p_1k_2)}\right]\sigma^-_{k_1}\sigma^-_{k_2} \ket{\Omega}\\
&=&\sum_{1\leq k_2<k_1\leq L}\left[A(p_1,p_2)e^{i(p_1k_2+p_2k_1)}+A(p_2,p_1)e^{i(p_2k_2+p_1k_1)}\right]\sigma^-_{k_1}\sigma^-_{k_2} \ket{\Omega}\;. \nonumber
\end{eqnarray}
This ansatz is clearly valid when $k_1\ll k_2$, since the action of the hamiltonian (\ref{hamheis})
couples only two nearby sites.  The condition that it is valid everywhere determines the ratio of the two coefficients $A(p_2,p_1)$ and $A(p_1,p_2)$ which defines the scattering matrix of two magnons
\begin{equation}
\label{twomags}
S(p_2,p_1)\equiv \frac{A(p_2,p_1)}{A(p_1,p_2)}=-\frac{e^{ip_1+ip_2}-2e^{ip_2}+1}{e^{ip_1+ip_2}-2e^{ip_1}+1}
\end{equation}
while the invariance under translations which is implicit on the summation imposes that  
\begin{equation}
\label{scatmag}
e^{ip_2L}=S(p_2,p_1)\;, \quad e^{ip_1L}=S(p_1,p_2)\;.
\end{equation}
The equations (\ref{scatmag}) are valid for any spin model with local interaction. What is remarcable
about the model (\ref{hamheis}) is that an ansatz of the type (\ref{Banz}) is valid for any number $M$ of magnons. 
The wave function in a different chamber $k_{P(1)}<k_{P(2)}<\ldots<k_{P(M)}$ can be obtained from the one
in $k_1<k_2<\ldots<k_M$ just by the inverse permutation of the momenta $p_1, p_2,\ldots p_M \to p_{P^{-1}(1)}, p_{P^{-1}(2)},\ldots p_{P^{-1}(M)}$. Since any permutation $P$ can be written as a product of transpositions, the elementary object is the two magnon scattering matrix (\ref{twomags}).
The magnon momenta are determined by the equations
\begin{equation}
\label{BAE}
e^{ip_nL}=\prod_{m\neq n=1}^M S(p_n,p_m)\;, \quad n=1,\ldots M
\end{equation}
The property of factorized scattering is related, at least for the systems with a finite number of degrees of freedom, to the existence of a number of conserved quantities equal to the number of degrees of freedom, that is, to integrability.

A different way to obtain the Bethe ansatz equations (\ref{BAE}), as well as the conserved quantities, is via the algebraic Bethe ansatz. For a pedagogical review of different aspects of the algebraic Bethe ansatz, see \cite{faddeev}. One starts with a matrix  $R(u)$ acting on $V\otimes V$, with $V=\MC^2$
which satisfies the Yang-Baxter equation
\begin{equation}
\label{YBE}
R_{12}(u-v)R_{13}(u)R_{23}(v)=R_{23}(v)R_{13}(u)R_{12}(u-v)
\end{equation}
where the subscript indicates the space where the matrix act.
In the $su(2)$ case, such a solution is provided by
\begin{equation}
R_{12}(u)=u+iP_{12}
\end{equation}
where $P_{12}$ is the permutation operator. For $u=\pm i$ the matrix $R(u)$ is proportional to a projector.
Starting with the $R$ matrix, one can construct the monodromy matrix $T_0(u)$ by
\begin{equation}
T_0(u)={\rm L}_1(u){\rm L}_2(u)\ldots {\rm L}_L(u) \;, \quad {\rm with}\quad {\rm L}_n(u)\equiv R_{0n}(u)
\end{equation}
which satisfies the Yang-Baxter equation in the form 
\begin{equation}
\label{YBET}
R_{00'}(u-v)\,T_{0}(u)\,T_{0'}(v)=T_{0'}(v)\,T_{0}(u)\,R_{00'}(u-v).
\end{equation}
A consequence of the equation (\ref{YBET}) is that the trace of the monodromy matrix, called the transfer matrix, commute with itself for different values of the spectral parameter $u$
\begin{equation}
\label{tcomm}
[T(u), T(v)]=0\;, \qquad T(u)\equiv \Tr _0\, T_0(u)\;,
\end{equation}
such that the transfer matrix $T(u)$ generates the conserved charges.  A special point is $u=0$, where the the conserved charges generated by the logarithm of transfer matrix are local, {\it e.g.}
\begin{equation}
\frac{1}{2g^2}(L-H_2)=i\frac{d}{du}\ln T(u)\vert_{u=0}\;.
\end{equation}
The common eigenvectors of the conserved quantities can be constructed from the reference vector $\ket{\Omega}$ as follows. The monodromy matrix can be written as
\begin{equation}
T_0(u)=\left(\begin{array}{cc}A(u) & B(u) \\C(u) & D(u)\end{array}\right)
\end{equation}
The conserved quantities are generated by
\begin{equation}
\label{consquant}
T(u)=A(u) + D(u) 
\end{equation}
while the eigenvectors are build as
\begin{equation}
\label{vectABA}
\ket{u_1,u_2,\ldots u_M}=B(u_1) B(u_2)\ldots B(u_M)\ket{\Omega}\;.
\end{equation}
The vectors (\ref{vectABA}) are eigenvectors of the operators (\ref{consquant}) if the rapidities $u_1,u_2,\ldots u_M$ satisfy the equations
\begin{equation}
\label{AlgBA}
\(\frac{u_k+i/2}{u_k-i/2}\)^L=\prod_{l\neq k=1}^M \frac{u_k-u_l+i}{u_k-u_l-i}
\end{equation}
which are identical with the set of equations (\ref{BAE}) upon the identification
\begin{equation}
\label{specpar}
u=\frac{1}{2}\cot\frac{p}{2}\;.
\end{equation}
The energy is given, in these variables, by 
\begin{equation}
E_2^{su(2)}=2g^2\sum_{k=1}^M\frac{1}{u_k^2+1/4}\;.
\end{equation}

\subsection{The $so(6)$ sector at one loop}

Minahan and Zarembo computed the one loop dilatation operator for all traces involving operators
in the $so(6)$ sector\footnote{At higher loop order this sector is not closed and it mixes with sectors containing fermions.}, that is, involving any of the bosonic operators $\Phi_{i}$ with $i=1,\ldots,6$,
\begin{equation}
\Tr\(\Phi_{i_1}\Phi_{i_2}\ldots \Phi_{i_L}\)(x)\;.
\end{equation}
The dilatation operator acts as a Hamiltonian for a $so(6)$ spin chain with the spins in the fundamental representation,
\begin{equation}
\label{so6oneloop}
H_2^{so(6)}=g^2\sum_{l=1}^L\(K_{l,l+1}+2-2P_{l,l+1}\)
\end{equation}
where the trace operator $K_{l,l+1}$ contracts the indices on the nearby sites $l$ and $l+1$
\begin{equation}
K_{i_l,i_{l+1}}^{j_l,j_{l+1}}=\delta_{i_l,i_{l+1}}\delta^{j_l,j_{l+1}}
\end{equation}
while the permutation operator  $P_{l,l+1}$ permutes the indices at the sites $l$ and $l+1$
\begin{equation}
P_{i_l,i_{l+1}}^{j_l,j_{l+1}}=\delta_{i_l}^{j_{l+1}}\delta^{j_l,}_{i_{l+1}}\;.
\end{equation}
This spin chain is integrable and it solution was obtained by Reshetikhin \cite{Reshetikhin:1986vd}.
It also can be obtained \cite{deVega:1987sh} by the method of nested Bethe ansatz, proposed by Kulish and Reshetikhin \cite{Kulish:1983rd}. 
The algebraic Bethe ansatz method from the previous section can be extended to algebras of higher rank and it associates a set of creation operators $B^{(r)}(u_r)$ with  each node $r$ of the Dynkin diagram. It is called {\it nested} because at each step of the procedure the rank of the algebra is reduced. 
The Hamiltonian (\ref{so6oneloop}) is diagonalized in terms of the rapidities satisfying the following system of nested Bethe ansatz equations
\begin{eqnarray}
\label{NBA}
1&=&\prod_{l\neq k}^{M_1} \frac{u_{1,k}-u_{1,l}+i}{u_{1,k}-u_{1,l}-i}\;
\prod_{l\neq k}^{M_2} \frac{u_{1,k}-u_{2,l}-i/2}{u_{1,k}-u_{2,l}+i/2}\\ \nonumber
\(\frac{u_{2,k}+i/2}{u_{2,k}-i/2}\)^L&=&\prod_{l\neq k}^{M_2} \frac{u_{2,k}-u_{2,l}+i}{u_{2,k}-u_{2,l}-i}\;
\prod_{l\neq k}^{M_1} \frac{u_{2,k}-u_{1,l}-i/2}{u_{2,k}-u_{1,l}+i/2}\;
\prod_{l\neq k}^{M_3} \frac{u_{2,k}-u_{3,l}-i/2}{u_{2,k}-u_{3,l}+i/2}\\ \nonumber
1&=&\prod_{l\neq k}^{M_3} \frac{u_{3,k}-u_{3,l}+i}{u_{3,k}-u_{3,l}-i}\;
\prod_{l\neq k}^{M_2} \frac{u_{3,k}-u_{2,l}-i/2}{u_{3,k}-u_{2,l}+i/2}\;,
\end{eqnarray}
The eigenvalues for the energy are given by
\begin{equation}
E_2^{so(6)}=2g^2\sum_{i=1}^{M_2}\frac{1}{u_{2,i}^2+1/4}\;.
\end{equation}
The $su(2)$ Bethe equations can be obtained from (\ref{NBA}) by restricting $M_1=M_3=0$ and $M_2=M$.
Due to the invariance of the trace which defines the local operators in the gauge theory, one has to retain only the states of the spin chain with zero total momentum, which are therefore invariant by translation. This condition can be reformulated as
\begin{equation}
\prod_{k=1}^{M_2}\frac{u_{2,k}+i/2}{u_{2,k}-i/2}=1\;,
\end{equation}
and has to be supplemented to the equations (\ref{NBA}).

The structure of the nested Bethe ansatz (\ref{NBA}) is closely related to the data from the Dynkin diagram. The equations were extended to any simple Lie algebra in any representation by Ogievetsky and Wiegmann  \cite{Ogievetsky:1986hu}. For any simple root $\alpha_r$ in the Dynkin diagram there is a set of rapidities $u_{r,k}$, with $k=1\ldots M_r$. These rapidities obey the set of equations
\begin{equation}
\label{GNBA}
\(\frac{u_{r,k}+i \al_r\cdot w/2}{u_{r,k}-i\al_r\cdot w/2}\)^L=
\prod_{r'}\prod_{l}^{M_{r'}} \frac{u_{r,k}-u_{r',l}+i\al_r\cdot \al_{r'}/2}{u_{r,k}-u_{r',l}-i\al_r\cdot \al_{r'}/2}\;\;
\end{equation}
where $w$ is the vector characterizing the representation, encoding the Dynkin labels  of the highest weight vector.
The corresponding energy is given, up to a multiplicative and additive constant, by
\begin{equation}
E=\sum_{r}\sum_{k=1}^{M_r}\(\frac{i}{u_{r,k}+i \al_r\cdot w/2}-\frac{i}{u_{r,k}-i \al_r\cdot w/2}\)\;.
\end{equation}

\begin{figure}
  \centering
        \includegraphics[width=0.2
        \textwidth]{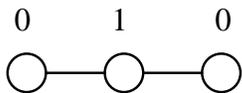}
  \caption{The Dynkin diagram and the Dynkin labels $\al_r\cdot w$ corresponding to the $so(6)$ Bethe ansatz (\ref{NBA}).}
  \label{sodynkin}
\end{figure}

The Bethe equation (\ref{NBA}) can be alternatively seen as equations for vector representation $so(6)$ with simple roots
$ \e_1-\e_2,\ \e_2-\e_3,\ \e_2+\e_3$ and weight $w=\e_1$, or for the antisymmetric representation
of $su(4)$ with roots $\e_1-\e_2,\ \e_2-\e_3,\ \e_3-\e_4$ and weight $w=\e_1+\e_2$. The corresponding Dynkin diagram is represented in figure \ref{sodynkin}.

\subsection{The full $psu(2,2|4)$ Bethe ansatz at one loop}
\label{fulloneloop}

\begin{figure}
  \centering
    \reflectbox{%
    \includegraphics[width=0.7\textwidth]{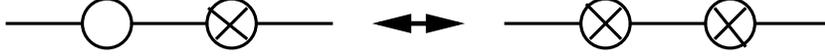}}
  \caption{Duality transformation on a piece of Dynkin diagram . }
  \label{fig:dualDyn}
\end{figure}

The action of the one loop dilatation operator for the whole $psu(2,2|4)$ states was computed by Beisert
\cite{Beisert:2003jj}. The main ingredient is the computation of the one-loop dilatation operator in the $sl(2)$ sector, which then can be lifted to an action on the whole $psu(2,2|4)$.
In terms of the $\CN =4$ fields, the $sl(2)$ sector corresponds to one of the complex scalar fields, say $Z$,
acted upon by an arbitrary numbers of covariant derivatives $\CD$. The corresponding representation
has weight $-1$ or spin $s=-1/2$, and it is a infinite dimensional heighest weight representation of $sl(2)$.
The $sl(2)$ Hamiltonian was determined by Beisert, both from  diagramatics \cite{Beisert:2003jj} and from symmetry consideration \cite{Beisert:2004yq}
\begin{equation}
 \label{hamsl2}
H_2^{sl(2)}=2g^2\sum_{i=1}^L\sum_{j=0}^\infty 2h(j)\;\CP_{i,i+1}^{j}\;, 
\end{equation}
where $\CP_{i,i+1}^{j}$  is the projector onto the spin $-1-j$  representation appearing in the tensor product of two spin $-1/2$ representations and the harmonic sum $h(j)$ is defined by
\begin{equation}
h(j)=\sum_{k=1}^j\frac{1}{k}=\psi(j+1)-\psi(1)\;.
\end{equation}
where $\psi(x)=d/dx \ln \Gamma(x)$.
This is the $s=-1/2$ case of the integrable system constructed in \cite{Kulish:1981gi} and explained in \cite{faddeev}.
The Hamiltonian (\ref{hamsl2}) is diagonalized by the following Bethe ansatz
\begin{equation}
\label{BEsl2}
\(\frac{u_k-i/2}{u_k+i/2}\)^L=\prod_{l\neq k}^{M} \frac{u_k-u_l+i}{u_{k}-u_{l}-i}\;.
\end{equation}
Finally, the $psu(2,2|4)$ Hamiltonian at one loop can be diagonalized by the following Bethe ansatz 
\begin{eqnarray}
\label{FNBA}
1&=&\prod_{l=1}^{M_2} \frac{u_{1,k}-u_{2,l}+i/2}{u_{1,k}-u_{2,l}-i/2}\\ \nonumber
1&=&\prod_{l\neq k}^{M_2} \frac{u_{2,k}-u_{2,l}-i}{u_{2,k}-u_{2,l}+i}\;
\prod_{l=1}^{M_1} \frac{u_{2,k}-u_{1,l}+i/2}{u_{2,k}-u_{1,l}-i/2}\;
\prod_{l=1}^{M_3} \frac{u_{2,k}-u_{3,l}+i/2}{u_{2,k}-u_{3,l}-i/2}\\ \nonumber
1&=&
\prod_{l=1}^{M_4} \frac{u_{3,k}-u_{4,l}-i/2}{u_{3,k}-u_{4,l}+i/2}\;
\prod_{l=1}^{M_2} \frac{u_{3,k}-u_{2,l}+i/2}{u_{3,k}-u_{2,l}-i/2}\\ \nonumber
\(\frac{u_{4,k}+i/2}{u_{4,k}-i/2}\)^L&=&\prod_{l\neq k}^{M_2} \frac{u_{4,k}-u_{4,l}+i}{u_{4,k}-u_{4,l}-i}\;
\prod_{l=1}^{M_3} \frac{u_{4,k}-u_{3,l}-i/2}{u_{4,k}-u_{3,l}+i/2}\;
\prod_{l=1}^{M_5} \frac{u_{4,k}-u_{5,l}-i/2}{u_{4,k}-u_{5,l}+i/2}\\
 \nonumber
1&=&
\prod_{l=1}^{M_4} \frac{u_{5,k}-u_{4,l}-i/2}{u_{5,k}-u_{4,l}+i/2}\;
\prod_{l=1}^{M_6} \frac{u_{5,k}-u_{6,l}+i/2}{u_{5,k}-u_{6,l}-i/2}\\ \nonumber
\nonumber
1&=&\prod_{l\neq k}^{M_6} \frac{u_{6,k}-u_{6,l}-i}{u_{6,k}-u_{6,l}+i}\;
\prod_{l=1}^{M_7} \frac{u_{6,k}-u_{7,l}+i/2}{u_{6,k}-u_{7,l}-i/2}\;
\prod_{l=1}^{M_5} \frac{u_{6,k}-u_{5,l}+i/2}{u_{6,k}-u_{5,l}-i/2}\\ \nonumber
1&=&\prod_{l=1}^{M_6} \frac{u_{7,k}-u_{6,l}+i/2}{u_{7,k}-u_{6,l}-i/2}\:.
\end{eqnarray}
with energy
\begin{equation}
E_2=2g^2\sum_{i=1}^{M_4}\frac{1}{u_{4,i}^2+1/4}\;.
\end{equation}
In order to avoid Bethe roots at infinity, the magnon numbers obey $M_1<M_2<M_3<M_4>M_5>M_6>M_7$. 
The equations (\ref{FNBA}) are written for the Dynkin diagram from Fig. 1b. For the Lie superalgebras, since the
Dynkin diagram is not unique, to different Dynkin diagram correspond different sets of nested Bethe ansatz equations. Of course, the different sets of Bethe ansatz equations should be equivalent to each other. This can be checked on the $sl(2|1)$ case on the Bethe equations for the supersymmetric t-J model \cite{Bares, GSeel} by performing the so-called duality transformations on the Bethe roots. The general case for an arbitrary root system of $su(N|M)$ was treated by Tsuboi \cite{Tsuboi98}.
A duality transformation act on
a sequence of two nodes as shown in the figure \ref{fig:dualDyn}.
%
Let us exemplify it first for the case $M_1=M_2=M_5=M_6=M_7=0$, with equations
\begin{eqnarray}
\label{DNBA}
1&=&
\prod_{l=1}^{M_4} \frac{u_{3,k}-u_{4,l}-i/2}{u_{3,k}-u_{4,l}+i/2}\\ \nonumber
\(\frac{u_{4,k}+i/2}{u_{4,k}-i/2}\)^L&=&\prod_{l\neq k}^{M_4} \frac{u_{4,k}-u_{4,l}+i}{u_{4,k}-u_{4,l}-i}\;
\prod_{l=1}^{M_3} \frac{u_{4,k}-u_{3,l}-i/2}{u_{4,k}-u_{3,l}+i/2}
\end{eqnarray}
which correspond to a $sl(2|1)$ subsector and is the case considered in \cite{GSeel}.
Let us consider the polynomial $P(v)$ of degree $M_4-1$
\begin{equation}
P(v)\equiv\prod_{l=1}^{M_4} (v-u_{4,l}-i/2)-\prod_{l=1}^{M_4} (v-u_{4,l}+i/2)=\prod_{k=1}^{M_3}(v-u_{3,k})\prod_{j=1}^{M_4-M_3-1}(v-v_{3,j})
\end{equation}
As the second equality sign suggests, the magnon rapidities $u_{3,k}$ are among the roots of $P(v)$,
due to the first set of equations in  (\ref{DNBA}). Since $M_3\leq M_4-1$, $P(v)$ may have other roots, denoted by 
$v_{3,j}$. 
In the same time we have
\begin{eqnarray}
\nonumber
P(u_{4,k}+i/2)&=&-i\prod_{l\neq k}(u_{4,k}-u_{4,l}+i)=\prod_{l=1}^{M_3}(u_{4,k}-u_{3,l}+i/2)\prod_{j=1}^{M_4-M_3-1}(u_{4,k}-v_{3,j}+i/2)\\ \nonumber
P(u_{4,k}-i/2)&=&-i\prod_{l\neq k}(u_{4,k}-u_{4,l}-i)=\prod_{l=1}^{M_3}(u_{4,k}-u_{3,l}-i/2)\prod_{j=1}^{M_4-M_3-1}(u_{4,k}-v_{3,j}-i/2)\;.
\end{eqnarray}
We conclude that the equations (\ref{DNBA}) are equivalent to the set of equations
\begin{eqnarray}
\label{DDNBA}
1&=&
\prod_{l=1}^{M_4} \frac{v_{3,j}-u_{4,l}-i/2}{v_{3,j}-u_{4,l}+i/2}\\ \nonumber
\(\frac{u_{4,k}+i/2}{u_{4,k}-i/2}\)^L&=&
\prod_{l=1}^{M_4-M_3-1} \frac{u_{4,k}-v_{3,l}+i/2}{u_{4,k}-v_{3,l}-i/2}
\end{eqnarray}
In particular, when $M_4-1=M_3$, the equations (\ref{DDNBA}) become the $gl(1|1)$, purely fermionic
equations
\begin{equation}
\(\frac{u_{4,k}+i/2}{u_{4,k}-i/2}\)^L=1
\end{equation}
which are equivalent with the Bethe ansatz equations  for the $XY$ model. Another interesting case, which can be treated analogously, is the one with $M_4-1=M_3=M_5$. Doing a duality transformation on both sides of the central node $4$ we obtain
\begin{equation}
\(\frac{u_{4,k}+i/2}{u_{4,k}-i/2}\)^L=\prod_{l\neq k}^{M_2} \frac{u_{4,k}-u_{4,l}-i}{u_{4,k}-u_{4,l}+i}
\end{equation}
which are nothing else than the Bethe equations (\ref{BEsl2}) for the $sl(2)$ sector. 
This is a consistency check for the one loop Hamiltonian (\ref{hamsl2}).

\section{Perturbative integrability in ${\cal N}=4$ SYM}

\subsection{The notion of perturbative integrability and the asymptotic Bethe ansatz.}

Once that the integrability was established at the level of one loop in the gauge theory and 
at the classical level in string theory, the question arose whether the two integrable structures
are not manifestations of the same integrable structure, which would exist at any order in perturbation theory in the gauge theory, and at the quantum level in string theory.
The first piece of evidence that this might be the case was put forward by Beisert, Kristjansen and Staudacher \cite{Beisert:2003tq}, who computed the two loop order dilatation operator in the $so(6)$ sector. 
Upon diagonalization, the spectrum of anomalous dimensions, in the planar limit, showed degeneracies which could not be explained by any symmetry of the theory and which are characteristic of integrable hamiltonians. The existence of this degeneracy led the authors of \cite{Beisert:2003tq} to conjecture that the dilatation operator is integrable at any loop order.

In the $su(2)$ sector, the two loop dilatation operator of \cite{Beisert:2003tq} can be written again in terms of the permutation operators as
 \begin{equation}
 \label{twodil}
D_4=-2\sum_{i=1}^L(P_{i,i+2}-1)+8\sum_{i=1}^L(P_{i,i+1}-1)\;.
\end{equation}
and it connects the next-to-nearest neighbors.
From planarity of the Feynman diagrams, we infer that at $n$th loop order the spin can interact at
distance at most $n$, therefore the dilatation operator would be a long range interacting spin chain with the range of the interaction controlled by the order in perturbation theory.
It is clear that $D_4$ is not an integrable Hamiltonian by itself, and it does not commute with $D_2$
so the two pieces cannot be diagonalized simultaneously. However, one can interpret $D_4$ as 
a piece of the deformation of the {\it integrable} Hamiltonian $D_2$. The notion of {\it perturbative} integrability can be understood as follows. The integrability of $D_2$ implies that there are other conserved charges, generated by the expansion of the transfer matrix (\ref{tcomm}). Let us call
$I^{(n)}$ with $n=3,\ldots, L$ the extra conserved charges, in addition to the shift operator and the Hamiltonian $D_2\equiv I^{(2)}$, 
 \begin{equation}
[I^{(n)},I^{(m)}]=0\;.
\end{equation} 
 When the coupling constant is turned on, the hamiltonian and the other conserved charges
 get deformed into $(D(g)-D_0)/g^2$ and $I^{(n)}(g)$
such that
 \begin{equation}
[I^{(n)}(g),I^{(m)}(g)]=0\;.
\end{equation} 
When expanded\footnote{In the $su(2)$ sector, the dilatation operator expands in even powers of $g$ only, because it conserves the parity of the number of scalar fields. This should be also the case for the other conserved quantities.} in powers of $g^2$, the conserved 
quantities are not conserved order by order in $g$. A practical way to determine the conserved quantities recursively order by order in perturbation theory is to impose
 \begin{equation}
 \label{pertint}
[\sum_{l=1}^k I^{(n)}_{2l}g^{2l-2},\sum_{l=1}^k I^{(m)}_{2l}g^{2l-2}]=0+\CO(g^{2k})\;.
\end{equation} 
The authors of \cite{Beisert:2003tq} used the definition (\ref{pertint}) and the   planar structure of the Feynman graph to predict what would be the three loop dilatation operator, in the case
this would be integrable. Their proposal is
 \begin{equation}
 \label{threedil}
D_6=-4\sum_{i=1}^L(P_{i,i+3}P_{i+1,i+2}-P_{i,i+2}P_{i+1,i+3})  +16\sum_{i=1}^L(P_{i,i+2}-1)-56\sum_{i=1}^L(P_{i,i+1}-1)\;.
\end{equation} 
It was later confirmed by Beisert \cite{Beisert:2003ys} that this expression is completely fixed by supersymmetry. The three loop dilatation operator was therefore proven to be integrable.
The most advanced to date perturbative computation which confirms that the dilatation operator
is integrable up to four loop is due to Beisert, McLoughlin and Roiban \cite{Beisert:2007hz}.

Several questions can be asked once we have determined that the dilatation operator is integrable
beyond the one loop order. One of them is how to diagonalize the higher order hamiltonians. The other is how to determine the interaction at orders $n\geq L$, when the Feynman graphs can be wrapped
around the spin chain.

A partial answer to the first question came \cite{Serban:2004jf} from the already known spin chains with long range interaction. The spin chain which can reproduce the structure in (\ref{twodil}) and (\ref{threedil}) is the Inozemtsev spin chain \cite{Inozemtsev:2002vb}. Unfortunately, the Inozemtsev spin chain involves only two-spin interactions and is not general enough to reproduce the many-spin interaction which proliferate in the dilatation operator at higher loops. However, using the higher order conserved Hamiltonians of the Inozemtsev model, it was possible to match the integrable structure of the dilatation operator up to three loop. In particular, it was possible to see that the dilatation operator can be still diagonalized by Bethe ansatz, { provided} that the chain is very long. The term {\it asymptotic} Bethe ansatz was invented to describe this situation. Unlike other integrable spin chains, in this spin chain the
magnon energies are not additive; the correction to the two-magnon energy vanishes exponentially with the length of the chain.

The comparison with the Inozemtsev model also allowed to check that the Yangian symmetry
survives at higher loops, at least asymptotically  \cite{Serban:2004jf}. The Yangian is a symmetry for the Heisenberg spin chain \cite{Bernard:1992ya} in the 
limit of infinite length, and an exact symmetry for the Haldane-Shastry model. Dolan, Nappi and Witten  \cite{Dolan:2003uh} emphasized the link between integrability and the Yangian in the context of AdS/CFT.  Agarwal and Rajeev \cite{Agarwal:2004sz} checked that the two-loop Yangian coming from the Inozemtsev model commutes with the two loop dilatation operator. Beisert and Erkal \cite{Beisert:2007jv} studied the higher loop deformations for the long-range $gl(n)$ spin chains proposed by Beisert and Klose
\cite{Beisert:2005wv}.

Later, we are going to see that the dilatation operator, up to three loops, can also be reproduced by a projection of the Hubbard model onto the spin sector.  However, to date there is no closed expression for the spin chain which corresponds to the spin part of the Hubbard model. 

Different long-range integrable deformations of the nearest-neighbor Heisenberg spin chaine were considered in the literature, see\cite{Beisert:2005wv} \cite{Bargheer:2008jt} \cite{Bargheer:2009xy}.

\subsection{The Inozemtsev model}

The Inozemtsev model \cite{Inozemtsev:2002vb} describes a periodic chain of $L$ spins in the $su(2)$ spin $1/2$ representation which continuously interpolates between the short range Heisenberg model and the long range Haldane-Shastry \cite{Haldane:1987gg,shastry} model. The Inozemtsev Hamiltonian is defined as
\begin{equation}
\label{ino}
H_{I}=\sum_{j=1}^L  \sum_{n=1} ^{L-1} {\cal P}_{L,\pi/\kappa}(n) (1-P_{j,j+n})\;,
\end{equation}
where the interaction strength ${\cal P}_{L,\pi/\kappa}(n)$ is given by the Weierstrass elliptic function
with periods $L$ and $i\pi/\kappa$,
\begin{equation}
{\cal P}_{L,\pi/\kappa}(z)=\frac{1}{ z^2} +\sum'_{m,n}\left( 
\frac{1}{ (z-mL-i n\pi/\kappa)^2}-
\frac{1}{ (mL+i n\pi/\kappa)^2}\right)\;,
\end{equation}
where in the primed sum the terms with $m=n=0$ is omitted.
When one of the periods become infinite, the Weierstrass function degenerates either to a
trigonometric or to a hyperbolic function
\begin{eqnarray}
\lim_{\kappa\to 0}{\cal P}_{L,\pi/\kappa}(z)&=&\left(\frac {\pi }{L}\right)^2
\left( \frac{1}{ \sin^2 \pi z /L}-\frac{1}{ 3}\right)\;,\\ 
\lim_{L\to \infty}{\cal P}_{L,\pi/\kappa}(z)&=&\kappa^2
\left(\frac {1}{ \sinh^2 \kappa z }+\frac{1}{ 3}\right)\;.
\end{eqnarray} 
The first case corresponds to the Haldane-Shastry model, which played a special role in the landscape of integrable models with long range interaction. This is one of the few cases where the Yangian is an exact symmetry of the Hamiltonian, even for systems of finite length, and it is also an example of system with purely statistical interaction. The second case is the one we will be concerned with in the following.

The Heisenberg model can be obtained in the limit when the imaginary period vanishes, $\kappa\to \infty$. We define 
\begin{equation}
t\equiv e^{-2\kappa}
\end{equation}
such that the interaction strength becomes, after removing an additive and multiplicative constant
 \begin{equation}
 \label{strength}
h_{t,L}(n) \equiv \frac{{\cal P}_{L,\pi/\kappa}(n)}{4\kappa^2}-\frac{1}{12}=
\sum_{m=-\infty}^{\infty}\frac {t^{n+Lm}}{ (t^n-t^{Lm})^2}\;.
\end{equation}
The interaction strength (\ref{strength}) has the property, characteristic of the $\CN=4$ SYM dilatation operator,  that $h_{t,L}(n)\sim t^{-n}$.
In particular, we have 
  \begin{equation}
\lim_{t\to 0} h_{t,L}(n)/t=\delta_{n,1}+\delta_{n, L-1}
\end{equation}
The full solution of the Inozemtsev model is highly involved and the wave functions can be found via a correspondence \cite{Inozemtsev:2002vb} with the elliptic Calogero-Moser system. The diagonalization cannot be done via Bethe ansatz and the eigenenergies are not the sum of the energies of individual magnons. The situation simplifies when the length of the chains is very large and
 \begin{equation}
h_{t,L}(n) =\frac {t^{n}}{ (t^n-1)^2}=\frac{1}{4\sinh^2\kappa n}\;.
\end{equation}
In this situation, the solution can be obtained by Bethe ansatz, with
\begin{equation}
\label{InBA}
e^{ip_kL}=\prod_{l\neq k=1}^M \frac{\varphi(p_k)-\varphi(p_l)+i}{\varphi(p_k)-\varphi(p_l)-i}
\end{equation}
with the rapidity $\phi(p)$ and magnon dispersion relation $\varepsilon(p)$
\begin{eqnarray}
\label{phaseshift}
\varphi(p)&=&\frac{p}{2\pi i \kappa}\zeta_1\left(\frac{i\pi}{2\kappa}
\right)-\frac{1}{2 i \kappa}\zeta_1\left(\frac{i p}{2\kappa}
\right)\;,\\
\varepsilon(p)&=&
\frac{1}{i\pi\kappa}\zeta_1\left(\frac{i\pi}{2\kappa}
\right)-\frac{1}{4\kappa^2}{\cal P}_1\left(\frac{i p}{2\kappa}
\right)-\varphi^2(p)\;
\end{eqnarray}
where ${\cal P}_1(z)={\rm d} \zeta_1(z)/{\rm d}z$ denotes the Weierstrass elliptic functions 
of periods $1$ and $i\pi/\kappa$. The non-additivity of the spectrum can be compared with that of the Y system (\ref{enconj}), where at finite length there are corrections to the energies of the individual magnons.

The hyperbolic Inozemtsev model posseses a Yangian symmetry which is a deformation of the
Yangian symmetry of the (infinite length)  Heisenberg chain. The generators of the Yangian symmetry 
can be obtained by analytical continuation of the ones \cite{Haldane:1992sj,Bernard:1993uz} of the Haldane-Shastry spin chain, by replacing
$z_k=e^{2\pi i k/L}$ by $z_k=t^{-k}$
\begin{eqnarray}
\label{IY}
\CQ_0^a&=&\sum_j\sigma^a_j\;,\cr
\CQ_1^a&=&i\varepsilon^{abc}\sum_{j<k}\frac{z_j+z_k}{z_j-z_k}\;
\sigma^b_j\,\sigma^c_k\;,
\end{eqnarray}
with $\sigma^a_j$ being the Pauli matrices associated to site $j$. 
The existence of a monodromy matrix, associated to the Yangian (\ref{IY}) allows to compute \cite{Haldane:1992sj,HT},
although relatively painfully, the higher order conserved charges of the hyperbolic Hamiltonian 
\begin{eqnarray}
 \label{conserved}
H_I^{(2)}&=&\sum'_{ij}\frac{z_iz_j}{z_{ij}z_{ji}}(P_{ij}-1) \;,\cr
H_I^{(3)}&=&\sum'_{ijk}\frac{z_iz_jz_k}{z_{ij}z_{jk}z_{ki}}(P_{ijk}-1)\;,\cr
H_I^{(4)}&=&\sum'_{ijkl}\frac{z_iz_jz_kz_l}{z_{ij}z_{jk}z_{kl}z_{li}}(P_{ijkl}-1)-
 2\sum'_{ij}\left(\frac{z_iz_j}{z_{ij}z_{ji}}\right)^2(P_{ij}-1) \;,
\end{eqnarray}
where the prime indicates that the sum is over distinct summation
indices, $z_{ij}=z_i-z_j$, and  $P_{ijk}$, etc. represent cyclic
permutations of spins.

\subsection{The three loop dilatation operator and Inozemtsev model}

Since the fourth order conserved charge $H_I^{(4)}$ of the Inozemtsev model contains a four-spin interaction, we can try to use it in order to match the first few orders of the dilatation operator
\begin{equation}
\label{dilat}
D(g)= L+f_1(g)H_I^{(2)}(t)+f_2(g)H_I^{(4)}(t)+...
\end{equation} 
Explicit computation shows that if we relate the coupling constant $g^2$ to 
\begin{equation}
\label{lambdat}
t=g^2-3g^4+
14g^6+\ldots
\end{equation} 
one can reproduce the terms (\ref{twodil}) and (\ref{threedil}). 
Using the rapidity (\ref{phaseshift}) representation in terms of the infinite sum
\begin{eqnarray}
\label{uphi}
\varphi(p)&=&\frac{1}{2}\cot\frac{p}{2}+\frac{1}{2}
\sum_{n>0}\left[\cot(\frac{p}{2}-i\kappa n)+\cot(\frac{p}{2}+i\kappa n)
\right] \\
&=&\frac{1}{2}\cot\frac{p}{2}  
+2\sum_{n>0}\frac{t^n \sin p}{(1-t^{n})^2+4t^n\sin^2 (p/2)}\;. \nonumber
\end{eqnarray}
one can use the Bethe ansatz (\ref{InBA}) to compute perturbatively the anomalous dimensions up to three loops in the $su(2)$ sector.

We are going to see later that the same problem can be solved starting from the Bethe ansatz of the Hubbard model. Neither the Inozemtsev nor the Hubbard model are able to reproduce
the fourth order dilatation operator.

A somehow different approach was taken by Staudacher \cite{Staudacher:2004tk}, who proposed to diagonalize the operators (\ref{twodil}) and (\ref{threedil}) on an infinite chain by generalizing the procedure employed in the coordinate Bethe ansatz for the Heisenberg model (\ref{Banz}) to a case where the interaction involves several sites. This procedure, which he named {\it perturbative asymtptotic Bethe ansatz} allowed to reproduce the scattering phase up to the third loop order. 
This point of view encouraged to concentrate more on the concept of scattering matrix and factorized scattering than on the notions of $R$ matrix and monodromy matrix which are central to the algebraic Bethe ansatz approach, and allowed to circumvent the difficulties related to change of length which occurs once larger sectors, like $su(2|3)$ are taken into account.

In the scattering image, one considers the reference state with a very large number of up spins - or 
operators of type $Z$ in the $\CN=4$ SYM case
\begin{equation}
\ket{\Omega}=\ket{ZZZZZ\ldots ZZZZZ}
\end{equation}
An elementary excitations above this "vacuum" is a magnon, which corresponds to replacing
of one of the $Z$ by another field, for example $X$. One therefore reduces the problem of solving 
completely the problem of $L$ spins to that, simpler, of scattering of magnons above the (infinite)
sea of $Z$ fields. Generically, the magnons carry quantum numbers and their scattering is described by a matrix, while in the particular $su(2)$ case the scattering matrix is just a phase factor. For the $su(2|3)$ chain the scattering matrix has  $su(2|2)$ symmetry, while for the full $psu(2,2|4)$ the scattering matrix 
has  $su(2|2)\otimes su(2|2)$ symmetry.

\section{Strings on $AdS_5\times S^5$}

String theory on $AdS_5\times S^5$ can be defined via a sigma model on a coset space. This construction is due to Metsaev and Tseytlin \cite{Metsaev:1998it}, following the corresponding construction of superstring action
in the ten dimensional flat space by Henneaux and Mezincescu \cite{Henneaux:1984mh}.
The coset space is given by 
\begin{equation}
\frac{PSU(2,2|4)}{SO(4,1)\otimes SO(5)}\;.
\end{equation}
The bosonic part of $PSU(2,2|4)$ is given by $SU(2,2)\otimes SU(4)$, which is locally isomorphic to $SO(5,2)\otimes SO(6)$, such that the bosonic part of the coset space is given by $AdS_5=SO(5,2)/SO(4,1)$ times $S^5=SO(6)/SO(5)$. The coset model provides a natural way to couple strings to the Ramond-Ramond fields.

In order to be able to write down the action, it is necessary to introduce a few basic facts. The presentation here follows the general lines of the pedagogical reference \cite{Arutyunov:2009ga}.

The superalgebra  $su(2,2|4)$ is a non-compact real form of the algebra $sl(4|4)$, which can be thought as the algebra of $4|4 \times 4|4$ supermatrices over the complex numbers,  with zero supertrace
\begin{equation}
M=\left(\begin{array}{cc} a & \theta \\\eta & b \\ \end{array}\right)
\end{equation}
where the $4\times 4$ elements $a$ and $b$ are even and $\theta$ and $\eta$ are matrices with odd (grassmannian) elements and $\STr M=\Tr a-\Tr b=0$.
The non-compact real form is specified by the relation
\begin{equation}
M^\dagger H+HM=0\;,
\end{equation}
where
\begin{equation}
H=\left(\begin{array}{cc|c}1_2 & 0 & 0 \\0 & -1_2 & 0 \\\hline 0 & 0 & 1_4\end{array}\right)
\end{equation}
It should be noticed that the identity belongs to the $su(2,2|4)$ superalgebra, since $\STr 1=4-4=0$ and that it is a central element, since it commutes with all the other elements of the algebra. Removing the identity reduces the  $su(2,2|4)$ superalgebra to $psu(2,2|4)$. The latter algebra has no  representation in terms of supermatrices.

The quotient algebra can be characterized with the help of the automorphism  generated by  the supertransposition via
 \begin{equation}
 \label{autom}
 \Omega(M)=-K M^{st}K^{-1}\;.
 \end{equation}
 The supertranspose $M^{st}$ of $M$ is defined by
\begin{equation}
M^{st}=\left(\begin{array}{cc}a^t & -\eta^t \\\theta^t & b^t\end{array}\right)\;,
\end{equation}
such that
\begin{equation}
(M^{st})^{st}=\left(\begin{array}{cc}a & -\theta \\-\eta & b\end{array}\right)\;
\end{equation}
and 
\begin{equation}
K={\rm diag} (\sigma,\sigma,\sigma,\sigma)\;, \quad \sigma= i^{-1}\sigma_2=\left(\begin{array}{cc}0 & -1 \\1 & 0\end{array}\right)\;.
\end{equation}
We deduce that $\Omega^4=1$, and on the even subalgebra this automorphism is of order $2$. 
One can decompose the superalgebra $sl(4|4)$ into a direct sum of graded spaces
\begin{equation}
\CG=\CG^{(0)}+\CG^{(1)}+\CG^{(2)}+\CG^{(3)}
\end{equation}
where $\CG^{(k)}$ is the subspace of $\CG=sl(4|4)$ corresponding to $\Omega(M)=i^k M$.
The fixed point  $\CG^{(0)}$ of the automorphism $\Omega$ coincides with the even subalgebra 
$so(4,1)\oplus so(5)$. 

\subsection{The Lagrangian}

We consider $\mathfrak{g}$ an element from the supergroup $SU(2,2|4)$ and we introduce the
associated one-form with values in the $su(2,2|4)$ algebra
\begin{equation}
A=-\mathfrak{g}^{-1} {\rm d} \mathfrak{g}=A^{(0)}+A^{(1)}+A^{(2)}+A^{(3)}\;.
\end{equation}
This definition insures that $A$ is a flat connection 
\begin{equation}
\d_\al A_\be-\d_\be A_\al -[A_\al,A_\be]=0\;,
\end{equation}
where the indices $\al, \be$ take one of the values $\sigma$ or $\tau$.
The Lagrangian density is postulated to be
\begin{equation}
\label{stringlag}
\CL=-g\left[\sqrt{-h}h^{\al\be}\STr(A_\al^{(2)}A_\be^{(2)})+\kappa \e^{\al\be}\STr(A_\al^{(1)}A_\be^{(3)})\right]
\end{equation}
with the coupling constant being related to the 't Hooft coupling constant by
\begin{equation}
g=\frac{\sqrt{\lambda}}{4\pi}
\end{equation}
$h^{\al\be}$ the world-sheet metric and $\e^{\tau\sigma}=1$. In the conformal gauge $h^{\al\be}={\rm diag}(-1,1)$.
Under right multiplication of $\mathfrak{g}$ with an element $\mathfrak{h}$ from $SO(4,1)\otimes SO(5)$, $\mathfrak{g}\to\mathfrak{g}\mathfrak{h} $, all the elements $A^{(i)}$ with $i=1,2,3$
transform by adjoint action $A^{(i)}\to \mathfrak{h}^{-1}A^{(i)}\, \mathfrak{h}$, while $A^{(0)}\to \mathfrak{h}^{-1}A^{(0)}\, \mathfrak{h}-\mathfrak{h}^{-1}{\rm d} \mathfrak{h}$. This means that 
the string lagrangian is invariant under local transformations in $SO(4,1)\otimes SO(5)$. Moreover, the lagrangian is invariant by a shift involving the identity matrix, so it depends 
on the coset elements from $PSU(2,2|4)/SO(4,1)\otimes SO(5)$.

The second term in (\ref{stringlag}) can be interpreted as a topological term and it contains the fermionic degree of freedom. 
In addition to the local symmetry generated by the right multiplications, the action possesses another
symmetry, generated by left multiplication with elements associated with the odd part of the $psu(2,2|4)$
algebra. This symmetry is the so-called kappa symmetry and it can be used to eliminate 16 non-physical
fermionic degrees of freedom and leave only 16 physical fermionic degrees of freedom. The kappa symmetry requires the coefficient in front of the Wess-Zumino term to be $\kappa=\pm 1$. For more details on the kappa symmetry we refer the reader to \cite{Arutyunov:2009ga}.

\subsection{Integrability of the classical sigma model}
\label{intclass}

Classical integrability can be formulated as the compatibility condition for a system of 
partial differential equations in two dimensions $(\sigma,\tau)$
\begin{eqnarray}
\label{lax}
\d_\sigma \Psi&=&L_\sigma(\sigma, \tau, z) \Psi \\\nonumber
\d_\tau \Psi&=&L_\tau(\sigma, \tau, z) \Psi\;.
\end{eqnarray}
Here, $\Psi$ is a vector of dimension $n$ and $L_{\sigma,\tau}$ are $n\times n$ matrices. Both $\Psi$ and $L$ depend on an extra complex parameter $z$, the spectral parameter.
The system (\ref{lax}) is compatible provided that the matrices $L_{\sigma,\tau}$ obey the
zero curvature condition 
\begin{equation}
\label{laxconn}
\d_\sigma L_\tau-\d_\tau L_\sigma-[L_\sigma, L_\tau]=0\;.
\end{equation}
The existence of the Lax connection (\ref{laxconn}) $L_{\sigma,\tau}$ allows to define the monodromy matrix
\begin{equation}
\label{monmat}
T(z)=P {\rm exp}  \int _0^{2\pi} d\sigma L_\sigma(z)
\end{equation}
where $P$ is path ordering operator and the system is considered to be on a cylinder of circumference $2\pi$. Due to the zero curvature condition, the trace of the monodromy matrix is independent 
of $\tau$.  

\vskip .2cm

{\it Construction of the Lax connection}

\vskip .2cm

The classical integrability of the coset model was established by Bena, Polchiski and Roiban \cite{Bena:2003wd},
who proved the existence of the Lax connection. This is written as an linear combination of the different $\MZ_4$ components of the flat connection $A_\al$
\begin{equation}
L_\al=c_0 A_\al^{(0)}+c_1 A_\al^{(2)}+c_2\gamma_{\al\be}\e^{\be\rho}A_\rho^{(2)}+c_3A_\al^{(1)}+c_4A_\al^{(3)}
\end{equation}
Imposing the  condition that the connection $L_\al$ is flat is equivalent to the equations of motion if
the condition for the kappa symmetry is obeyed
\begin{equation}
\kappa^2=1
\end{equation}
and the coefficients $c_i$ are determined by a single parameter $z$
\begin{equation}
c_0=1\;, \quad c_1=\frac{1}{2}\(z^2+\frac{1}{z^2}\)\;, \quad c_1=-\frac{1}{2\kappa}\(z^2-\frac{1}{z^2}\)\;, \quad c_3=\frac{1}{c_4}=z\;.
\end{equation}
The action of the automorphism $\Omega$ translates into
\begin{equation}
\Omega(L_\al(z))=L_\al (iz)
\end{equation}
A different presentation  of the spectral parameter, which will be used in the following   is given by
\begin{equation}
z^2=\frac{x+1}{x-1}
\end{equation}
such that 
\begin{equation}
\Omega(L_\al(x))=L_\al (1/x)
\end{equation}
The Lax connection becomes singular at the points $x=\pm1$, or $z=0,\infty$. 
The monodromy matrix $T(z)$, defined by the equation (\ref{monmat}) depends only on the point where the contour is cut open. Changing this point will only change $T(z)\to \gamma(z)T(z)\gamma^{-1}(z)$, such that the eigenvalues of the monodromy matrix encode the physical information.
Diagonalizing the monodromy matrix we obtain \cite{Beisert:2005bm}
\begin{equation}
u(z)T(z)u^{-1}(z)={\rm diag}(e^{i\tilde p_1(z)},e^{i\tilde p_2(z)},e^{i\tilde p_3(z)},e^{i\tilde p_4(z)} | e^{i\hat p_1(z)},e^{i\hat p_2(z)},e^{i\hat p_3(z)},e^{i\hat p_4(z)})
\end{equation}
where the two groups of eigenvalues correspond to the two gradings. Unimodularity imposes ${\rm SDet} T(z)=1$, which translates into the condition
 \begin{equation}
\tilde p_1+\tilde p_2+\tilde p_3+\tilde p_4-\hat p_1-\hat p_2-\hat p_3-\hat p_4\in 2\pi \BZ
\end{equation}
The eigenvalues $e^{i\tilde p_j(z)}$, $e^{i\hat p_k(z)}$ are analytic with respect to $z$ except at the points $z=0, \infty$ and the points $z_a$  where eigenvalues become degenerate. The typical case is that when two eigenvalues cross and it can be analyzed on a $2 \times 2$ matrix 
\begin{equation}
\Gamma=\left(\begin{array}{cc}a & b \\c & d\end{array}\right)
\end{equation}
with eigenvalues 
\begin{equation}
\gamma_{1,2}=\frac{1}{2}\left(a+d\pm\sqrt{(a-d)^2+4bc}\right)
\end{equation}
The singular point $z_a$ corresponds to the point where $(a-d)^2+4bc=0$, and expanding in $z$ around $z_a$ one gets
\begin{equation}
\label{brcut}
\gamma_{2}=\gamma_1\left(1-\alpha_a\sqrt{z- z_a}+\CO(z-z_a)\right)\;.
\end{equation}
We conclude that the typical singularity when two eigenvalues of similar grading collide is of the type square root. When two eigenvalues of different grading come together, we have to analyze the case
when $\Gamma$ is a $2\times 2$ supermatrix, with eigenvalues
\begin{equation}
\gamma_{1}=\frac{bc}{d-a}+a\;, \quad  \gamma_{2}=\frac{bc}{d-a}+d
\end{equation}
where $b$ and $c$ are grassmannian variables with $(bc)^2=0$.
When $a$ approaches $d$, the two eigenvalues become close to each other.  So the typical singularity when two eigenvalues of different grading approach to each other is a pole.
At $z=0,\infty$, the functions $e^{i\tilde p_j(z)}$, $e^{i\hat p_k(z)}$ have essential singularities of the type $e^{\alpha_0/z^2}$ and $e^{\al_\infty z^2}$. In order to remove this singularities, one can consider the logarithmic derivative of the monodromy matrix
$$Y(z)=-iz\frac{\d}{\d z}\,\ln T(z)$$
whose eigenvalues $y_k=z\d_z \, p_k$ lie on an algebraic curve, since their only singularities can be poles or branch cuts.

The automorphism (\ref{autom}) acts on the matrix 
\begin{equation}
\label{algcur}
y(z)={\rm diag}(\tilde y_1(z), \tilde y_2(z),\tilde y_3(z),\tilde y_4(z)|\,\hat y_1(z), \hat y_2(z), \hat y_3(z), \hat y_4(z))
\end{equation}
by
\begin{equation}
\Omega(y(z))=y(iz)=-{\rm diag}(\tilde y_2(z), \tilde y_1(z),\tilde y_4(z),\tilde y_3(z)|\,\hat y_2(z), \hat y_1(z), \hat y_4(z), \hat y_3(z))
\end{equation}
since the conjugation with the matrix $\sigma=i^{-1}\sigma_2$  exchanges the diagonal elements. 
We deduce that 
\begin{equation}
\tilde y_k(-z)=\tilde y_k(z)\;, \quad \hat y_k(-z)=\hat y_k(z)
\end{equation} 
so that the algebraic curve can be parametrized by
 \begin{equation}
x=\frac{z^2+1}{z^2-1}\;.
\end{equation} 
Under this parametrization, the sheets of the algebraic curve are permuted by the transformation $x\to 1/x$ as follows
 \begin{eqnarray}
\tilde y_1(x)=\tilde y_2(1/x)\;, \quad \hat y_1(x)=\hat y_2(1/x)\\ \nonumber
\tilde y_3(x)=\tilde y_4(1/x)\;, \quad \hat y_3(x)=\hat y_4(1/x)
\end{eqnarray} 
It is clear that at  the fixed points of the transformation $x\to 1/x$, that is $x=\pm 1$, the algebraic curve has singularities.  A particular case where these singularities collide with the branch cuts of the type (\ref{brcut})
will be studied in section \ref{sec:strg}.


\section{Comparison between the gauge theory and the string results}

An important point in verifying the hypothesis of integrability at any value of the coupling constant was 
the comparison between the results obtained in the weak coupling regime, corresponding to perturbative gauge theory, and the strong coupling regime, corresponding to perturbative string theory.
A priori, the two regimes do not superpose with each other, so it looks like the results cannot be compared. However, Berenstain, Maldacena and Nastase  \cite{Berenstein:2002jq} proposed to study a regime where 
a string moves very fast along the equator of $S_5$, with angular momentum $J$. In terms of the spin chain, $J=L-M$. In this regime the effective coupling constant is the so-called BMN coupling constant
\begin{equation}
\lambda'=\frac{\lambda}{J^2}\qquad {\rm or }\qquad g'=\frac{g}{J}
\end{equation}
If one considers $\lambda$ and $J$ large but $\lambda'$ small, then one may compare the results 
obtained from the regime $\lambda$ small, $J$ large. Of course, nothing insures that there will not be
order of limits problems, such that the results in the two regimes may differ. Luckily, it occurred that the 
results coincide up to order $\lambda'^2$, and these results were enough to allow conjecturing the 
general structure of the Bethe ansatz equations.  Starting with the third order in $\lambda'$, the string and gauge theory results are different and at four loop order the dilatation operator does not have a well defined BMN limit anymore.

\subsection{Strings in the plane-wave background. The BMN result}

The plane-wave geometry is the geometry seen by a particle moving at a speed close to the speed of light  \cite{Penrose}.
In the case of the $AdS_5 \times S^5$ geometry with metric
\begin{equation}
ds^2_{
AdS_5\times S^5} = R^2 [-dt^2 \cosh^2 \rho + d\rho^2 + \sinh^2 \rho\; d\Omega_3^2 + d\psi^2 \cos^2 \theta + d\theta^2 + \sin^2 \theta d\Omega^{'2}_3
 ]
\end{equation}
one can consider a massless particle moving on the light-like trajectory parametrized by
\begin{equation}
\rho=0\;, \quad \theta=0\;, \quad t=t(\phi) \;, \quad \psi=\psi(\phi)
\end{equation}
Strictly speaking, the string background is $AdS_5 \times S^5$ with the otherwise compact
coordinate $t$ decompactified.
By introducing the light-cone coordinates $\tilde x^\pm=(t\pm \psi)/2$ and rescaling
\begin{equation}
x^+=\frac{\;\tilde x^+}{\mu}\;, \quad x^-=\mu R^2\;  \tilde x^-\;, \quad r=R\; \rho\;, \quad y=R\; \theta 
\end{equation}
with $\mu$ a mass parameter necessary to keep the dimensions right. Upon sending the
parameter $R$ to infinity, the $AdS_5 \times S_5$ metric becomes the pp wave metric \cite{Penrose}
\begin{equation}
\label{metricpw}
ds^2_{pw} =-4dx^+ dx^--\mu^2(\vec{y}^2+ \vec{r}^2)(dx^+)^2+d\vec{y}^2+d\vec{r}^2
\end{equation}
with $\vec{r}$ and $\vec{y}$  two four-vectors embedded in $AdS_5$ and $S_5$ respectively.
The mass parameter $\mu$ is related to value of the self-dual five-form by
\begin{equation}
F_{+1234}=F_{+5678}=4\mu\;
\end{equation}
which breaks the symmetry of the transverse modes from $SO(8)$ to $SO(4)\times SO(4)$.
When the five-form field strength vanishes, $\mu \to 0$, the plane-wave metric becomes the ten dimensional Minkowski metric.
The energy $E=i\partial_t$ and angular momentum $J=-i\partial_\psi$ get translated, in the light cone variables $x^\pm$ into
\begin{eqnarray}
\label{pminus}
2p^-&=&i\partial_{x^+}=i\mu( \partial_t+\partial_\psi)=\mu(E-J)\\ \label{pplus}
2p^+&=&i\partial_{x^-}=\frac{i}{\mu R^2}( \partial_t-\partial_\psi)=\frac{E+J}{\mu R^2}
\end{eqnarray}
The light cone momentum $p^+$ is non-vanishing provided $J$ grows in a correlated way with
$R^2$, which means $J\sim \alpha'\sqrt{\lambda}$. The light cone Hamiltonian is given by $\CH_{lc}=2p^-$. Under the AdS/CFT correspondence, the energy $E$ should correspond to the conformal dimension $\Delta$ of some local operator in the $\CN=4$ SYM theory, while the angular momentum $J$, which is a $SO(6)$ charge, should correspond to one of the  R-charges. In particular, 
the case $E=J$
corresponds to a point-like string moving on the equator of the sphere $S^5$. This is a BPS state with $\CH_{lc}=0$ and its gauge theory counterpart is a state of the type $\Tr Z^J$. 

The gauge freedom for the reparametrization of the string can be fixed by choosing the light cone gauge $x^+=\tau$, with $\tau$ the world-sheet time.

The action of superstrings in the plane-wave background in light-cone gauge was written down by Metsaev \cite{Metsaev:2001bj}, who showed that it corresponds to eight massive bosons and eight massive fermions, whose masses are equal due to supersymmetry.
As it can be deduced from (\ref{metricpw}), this action reads
\begin{equation}
\label{BMNac}
S_{pw}=\frac{1}{2\pi \alpha'}\int  d\tau\int_0^{\ell} d \sigma\left[\frac{1}{2} (\d_\tau \vec{z})^2-\frac{1}{2}(\d_\sigma \vec{z})^2-\frac{1}{2}\mu^2 \vec{z}^{\;2} + {\rm fermions}\right]
\end{equation} 
with the eight-dimensional vector $\vec{z}\equiv(\vec{r},\vec{y})$ and $\ell=2\pi \alpha' p^+$ \cite{Polchinski}.
The action (\ref{BMNac}) can be quantized similarly to the free bosonic string action, its spectrum being
spanned by massive bosonic (fermionic) oscillators $\alpha^i_n,\ \tilde \alpha^i_n\ (\theta_n^i,\ \tilde \theta_n^i)$ with energy
\begin{equation}
\label{BMNen}
E_{n}=\mu \sqrt{1+\frac{n^2}{(\mu \alpha'p^+)^2}}\;.
\end{equation} 
Including the zero modes which arise in the curved background, the energy of the string in the light cone gauge is given by
\begin{equation}
E_{lc}=N_0 E_0 +\sum_{n\geq 1}(N_n+\tilde N_n)E_n\;.
\end{equation} 
where the occupation numbers $N_n$, $\tilde N_n$ obey the level-matching condition (or Virasoro constraint)
\begin{equation}
\sum_{n\geq 1}(N_n-\tilde N_n)=0\;.
\end{equation} 
Taking into account the fact that $E\simeq J$, and $R^4=\lambda \alpha'^2$, we obtain 
from (\ref{pplus}) 
\begin{equation}
\frac{1}{(\alpha' p^+ \mu)^2}\simeq \frac{\lambda}{J^2}\equiv {\lambda'}
\end{equation} 
such that the dispersion relation of the impurities (\ref{BMNac}) take the so-called BMN form
\begin{equation}
\label{BMNdisp}
E_{n}/\mu= \sqrt{1+\lambda'{n^2}}=\sqrt{1+\frac{\lambda n^2}{J^2}}\;.
\end{equation} 
Let us consider the two-impurity case with $N_n=\tilde N_n=1$ for some mode number $n$ and compare the leading result in $\lambda'$ for the energy of the string and the  corresponding result in gauge theory. From (\ref{pminus}) we conclude that
\begin{equation}
\label{ols}
\Delta-J=E_{lc}/\mu=2 \sqrt{1+\lambda'{n^2}}=2+{\lambda' n^2}-\frac{\lambda^{'2} n^4}{4}+\frac{\lambda^{'3}n^6}{8}-\CO(\lambda'^2)
\end{equation} 
In the same time, from the Bethe ansatz equations (\ref{NBA}) one has, for $u_1=-u_2=u$, or equivalently $p_1=-p_2=p$,
\begin{equation}
\(\frac{u+i/2}{u-i/2}\)^{L-1}=1\;, \quad {\rm or} \quad e^{i p(L-1)}=1\;,
\end{equation}
with the solution 
\begin{equation}
p=\frac{2\pi n}{L-1}
\end{equation}
The one-loop prediction in gauge theory is then
\begin{equation}
\label{olg}
\Delta=L+\frac{\lambda}{\pi^2}\sin^2\frac{\pi n}{L-1}= L+\frac{\lambda n^2}{(L-1)^2}+\CO(L^{-4})
\end{equation}
where the second relation holds for $n\ll L$.
Comparing (\ref{ols}) and (\ref{olg}) we conclude that $J=L-2$ for two impurities and the two expressions agree up to terms of order $J^{-3}$ at one loop. 
One can go beyond the one-loop order in computing the conformal dimension of the two-impurity operator, either by direct diagonalization of (\ref{twodil}) and (\ref{threedil}) \cite{Beisert:2003tq},
or by using the Inozemtsev asymptotic Bethe ansatz \cite{Serban:2004jf}. The result is
\begin{equation}
\label{olgp}
\Delta(\lambda)=J+2+\lambda E^{(1)}+\lambda^2 E^{(2)}+\lambda^3E^{(3)}+\ldots
\end{equation}
with
\begin{eqnarray}
\label{twoimp}
E^{(1)}&=&\frac{1}{\pi^{2}}\sin^{2}\frac{p_n}{2}\;, \\ \nonumber
E^{(2)}&=&-\frac{1}{\pi^{4}}\sin^{4}\frac{p_n}{2}\left(\frac{1}{4}+
\frac{1}{J+1}\cos^2\frac{p_n}{2} \right)\;,  \\  \nonumber
E^{(3)}&=&\frac{1}{\pi^{6}}\sin^{6}\frac{p_n}{2}\left(\frac{1}{8}+
\frac{\frac{1}{2} \cos^4\frac{p_n}{2}+\frac{3}{4}\cos^2\frac{p_n}{2}}{J+1}+\frac{\frac{5}{2} \cos^4\frac{p_n}{2}-\frac{3}{4}\cos^2\frac{p_n}{2}}{(J+1)^2}
\right)\;.
\end{eqnarray}
which agrees with (\ref{ols}) in the limit $n\ll J$ and at leading order in $J$. 
The $1/J$ corrections to quantum strings beyond the plane-wave geometry were computed in \cite{Callan:2003xr,Callan:2004ev}. While the one and two-loop part of these corrections agree with  
 (\ref{twoimp}), the three-loop part disagrees. This was the first occurence of a mismatch beween the BMN limit in strings and gauge theory, and it was finally understood to be a consequence of the non-commuting sequences of limits which are considered.

The string solutions considered above are a very particular case, since just one of the angular momenta is large, and the string excitations are small fluctuations around the pointlike string solution. More general string solutions were considered by Frolov and Tseytlin \cite{Frolov:2002av,Frolov:2003qc,Frolov:2003tu,Frolov:2003xy}, where strings can have large angular momentum in some planes in $S^5$ or/and in $AdS_5$. Some of these solutions still depend on the BMN  coupling constant $\lambda'$, but now the approximation of dilute impurities which is implicit in the BMN computation is not valid anymore. The first comparison between the spinning string solutions and the one-loop Bethe ansatz equations was done in \cite{Beisert:2003xu} and
it was pursued in an important number of papers \cite{Frolov:2004bh,Park:2005ji,Hernandez:2005nf,Beisert:2005mq,SchaferNameki:2005tn,Beisert:2005cw,Minahan:2005qj}. The $1/J$ corrections were also extensively checked for these solution. 

An alternative way to compare the predictions of the string theory with the results from the spin
chain was supplied by the formulation of the (finite gap) solutions of the string sigma model
in terms of algebraic curves \cite{Kazakov:2004qf}. The continuum limit of the Bethe ansatz equations gives also rise 
to an algebraic curve, very similar to that of the sigma model. The comparison between 
the two curves was very fruitful for fixing the general structure of the all loop Bethe ansatz.
In the next section, this comparison is presented for the particular case of the $su(2)$ sector.  

\subsection{Algebraic curves and the continuum limit of the Bethe ansatz equations: the $su(2)$ principal chiral model}
\label{pcm}

The full string sigma model (\ref{stringlag}) is relatively complicated to study in full generality. 
At the classical level, one can reduce the dimensionality of the space in which the string moves
by a  restriction which is similar to the restriction to sectors in perturbative gauge theory. We are then left with a (string) sigma model on a subspace of the original coset space. While this reduction is classically consistent, it is not consistent quantum mechanically. The string, although it moves only inside the restricted space, fluctuates on all directions of the original space. However, we can forget for the moment about the quantum corrections and study strings
which move, for example, only on the sphere $S^3\subset S^5$ via the $S^3\times R$ sigma model \cite{Kazakov:2004qf}.  
Let us consider the coordinates $\Phi_1,...,\Phi_4$ with $\sum_i \Phi_i^2=1$ be the coordinates of the sphere $S^3$, which is the also the group manifold for $SU(2)$ , and $\Phi_0$ the timelike coordinate 
\begin{equation}
\mathfrak{g}=\left(\begin{array}{cc}\Phi_1+i\Phi_2 & \Phi_3+i\Phi_4 \\-\Phi_3+i\Phi_4 & \Phi_1-i\Phi_2\end{array}\right)\equiv \left(\begin{array}{cc}Z & X \\-\bar X & \bar Z \end{array}\right)\in SU(2)
\end{equation}
The string action is given by
\begin{eqnarray}
\label{acs3}
S_{S^3}&=&g\int_0^{2\pi}d\sigma\int d\tau \left[(\partial_\al \Phi_i)^2-(\partial_\al \Phi_0)^2\right]\\
&=&-g\int_0^{2\pi}d\sigma\int d\tau  \left[\frac{1}{2}\Tr (\mathfrak{g}^{-1}\partial_\al \mathfrak{g})^2+(\partial_\al \Phi_0)^2\right]
\end{eqnarray}
The global symmetry of the action is given by $SU(2)_L\otimes SU(2)_R\simeq SO(4)$. The two $SU(2)$ symmetries correspond to the left and right multiplication respectively, with currents given by $l_\al=\partial_\al \mathfrak{g}\mathfrak{g}^{-1}$ and $j_\al=\mathfrak{g}^{-1}\partial_\al \mathfrak{g}$. Under the first group, $(Z,-\bar X)$ and $(X,\bar Z)$ transform as doublets, while under the second $(Z,X)$ and $(-\bar X, \bar Z)$ are doublets. As the notation suggests, the two
$SU(2)$ groups correspond to the ones in perturbative gauge theory which permute the complex scalars, generated by combinations of the oscillators $c^4, c^1$ and $c^3, c^2$ respectively.
In the normalization where the charge of the states in the doublets are $\pm1$, an operator of the type $\Tr Z^{L-J}X^{J}+\ldots$ will have a charge $L$ under the $SU(2)_L$ and charge $L-2J$
under $SU(2)_R$.

This time we consider the action in the gauge $\Phi_0=\kappa \tau$ and the energy of the string
is given by
\begin{equation}
\label{deltakappa}
\Delta=2g\int_0^\infty d\sigma \partial_\tau \Phi_0=4\pi g\kappa\;.
\end{equation}
The equations of movement for the action (\ref{acs3}) can be written, in the light cone coordinates $\sigma_\pm=\frac{1}{2}(\tau\pm\sigma)$, $\d_\pm=\d_\tau\pm\d_\sigma$
\begin{eqnarray}
\nonumber
&\ &\d_+j_-+\d_-j_+=0\;; \quad \d_+\d_- \Phi_0=0\;\\ 
\label{eqmo}
&\ &\quad \d_+j_--\d_-j_++[j_+,j_-]=0\;,
\end{eqnarray}
where the first equation corresponds to the current conservation and the second line is the equation of motion.
These equations are to be supplemented with the Virasoro condition, which express the vanishing of the stress-energy tensor,
\begin{equation}
\frac{1}{2}\Tr\;j_+^2=\frac{1}{2}\Tr\;j_-^2=-\kappa^2\;.
\end{equation}
The equations  (\ref{eqmo}) can be reformulated as the zero curvature condition \cite{Zakharov:1978wc}
\begin{equation}
\d_+J_--\d_-J_++[J_+,J_-]=0
\end{equation}
for the connection
\begin{equation}
J_\pm=\frac{j_\pm}{1\mp x}\;.
\end{equation}
Of course, this connection is a particular case of the one presented in section (\ref{intclass}). The monodromy matrix
\begin{equation}
\Omega(x)=P \exp{\int_0^{2\pi} d\sigma \frac{1}{2}\(\frac{j_+}{1-x}-\frac{j_-}{1+x}\)}
\end{equation}
can be diagonalized in terms of the quasi-momentum $p(x)$ defined by
\begin{equation}
\Tr \Omega(x)=2\cos p(x)\;.
\end{equation}
Near the singular points $x=\pm1$, the quasi-momentum behaves as 
\begin{equation}
p(x)=-\frac{\pi \kappa}{x\pm1}+\ldots\;.
\end{equation}
while the asymptotics at $x=0,\infty$ are related to the conserved charges, see \cite{Kazakov:2004qf} for more details,
\begin{eqnarray}
p(x)=-\frac{L-2J}{2gx}+\ldots\quad x\to \infty \\ \nonumber
p(x)=2\pi m +\frac{L}{2g}x+\ldots\quad x\to 0\;.
\end{eqnarray}
This behavior suggests that the transformation $x\to 1/x$ exchanges the two $su(2)$ components.
In order to remove the poles at $x=\pm 1$, one introduces the resolvent
\begin{equation}
G(x)=p(x)+\frac{\pi \kappa}{x-1}+\frac{\pi \kappa}{x+1}\;.
\end{equation}
which is analytic on the physical sheet, with possible branch cuts. The resolvent can be 
represented in terms of a density $\rho(x)$ with support on the branch cuts of $G(x)$
\begin{equation}
G(x)=\int d y\; \frac{\rho(y)}{x-y}\;
\end{equation}
such that $2\pi i \rho(x)=G(x-i0)-G(x+i0)$.
The asymptotic conditions for $p(x)$ at $x=0$ and $\infty$ translate into normalization conditions on $\rho(x)$
\begin{eqnarray}
\label{normdens}
&\ &\qquad \int dx\; \rho(x)=\frac{\Delta+2J-L}{2g}\;,\\
&\ &\int dx\; \frac{\rho(x)}{x}=2\pi m\;, \qquad
\int dx\; \frac{\rho(x)}{x^2}=\frac{\Delta-L}{2g} \nonumber
\end{eqnarray}
The continuous part of $p(x)$ on the cut is 
\begin{equation}
p(x+i0)+p(x-i0)=2\pi n_C
\end{equation}
where $n_C$ is specific to the branch cut $C$, or, in terms of the resolvent
\begin{equation}
\label{stringcurve}
G(x+i0)+G(x-i0)=2\pint dy \frac{\rho(y)}{x-y}=\frac{4\pi \kappa x}{x^2-1}+2\pi n_C\;.
\end{equation}
This equation, together with the normalization conditions (\ref{normdens}) can be compared
with the continuum limit of the Bethe ansatz equations (\ref{AlgBA}). 
Let us take the logarithm of  (\ref{AlgBA}) and rescale the rapidities $u_k\to L u_k$
\begin{equation}
p_kL\equiv L\ln \(\frac{1+i/2Lu_k}{1-i/2Lu_k}\)=2\pi i\tilde n_C+\sum_{j\neq k}\ln\( \frac{1+i/L(u_k-u_j)}{1-i/L(u_k-u_j)}\)
\end{equation}
For $L$ large, expanding in $1/L$ one gets
\begin{equation}
\label{gaugecurve}
\frac{1}{u}=2\pi  \tilde n_C+2\pint dv \frac{\tilde \rho(v)}{u-v}\;,
\end{equation}
with $\int du\; \tilde \rho(u)=J/L$.
In order to compare with (\ref{stringcurve}), we make the redefinition $x\to xL/g$ and take into account (\ref{deltakappa}) that $4\pi \kappa=\Delta/g$ such that (\ref{stringcurve}) becomes
\begin{equation}
\label{stringcurvebis}
2\pint dy \frac{\rho(y)}{x-y}=\frac{x\Delta/L}{x^2-g^2/L^2}+2\pi n_C\;.
\end{equation}
This transformation introduces explicitly the BMN coupling constant 
$$\lambda'=16\pi^2 g^2/L^2\equiv 16\pi^2 g'^2\;.$$
Since $\Delta=L+\CO(\lambda')$, we see that at leading order in $\lambda'$ (\ref{stringcurvebis})
and (\ref{gaugecurve}) coincide. We can go with the comparison at higher loops, by using the
Inozemtsev asymptotic Bethe ansatz (\ref{InBA}). If we replace the Inozemtsev rapidity $\varphi \to Lu$, we obtain perturbatively the potential on the l.h.s. of (\ref{gaugecurve})
\begin{equation}
\label{Inopot}
pL=\frac{1}{u}+{2g'^2}\frac{1}{u^3}+{6g'^4}\frac{1}{u^5}+\ldots
\end{equation}
In this expansion, we dropped the higher powers of $1/L$. The terms under the dots do not have the so-called BMN scaling, in powers of $g'=g/L$ plus $1/L$ corrections, anymore.
On the other hand, from the first and the last normalization conditions (\ref{normdens}) we deduce that 
\begin{equation}
\label{normgs}
\int dx \(1-\frac{g'^2}{x^2}\)\rho(x)=\frac{J}{L}=\int du\; \tilde \rho(u)
\end{equation}
the last equality being the normalization condition for the density in the spin chain. The relation 
(\ref{normgs}) suggests that the Bethe ansatz rapidity $u$ and sigma model rapiditiy $x$ are related by 
\begin{equation}
\label{bdsrap}
u=x+\frac{g'^2}{x}\;, \qquad \rho(x)=\tilde \rho(u)\;
\end{equation}
at least perturbatively. Inserting this definition in the perturbative expansion (\ref{Inopot}) of the  potential we obtain
\begin{equation}
\label{Inopotp}
pL=\frac{1}{x}+{g'^2}\frac{1}{x^3}+{g'^4}\frac{1}{x^5}+\ldots\simeq \frac{x}{x^2-g'^2}\;.
\end{equation}
The proposals (\ref{bdsrap}) and (\ref{Inopot}) were made by Beisert, Dippel and Staudacher (BDS) \cite{Beisert:2004hm} in order to extend the Inozemtsev Bethe ansatz beyond the three loop order. The corresponding equations, written in the variable $x$, are
\begin{equation}
\label{gaugecurvefin}
2\pint dy \frac{ \rho_{BDS}(y)}{x-y}=\frac{x}{x^2-g'^2} +\frac{2g'^2}{x^2}\pint dy \frac{ \rho_{BDS}(y)}{(x-y)(1-g'^2/xy)}+2\pi  n_C\;,
\end{equation}
while the corresponding equation obtained from the sigma model reads
\begin{equation}
\label{stringcurvefin}
2\pint dy \frac{ \rho_s(y)}{x-y}=\frac{x}{x^2-g'^2} +\frac{2g'^2}{x^2}\pint dy \frac{ \rho_s(y)}{(x-y)(1-g'^2/x^2)}+2\pi  n_C\;.
\end{equation}
The two equations are almost identical, except for a term in the denominator in the integral in the r.h.s. Since the two equations are different, we have assigned different indices to the corresponding densities. When expanded in powers of $g'$, they agree up to order $g'^2$ and start to disagree from order $g'^4$. The mismatch will occur in the conformal dimensions at order  $g'^6$. At order $g^6$, both Inozemtsev and BDS Bethe ansatz are known to be correct, therefore the mismatch comes from a different order of limits taken to arrive at the two formulas. In the case of the string sigma model (\ref{stringcurvefin}), one takes $g\to \infty,\ L\to \infty$ with $g'=g/L$ fixed, then $g'\ll 1$ while 
in the case of the gauge theory spin chain (\ref{stringcurvefin}) one takes $g$ finite, $L\to \infty$ and then $g'=g/L\ll1$. Knowing the all-loop Bethe ansatz allows to remedy this mismatch. 

The comparison between the BDS chain and the $su(2)$ principal chiral model presented above involves only one of the $su(2)$ symmetry sectors. A spin model with two $su(2)$ types of spin  was proposed by Faddeev and Reshetikhin 
\cite{Faddeev:1985qu} 
as a discretization of the
principal chiral model, see also \cite{faddeev} . 

The algebraic curve for the whole  system was obtained in \cite{Beisert:2005bm}. Intermediate steps in obtaining this result consisted in analyzing the thermodynamical limit of the $so(6)$  \cite{Beisert:2004ag} and $psu(2,2|4)$  \cite{Beisert:2005di} one-loop Bethe ans\"atze.

\section{All loop integrability in ${\cal N}=4$ SYM}

The first step in extending the Bethe ansatz to all loop order was made by Beisert, Dippel and Staudacher \cite{Beisert:2004hm}, who proposed how to extend the three-loop $su(2)$ Bethe ansatz obtained from the Inozemsev model such that it matches at least partially with the prediction of the string sigma model and with the BMN dispersion relation (\ref{BMNdisp}). As it was explained in the previous chapter, the BDS ansatz is not able to reproduce the results at strong coupling. There is therefore a missing piece which allows to interpolate between weak and strong coupling, and this piece was called {\it dressing phase}, for reasons which will be clear later. It was by no means obvious that the all-loop ansatz proposed by BDS is consistent and that it corresponds to any spin chain, be it with multiple spin, long range interaction, except from the fact that it matched a 
fifth-order integrable extension of the dilatation operator. 

Shortly after, Beisert and Staudacher  \cite{Beisert:2005fw} conjectured the $su(1|1)$ symmetric $S$-matrix for the 
$su(2|1)$ sector at all loops, and, from the consistency relations which the nested Bethe ansatz has to satisfy, they conjectured the full $psu(2,2|4)$ Bethe ansatz at all loops. As in the case of Inozemtsev spin chain, the Bethe ansatz is only asymptotic, that is, is valid for $L\to \infty$.

A new step forward was made by Beisert \cite{Beisert:2005tm}, who analyzed the $su(2|2)$ symmetric
$S$-matrix of the $su(3|2)$ spin chain. This sector corresponds to one of the two branches of the Dynkin diagram and is also the smallest sector in which the spin chain can change its length. While on a finite chain this property seems to introduce an insurmontable difficulty, on a chain of infinite length it allows to modify the magnon symmetry group from $su(2|2)$ to centrally extended $su(2|2)$, the extra central charge being related to the magnon momentum. This symmetry allows to determine the magnon dispersion relation, which is identical to that proposed by BDS \cite{Beisert:2004hm}. The same symmetry imposed to the $S$-matrix, which is supposed to be factorizable, fixes the $S$-matrix up to an overall factor called the dressing factor.  
As explained in chapter (\ref{symmetries}), the choice of the reference state breaks the $psu(2,2|4)$
symmetry to $su(2|2)\otimes su(2|2)$, with the two factors playing symmetric roles. Therefore, imposing the symmetry is enough to find the full $S$-matrix, modulo the dressing factor. The Bethe ansatz associated with the $S$-matrix determined by Beisert  \cite{Beisert:2005tm} coincides with the one conjectured by Beisert and Staudacher \cite{Beisert:2005fw}. The only assumptions are that the system
{\it is} integrable and the asymptotic condition $L\to \infty$.

Although the $\CN=4$ SYM  Bethe ansatz equations where conjectured and then proven using symmetry considerations, a concrete realization of the corresponding spin chain was lacking. It was surprising to find that the BDS ansatz can be obtained by projecting the Hubbard model at half filling, in the way in which the Heisenberg model can be obtained from the Hubbard model at half filling in the limit of strong Coulomb interaction \cite{taka1}. The one dimensional Hubbard model is integrable and its solution in terms of Bethe ansatz was found by Lieb and Wu \cite{LW}. In the language of the Hubbard model, the $\CN=4$ SYM magnons are bound states of fermions. In this representation, it is straightforward to obtain the BMN/BDS dispersion relation for the magnons. 

The relation between the Hubbard model and the $\CN=4$ SYM magnons is deeper, and it fair to say that it is not yet completely understood. It was noticed by Staudacher \cite{Matthias} and elaborated by Beisert \cite{Beisert:2006qh} that the $S$-matrix  with centrally extended $su(2|2)$ symmetry is essentially the same as the $R$-matrix of the Hubbard model \cite{ShastryR}.

\subsection{The BDS conjecture}

The computation the dilatation operator at higher loops was done perturbatively up to five loops \cite{Beisert:2004yq} in the $su(2)$ sector. The constraint of integrability, the structure inherited from the perturbative gauge theory and the condition of BMN scaling fixed uniquely the Hamiltonian of the spin chain. However, it
became more and more clear\footnote{See the discussion in the previous chapter about the three loop mismatch.} that the BMN scaling should not hold at arbitrary loop order. If one relaxes the BMN constraint, more integrable spin chains exist. 

Let us concentrate first to the chain which obeys the BMN scaling at five loops. Beisert, Dippel and Staudacher \cite{Beisert:2004hm} conjectured that this spin chain exists at any loop order and it is diagonalized by the following Bethe ansatz
\begin{equation}
\label{BDSBA}
e^{ip_kL}=\prod_{l\neq k=1}^M \frac{u(p_k)-u(p_l)+i}{u(p_k)-u(p_l)-i}
\end{equation}
with
\begin{equation}
\label{BDSrap}
u(p)=\frac{1}{2}\cot \frac{p}{2}\sqrt{1+16g^2\sin^2\frac{p}{2}}
\end{equation}
and the energy given by
\begin{equation}
\label{BDSen}
E_{BDS}=\sum_{k=1}^M\(\sqrt{1+16g^2\sin^2\frac{p_k}{2}}-1\)
\end{equation}
The equations (\ref{BDSBA}) should be supplemented with the condition of translational invariance
$$\prod_{k=1}^Me^{ip_k}=1\;.$$
As we have seen in the previous chapter, one can introduce another rapidity variable $x$, related to $u$ by the relation\footnote{In order to avoid too cumbersome notations, we keep the same notation for the rapidities $u$ and $x$ in different normalizations. It should be clear from the context which normalization is used.} 
\begin{equation}
\label{Juk}
u=x+\frac{g^2}{x}
\end{equation}
In this notation, the momentum $p$ can be expressed as
\begin{equation}
\label{pJuk}
e^{ip}=\frac{x^+}{x^-}\;, \quad {\rm with}\quad x^{\pm}=x(u\pm i/2)\;,
\end{equation}
and the BDS equations can be rewritten as
\begin{equation}
\label{BDSBAx}
\(\frac{x_k^+}{x_k^-}\)^L=\prod_{l\neq k=1}^M \frac{u_k-u_l+i}{u_k-u_l-i}
\end{equation}
with the energy
\begin{equation}
\label{BDSenx}
E_{BDS}=g^2\sum_{k=1}^M\(\frac{i}{x_k^+}-\frac{i}{x_k^-}\)\;.
\end{equation}
The higher conserved charges of the BDS spin chain, which generalize the higher conserved charges
of the Heisenberg model, can be put in the form
\begin{equation}
\label{BDShc}
q_{r}=\sum_{k=1}^M\frac{i}{r-1}\left[\frac{1}{(x_k^+)^{r-1}}-\frac{1}{(x_k^-)^{r-1}}\right]\;.
\end{equation}
The limit $r\to 1$ gives the momentum $p$ and $r=2$ is, up to a constant, the energy $E_{BDS}$.
 
In addition to reproducing the known spectrum of the dilatation operator up to three loops, the BDS
ansatz has also the merit to reproduce the BMN dispersion relation (\ref{BMNdisp}). For two magnons 
with $p_1=-p_2=p\simeq 2\pi n/L$ with $n\ll L$, the energy (\ref{BDSen}) is given by 
\begin{equation}
E_{BMN}=2\(\sqrt{1+\lambda' n^2}-1\)\;.
\end{equation}
The dispersion relation fixes the two-spin part of the interaction, or more precisely it fixes the 
part of the interaction which is linear in the operators
$$Q_{i,i+n}\equiv P_{i,i+n}-1$$
as it was shown by Ryzhov and Tseytlin \cite{Ryzhov:2004nz}. The higher power in non-overlapping permutations $Q_{i,i+n}$ vanish on the one-magnon states. The corresponding interaction strength is 
simply the Fourier transform of the magnon dispersion relation (\ref{BDSen}) and it reads 
\begin{equation}
\label{TR}
h(n,g)=(4g^2)^n\frac{\Gamma(n-1/2)}{4\sqrt{\pi}\Gamma(n+1)} \ _2F_1\(n-\frac{1}{2},n+\frac{1}{2}, 2n+1, -16g^2\)
\end{equation}
As a function of the coupling constant $g$, the interaction strength has singularities at the points $g=0,\ \infty$ and $\pm i/4$. At these last two points, the interaction strength simplifies, to give
\begin{equation}
\label{intAF}
h(n,\pm i/4)=\frac{(-1)^n}{4\pi}\frac{1}{n^2-1/4}
\end{equation}
This is an antiferromagnetic interaction, as opposed to the ferromagnetic interaction we retrieve at $g^2>0$. This property is also confirmed by the magnon dispersion relation which become at this point
\begin{equation}
\label{enAF}
E(p)_{|_{g=\pm i/4}}=|\cos p/2|-1
\end{equation}
which has the minimum at $p=\pi$, which is the antiferromagnetic point. Let us also notice that the magnon energies (\ref{enAF}) are negative, a signal of instability of the ferromagnetic state, which will not be the ground state anymore.

Due to the transformation properties of the hypergeometric function, the strong coupling limit of the interaction (\ref{TR}) is similar to that at $g^2=-1/16$, except from the sign of the interaction
\begin{equation}
\label{intF}
h(n,g)_{|_{g\to \infty}}=\frac{g}{\pi}\frac{1}{n^2-1/4}
\end{equation}
while the magnon energy becomes, at least for $p\gg 1/g$,
\begin{equation}
\label{enF}
E(p)_{|_{g\to \infty}}=4g|\sin p/2|-1
\end{equation}

It is interesting that the interaction (\ref{intAF}) and  (\ref{intF}) ressembles the inverse-square interaction
which is familiar from the study of the spin chains like Haldane-Shastry \cite{Haldane:1987gg,shastry}. The expressions (\ref{intAF}) and  (\ref{intF}) hold for $L\to \infty$, while for a finite chain they should be periodized, such that, for example
\begin{equation}
\label{intAFL}
h(n,g,L)_{|_{g\to \infty}}=\frac{g}{\pi}\frac{\sin^2\frac{\pi}{L}}{\sin \frac{\pi(2n+1)}{2L}\sin \frac{\pi(2n-1)}{2L}}
\end{equation}
This interaction differs from the Haldane-Shastry interaction only by the point-splitting 
  {$$\footnotesize \sin^2 \frac{\pi n }{L}\to\sin \frac{\pi(2n+1)}{2L}\sin \frac{\pi(2n-1)}{2L}$$}
 which is responsible for changing the dispersion relation from quadratic to trigonometric.
 
The comments above hold for any spin chain with the dispersion relation (\ref{BDSen}), integrable or not and diagonalized by the BDS equations (\ref{BDSBA}) or not. 
In order to be integrable, a spin chain of the type (\ref{intAF}) or  (\ref{intF})  has to be supplemented with higher spin interaction. It is interesting to know whether such a model could be obtained from an 
integrable finite-difference Hamiltonian like Ruijsenaars-Schneider \cite{RuijSch} in the same way the Haldane-Shastry spin chain can be obtained \cite{Bernard:1993pm} from the Calogero-Sutherland model. 
Another common point between the BDS model at strong coupling and the Haldane-Shastry spin chain is given by the phase shift
\begin{equation}
\theta(p_1,p_2)=\frac{1}{i}\ln\frac{u(p_1) -u(p_2)+i}{u(p_1)-u(p_2)-i}\ \ {\buildrel  g\to\infty \over \longrightarrow}\ \ \pi \sgn(p_1-p_2)
\end{equation}
where we considered $p\in[0,2\pi]$ and considered the determination of the logarithm such that the  phase shift is antisymmetric, $\theta(p_1,p_2)=-\theta(p_2,p_1)$.

\subsection{BDS spin chain and the Hubbard model}
\label{HubBDS}

A computation which can be easily done for the BDS system in the thermodynamic limit $L\to \infty$ is that of the  energy for the antiferromagnetic state, the state with the maximum number of magnons $M=L/2$. This is the equivalent of the problem solved by Hulth\'en \cite{Hulthen} in 1938 for the Heisenberg model and amounts to solving the integral equation for the density of magnons
\begin{equation}
\label{hultheneq}
-\frac{d p(u)}{d u}=2\,\pi\,\rho(u)+ 2\,\int_{-\infty}^{\infty} du'\, 
\frac{\rho(u')}{(u-u')^2+1}\, ,
\end{equation}
with
\begin{equation}
\label{dp}
-\frac{d p(u)}{d u}=i \frac{d}{du}\log\frac{x^+(u)}{x^-(u)}=
\frac{i}{\sqrt{u_+^2-4g^2}}
-\frac{i}{\sqrt{u_-^2-4g^2}}\, .
\end{equation}
The solution to the equation (\ref{hultheneq}) can be found by Fourier transform and it reads
\begin{equation}
\label{density1}
\rho(u)=
\int_0^{\infty}\,\frac{dt}{2 \pi}\,
\frac{\cos\left(t u\right)\,
J_0({2} g t)}{\cosh\left(\frac{t}{2}\right)}\, ,
\end{equation}
wich gives for the energy of the antiferromagnetic state
\begin{equation}
\label{antiferroenergy}
E_{AF}(g)=2gL\,
\int_0^{\infty}\,\frac{dt}{t}\,
\frac{J_0({2} g t)\,J_1({2} g t)}{1+e^t}\, .
\end{equation}
Up to a sign and an overall constant, this is the energy of the antiferromagnetic ground state of the Hubbard model at half filling computed by Lieb and Wu \cite{LW}. This suggests that there is a link between the BDS spin chain and Hubbard model at coupling $U=1/g$ and  $t=-g$. 

The Hubbard model \cite{Hubbard} is a lattice model for itinerant fermions with spin 1/2. The Coulomb interaction is modelled by an onsite repulsion. In one dimension we have
\begin{equation}
\label{H}
H_{{\rm Hubbard}}=
-t\, \sum_{i=1}^L \sum_{\sigma=\up,\don}
\left(c^\dagger_{i,\sigma} c_{i+1,\sigma}+
c^\dagger_{i+1,\sigma} c_{i,\sigma}\right)+
t\,U\, \sum_{i=1}^L 
c^\dagger_{i,\up} c_{i,\up}c^\dagger_{i,\don} c_{i,\don}\, .
\end{equation}
The operators $c^\dagger_{i,\sigma}$ and $c_{i,\sigma}$
are canonical Fermi operators satisfying the anticommutation relations
\begin{eqnarray}
\{c_{i,\sigma},c_{j,\sigma'}\}&=&
\{c^\dagger_{i,\sigma},c^\dagger_{j,\sigma'}\}=0\, ,
\\
\{c_{i,\sigma},c^\dagger_{j,\sigma'}\}&=&
\delta_{i j}\,\delta_{\sigma \sigma'}\, .
\nonumber
\end{eqnarray}
Due to the Pauli principle, there are four states by site: with no fermions ($\hole$), with one fermion with
spin up ($\up$), with one fermion with spin down ($\don$) and with two fermions of different spins ($\updn$). Therefore, the Hilbert space of the Hubbard model is of dimension $4^L$, much larger than the Hilbert space of the  spin chain, of dimension $2^L$. A way to separate the spin states $\up,\ \don$, from the others is to go at infinite Coulomb interaction and half-filling, $N_{{\rm fermions}}=L$. In this limit, the states with exactly one fermion of either spin per site have finite energy, while the states where there is 
at least one double occupancy $\updn$ have infinite energy. Following these states by continuity in the perturbation parameter $g=1/U$  will allow to give an all-loop definition to the BDS spin chain.

Explicit computation show that the projection of the Hubbard hamiltonian (\ref{H}) coincides with the BDS spin chain only up to order $g^{L}$ on chains with even length. This is in contrast with the expectation that the BDS ansatz is correct up order $g^{2L}$. This slight mismatch can be corrected by introducing
twisted boundary conditions for the fermion operators or, equivalently, introducing an Aharonov-Bohm flux through the system,
\begin{equation}
\label{Hflux}
H=
g\, \sum_{i=1}^L \sum_{\sigma=\up,\don}
\left(e^{i \phi_\sigma}\,c^\dagger_{i,\sigma} c_{i+1,\sigma}+
e^{-i\phi_\sigma}\,c^\dagger_{i+1,\sigma} c_{i,\sigma}\right)
-\, \sum_{i=1}^L 
c^\dagger_{i,\up} c_{i,\up}c^\dagger_{i,\don} c_{i,\don}\, ,
\end{equation}
with the twist given by
\begin{equation}
\label{twist}
\phi_\sigma=\phi=\frac{\pi(L+1)}{2L}\;, \qquad \sigma=\up,\don\;
\end{equation}
The overall sign of $H$ is opposite to the sign of the usual repulsive Hubbard model; in our case we are interested in getting a ferromagnetic ground state, at least in the spin sector,
\begin{equation}
\label{BPSgs}
|Z^L\rangle = |\up \up \ldots \up \up \rangle=
c^\dagger_{1\up} c^\dagger_{2\up} \ldots 
c^\dagger_{L-1 \up} c^\dagger_{L \up}\,|0\rangle
\end{equation}
We ignore for the moment the fact that the doubly occupied states have, for this choice of sign of the Hamiltonian, lower energy than the singly occupied ones. What is important for us is that, at least for $g\to 0$, the spin states are clearly separated from the states with double occupancy. 

The Hamiltonian (\ref{Hflux}) is covariant under particle-hole transformations on the fermi- onic operators
of either spin \cite{Woynar}. Such a transformation is particularly convenient for states close to the BPS ground state (\ref{BPSgs}), where it largely diminish the number of fermions. We therefore choose to transform
\begin{eqnarray}
\label{particlehole}
\hole &\Longleftrightarrow& \up
\\
\don &\Longleftrightarrow& \updn
\end{eqnarray}
or, explicitly in terms of creation/annihilaton operators
\begin{eqnarray}
\label{Shiba}
c_{i,\hole}=c^\dagger_{i,\up}\;, &\ &
\qquad c^\dagger_{i,\hole}=c_{i,\up}\;,
\\
c_{i,\updn}= c_{i,\don}\;, &\ &
\qquad c^\dagger_{i,\updn}= c^\dagger_{i,\don}\;.
\end{eqnarray}
This transformation changes the grading of the 
states transforming a state with an odd number of fermions into a state with an even 
number of fermions, so we may call it a duality transformation in the sense of section (\ref{fulloneloop}).
In view of the link, not yet fully clarified, of the Hubbard model with a $su(2|2)$ symmetric 
model \cite{Beisert:2006qh}, this transformation can be viewed as a supersymmetry transformation.
The dual Hamiltonian is 
\begin{equation}
\label{Hdual}
H=
g\, \sum_{i=1}^L \sum_{\sigma=\hole,\updn}
\left(e^{i\phi_\sigma}\,c^\dagger_{i,\sigma} c_{i+1,\sigma}+
e^{-i\phi_\sigma}\,c^\dagger_{i+1,\sigma} c_{i,\sigma}\right)-
\sum_{i=1}^L 
(1-c^\dagger_{i,\hole} c_{i,\hole})c^\dagger_{i,\updn} c_{i,\updn}\, .
\end{equation}
where $\phi_\updn=\phi_\don$, while  $\phi_\hole=\pi-\phi_\up$.
Comparing the two expressions (\ref{Hflux}) and (\ref{Hdual}) we conclude that
under the duality transformation, the  Hamiltonian   (\ref{Hflux})  
transforms as
\begin{equation}
\label{dualrelation}
 H(g;\phi,\phi) \to -H(-g;\pi-\phi,\phi)-{M}
\end{equation}
where $M$ is the initial number of fermions with down spins.

The one dimensional Hubbard model was solved by Lieb and Wu \cite{LW}, who showed that the coordinate Bethe ansatz was essentialy the same as that for the two component fermion gas with delta interaction, problem which in turn was solved by Yang \cite{Yang} and Gaudin \cite{Gaudin}. The Lieb-Wu equations for the Hamiltonian in the form (\ref{Hflux}), specialized at half-filling, are
\begin{eqnarray}
\label{liebwu1}
&\ &e^{i\tilde q_nL}=\prod_{j=1}^M 
\frac{u_j-2g\sin (\tilde q_n+\phi) -i/2}
{u_j-2g\sin  (\tilde q_n+\phi)+i/2}\, ,
\qquad n=1,\ldots, L
\\
\label{liebwu2}
&\ &\prod_{n=1}^{L}
\frac{u_k-2g\sin  (\tilde q_n+\phi) +i/2}
{u_k-2g\sin  (\tilde q_n+\phi)- i/2}= 
\prod_{\textstyle\atopfrac{j=1}{j\neq k}}^M
\frac{u_k-u_j +i}{u_k-u_j-i}\, ,
\quad k=1,\ldots, M
\end{eqnarray}
where $\tilde q_k$ are the momenta of the $L$ fermions and $u_j$ are the rapidities of the $M$ reversed spins. The energy is given by
\begin{equation}
\label{entwist}
E=2g\sum_{k=1}^{N_f} \cos (\tilde q_k+\phi)\;.
\end{equation}
The dual Hamiltonian can be diagonalized, by the dual Lieb-Wu equations, which in the particular case of half filling read
\begin{eqnarray}
\label{liebwudual1}
&\ &e^{i q_nL}=\prod_{j=1}^M 
\frac{u_j-2g\sin (q_n-\phi) -i/2}
{u_j-2g\sin  (q_n-\phi)+i/2}\, ,
\qquad n=1,\ldots, 2 M
\\
\label{liebwudual2}
&\ &\prod_{n=1}^{2M}
\frac{u_k-2g\sin  ( q_n-\phi) +i/2}
{u_k-2g\sin  (q_n-\phi)- i/2}=
-\prod_{\textstyle\atopfrac{j=1}{j\neq k}}^M
\frac{u_k-u_j +i}{u_k-u_j-i}\, .
\quad k=1,\ldots, M
\end{eqnarray}
The number of fermions was reduced from $L$ to $M+L-(L-M)=2M$ and the energy is now given by
\begin{equation}
\label{liebwudualeng}
E=-2g\;
\sum_{n=1}^{2M}\cos  (q_n-\phi)-{M}\;.
\end{equation}
The minus factor after the equality sign in (\ref{liebwudual2}) comes from the particular value of the twist. The Lieb-Wu equations with arbitrary value of the twist are written in appendix C of \cite{Rej:2005qt}, using a result by Yue and Deguchi \cite{YueDeguchi}.

The equations (\ref{liebwudual1}) and (\ref{liebwudual2}) allow bound states, which were classified by Takahashi \cite{takabound}.
Here, we are interested by the particular case when a pair of  fermions and one magnon make a bound state. Let us first consider the case when just one magnon is present and put
\begin{equation}
\label{1ansatz1}
q_1-\phi=\frac{1}{2}\(\pi+p+i\,\beta\)\, ,
\qquad
q_2-\phi=\frac{1}{2}\(\pi+p-i\,\beta\)\, .
\end{equation}
with real $p$ and $\beta>0$. Because of the imaginary part of the momentum, at large $L$ the l.h.s. of (\ref{liebwudual1}) will be zero or infinite for $n=1,2$. This implies that the r.h.s. has a zero or a pole, respectively,
\begin{eqnarray}
 u-\frac{i}{2} ={2}g \sin ( q_1 -\phi) \;, \qquad  u+\frac{i}{2} ={2}g\, 
\sin (  q_{2} -\phi)\;.
 \end{eqnarray} 
This implies that the real and imaginary part of the momentum are related by
\begin{equation}
\label{pbeta}
\sin\frac{p}{2}\;\sinh\frac{\beta}{2}=\frac{1}{4g}
\end{equation}
and we obtain for the magnon rapidity
\begin{equation}
\label{uph}
u=\frac{1}{2}\cot\frac{p}{2}\sqrt{1+16g^2\sin^2\frac{p}{2}}\;.
 \end{equation} 
 The magnon energy is given by the sum of the two fermion energies and it also takes the BDS form
\begin{eqnarray}
\label{hubdsen}
E= {2}g\left(\sin \frac{p+i\beta}{2}+ \sin  \frac{p-i\beta}{2}\right)-1=\sqrt{1+16g^2\sin^2\frac{p}{2}}-1
 \end{eqnarray} 
The bound states are described by the position of their centers $u$ (\ref{uph}).
 When several such bound states are present, Lieb-Wu equations can be reduced to equations for their centers $u_k$. The first Lieb-Wu equation becomes
 \begin{equation}
 \label{bdsbis}
 e^{ip_kL}=
\prod_{\textstyle\atopfrac{j=1}{j\neq k}}^M\frac{u_k-u_j+i}{u_k-u_j-i}\;.
 \end{equation}
 which is nothing else than the BDS equation,
 while the second Lieb-Wu equation is identically satisfied. 
 The finite size correction are of the order $e^{-\beta L}$. At $g\to 0$, we get from (\ref{pbeta}) that $\beta\sim -2\ln g$ such that the finite size corrections are of the order $g^{2L}$, as expected.

In conclusion, the connection with the Hubbard model provides two key ingredients of the all-loop Bethe ansatz, which are the rapidity-momentum relation (\ref{uph}) and the magnon dispersion relation  (\ref{hubdsen}).

It was first noticed by Janik \cite{Janik:2006dc} that the variables $x^\pm$ defined in (\ref{Juk}) and (\ref{pJuk}) live on a torus. A related parametrization,  obtained by a Gauss-Landen transformation from that of \cite{Janik:2006dc}, appeared independently in 
\cite{Beisert:2006qh} \cite{Gomez:2007zr} and \cite{Kostov:2007kx}
 \begin{equation}
  \label{defU}
      u (s) = \frac{1}{ k}\, \frac{\cn s }{\sn s \ \dn s} 
 \end{equation}
 with the modulus $k$ of the torus 
 \begin{equation}
 k^2=\frac{16g^2}{1+16g^2}.
 \end{equation}
The torus degenerates at the points $k^2=0,1,\infty$, which correspond to the singular values
$g^2=0, \infty$ and $-1/16$ we already encountered in the previous section. Under this parametrization, the 
momentum and the size of the bound states (\ref{1ansatz1}) correspond to the elliptic amplitude function
 \begin{equation}
 \label{ellipar}
 p(s) = \pi - 2\, {\rm am} (\K-s, k),\ \ \b(s)= -i\pi
-2i\, {\rm am} (i\K' - s, k). 
 \end{equation}
and they are related by a translation of quarter of period in both directions on the torus
 \begin{equation}
 p(s+\K-i\K') =2\pi- i\beta(s). 
 \end{equation}

 \subsection{The dressing phase}
 
 As explained in section (\ref{pcm}), although the BDS proposal reproduces the spectrum of the dilatation operator up to three loops in the $su(2)$ sector, it does not reproduce the strong-coupling limit
obtained from the $su(2)$ principal chiral model, {\it cf.} (\ref{gaugecurvefin}) and (\ref{stringcurvefin}).
For the BDS ansatz to reproduce the strong coupling behaviour (\ref{stringcurvefin}), it has to be supplemented with an extra piece $\sigma^2(x_k,x_l)$ which was called {\it dressing factor} in \cite{Beisert:2005fw}
\begin{equation}
\label{BDSBAxs}
\(\frac{x_k^+}{x_k^-}\)^L=\prod_{l\neq k=1}^M \frac{u_k-u_l+i}{u_k-u_l-i}\; \sigma^2(x_k,x_l)
\end{equation}
At weak coupling, the dressing factor should be equal to one at least to the third loop order
\begin{equation}
 \sigma^2(x_k,x_l)=1+\CO(g^6)
\end{equation}
Arutynov, Frolov and Staudacher \cite{Arutyunov:2004vx} showed that , in order to reproduce the leading strong coupling behaviour, one can assume
\begin{equation}
\label{AFSdress}
 \sigma^2(x_k,x_l)=\exp{2i\sum_{r=0}^\infty g^{2r+4} \Big( q_{r+2}(x_k)q_{r+3}(x_l)- q_{r+3}(x_k)q_{r+2}(x_l)\Big)}\;,
\end{equation}
where $q_r(x)$ are the conserved charges defined in (\ref{BDShc}).
This cannot be the right answer at weak coupling, since the first corrections at weak coupling appear at  order $g^4$ and not $g^6$ as they should. The answer to the  problem of interpolation of the dressing phase between weak and strong will be given in section \ref{alldp}.
 
 \subsection{The all-loop asymptotic Bethe ansatz}
 
 \subsubsection{The conjecture}
 
 One of the key steps to extend the BDS ansatz to the full $psu(2,2|4)$ algebra was to write the equivalent ansatz
for the $su(1|1)$ sector, with spin states $Z$ and $\psi$. Beisert and Staudacher \cite{Beisert:2005fw} verified that the three loop dilatation operator computed by Beisert \cite{Beisert:2003ys} could be reproduced with the ansatz
 \begin{equation}
\label{BDSBsu11}
\(\frac{x_k^+}{x_k^-}\)^L=\prod_{l\neq k=1}^M \frac{1-g^2/x_k^+x_l^-}{1-g^2/x_k^-x_l^+}\;\sigma^2(x_k,x_l)
\end{equation}
Embedding the $su(1|1)$ sector into a larger $su(2|1)$ sector, with spin states 
$Z, X$ and $\psi$, allowed to check that corresponding, $su(1|1)$ symmetric, $S$ matrix satisfies the Yang-Baxter equation and therefore corresponds to some consistent integrable model.
Imposing the other consistency conditions coming from the $psu(2,2|4)$ algebra, Beisert and Staudacher \cite{Beisert:2005fw} arrived at the following ansatz, which generalizes the one-loop Bethe ansatz (\ref{FNBA})
\begin{eqnarray}
\label{allFNBA}
1&=&\prod_{l=1}^{M_2} \frac{u_{1,k}-u_{2,l}+i/2}{u_{1,k}-u_{2,l}-i/2}\prod_{l=1}^{M_4} \frac{1-g^2/x_{1,k}x^+_{4,l}}{1-g^2/x_{1,k}x^-_{4,l}}\\ \nonumber
1&=&\prod_{l\neq k}^{M_2} \frac{u_{2,k}-u_{2,l}-i}{u_{2,k}-u_{2,l}+i}\;
\prod_{l=1}^{M_1} \frac{u_{2,k}-u_{1,l}+i/2}{u_{2,k}-u_{1,l}-i/2}\;
\prod_{l=1}^{M_3} \frac{u_{2,k}-u_{3,l}+i/2}{u_{2,k}-u_{3,l}-i/2}\\ \nonumber
1&=&
\prod_{l=1}^{M_4} \frac{x_{3,k}-x^+_{4,l}}{x_{3,k}-x_{4,l}^-}\;
\prod_{l=1}^{M_2} \frac{u_{3,k}-u_{2,l}+i/2}{u_{3,k}-u_{2,l}-i/2}\\ \nonumber
1&=&\(\frac{x_{4,k}^-}{x_{4,k}^+}\)^L\prod_{l\neq k}^{M_4} \frac{u_{4,k}-u_{4,l}+i}{u_{4,k}-u_{4,l}-i}\;\sigma^2(x_{4,k},x_{4,l})\\ \nonumber
\quad &\ & \times \;
\prod_{l=1}^{M_1} \frac{1-g^2/x_{4,k}^- x_{1,l}}{1-g^2/x_{4,k}^+ x_{1,l}}\;
\prod_{l=1}^{M_3} \frac{x_{4,k}^--x_{3,l}}{x_{4,k}^+-x_{3,l}}\;
\prod_{l=1}^{M_5} \frac{x_{4,k}^--x_{5,l}}{x_{4,k}^+-x_{5,l}}\;
\prod_{l=1}^{M_7} \frac{1-g^2/x_{4,k}^- x_{7,l}}{1-g^2/x_{4,k}^+ x_{7,l}}\\ \nonumber
1&=&
\prod_{l=1}^{M_4} \frac{x_{5,k}-x_{4,l}^+}{x_{5,k}-x_{4,l}^-}\;
\prod_{l=1}^{M_6} \frac{u_{5,k}-u_{6,l}+i/2}{u_{5,k}-u_{6,l}-i/2}\\ \nonumber
1&=&\prod_{l\neq k}^{M_6} \frac{u_{6,k}-u_{6,l}-i}{u_{6,k}-u_{6,l}+i}\;
\prod_{l=1}^{M_7} \frac{u_{6,k}-u_{7,l}+i/2}{u_{6,k}-u_{7,l}-i/2}\;
\prod_{l=1}^{M_5} \frac{u_{6,k}-u_{5,l}+i/2}{u_{6,k}-u_{5,l}-i/2}\\ \nonumber
1&=&\prod_{l=1}^{M_6} \frac{u_{7,k}-u_{6,l}+i/2}{u_{7,k}-u_{6,l}-i/2}
\prod_{l=1}^{M_4} \frac{1-g^2/x_{7,k}x^+_{4,l}}{1-g^2/x_{7,k}x^-_{4,l}}\:.
\end{eqnarray}
with conserved charges 
\begin{equation}
\label{BDShcf}
q_{r}=\sum_{k=1}^{M_4}\frac{i}{r-1}\left[\frac{1}{(x_{4,k}^+)^{r-1}}-\frac{1}{(x_{4,k}^-)^{r-1}}\right]\;.
\end{equation}

\begin{figure}
  \centering
    {%
    \includegraphics[width=0.2\textwidth]{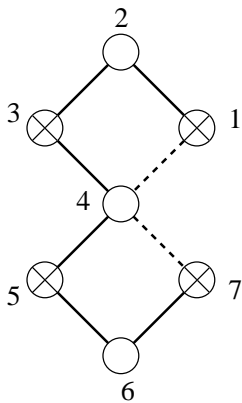}}
  \caption{Connectivity of the nodes of the Dynkin diagram for the all-loop Bethe ansatz. The dotted lines represent
  the nodes which are connected only at $g\neq 0$. }
  \label{fig:enhDyn}
\end{figure}

These equations should be again supplemented with the zero momentum condition
 \begin{equation}
\prod_{k=1}^{M_4}\frac{x_{4,k}^+}{x_{4,k}^-}=1
\end{equation}
It is interesting to note that the connectivity of the roots of the Dynkin diagram is  enhanced compared to the one-loop ansatz (\ref{FNBA}), since 
the roots number $1$ and number $7$ are connected to the root number $4$ by a factor which goes to
unity when the coupling constant vanishes, $g\to 0$, see figure \ref{fig:enhDyn}. The symmetry of the diagram is also enhanced, as the nodes $1$ and $3$ and $7$ and $5$ can be interchanged by transforming
the $x_1\leftrightarrow g^2/ x_3 $ and  $x_7\leftrightarrow g^2/ x_5 $, at the expense of exchanging 
$L\leftrightarrow L-M_1+M_3$ and $L\leftrightarrow L-M_7+M_5$ respectively\footnote{Extra factors of $-1$ appear for $M_4$ odd.}. In fact, as there is no self-interaction for the odd nodes, the roots can be transformed one by one from type $1$ to type $3$ and from type $7$ to type $5$, 
$x_{1,j}\to g^2/ x_{3,l} $ and $x_{7,j}\to g^2/ x_{5,l} $. 
Such an operation changes the length of the spin chain
$L\to L-1$. 

 \subsubsection{The proof}
 
The excitations corresponding to the Bethe ansatz (\ref{allFNBA}), above the vacuum $\ket{\Omega}=\ket{ZZZZ\ldots ZZZZ}$ are symmetric under $su(2|2)\otimes su(2|2)$, each of the branches which starts at the node $4$ corresponding to one of the $su(2|2)$ components. We can therefore restrain ourselves
to the study of one component, which corresponds to a $su(3|2)$ spin chain with states $Z,\ X,\ Y$ and $\psi^1,\psi^2$. This spin chain is {\it dynamic}, that is its length can vary, since the Hamiltonian connects 
states containing the following two sequences with different lenghts \cite{Beisert:2003jj}
 \begin{equation}
 \label{btof}
\ket{...XYZ...}\quad \leftrightarrow \quad \ket{...\psi^1\psi^2...}\;.
\end{equation}
Terms of this type appear in the Hamiltonian first at order $g^3$, which corresponds to loop order one and half at they do not exist at one-loop order.

The definition of a magnon is, at least at one loop order, a replacement of $Z$ by another field $X,\ Y,\ \psi^1$ or $\psi^2$, which are interchanged by the action of $su(2|2)$. 
At higher loop order,
because of the dynamical nature of the spin chain, the definition of a magnon is more subtle. 
It is rather easy to define the action of a $su(2|1)$ algebra on the magnons, which respects the structure of the interaction (\ref{btof}) and does not lead to states of fluctuating length. One considers the states $\ket{\phi}=\ket{X}$ and $\ket{\chi}=\ket{YZ}$ 
together with the fermions $\ket{\psi^{1,2}}$, which form a four-dimensional representation representation of $su(2|1)$. The finite-dimensional representations of $su(2|1)$ were classified in \cite{Scheunert:1976wj}, and the four dimensional representation generically depend on a continuous parameter ${\bf b}$ which is the charge for  the $u(1)$ part of the bosonic subalgebra. 
In such a representation, the generators act like \cite{Janik:2006dc}
 \begin{eqnarray}
R&=&\frac{1}{2}\ket{\phi}\bra{\phi}-\frac{1}{2}\ket{\chi}\bra{\chi}\\
L_\al^\be&=&\delta_{\ga\be}\ket{\psi^\al}\bra{\psi^\gamma}-\frac{1}{2}\delta^\al_{\be}\ket{\psi^\ga}\bra{\psi^\gamma}
 \end{eqnarray}
 and the odd generators act like
  \begin{eqnarray}
Q^1=a\ket{\psi^1}\bra{\phi}+b\ket{\chi}\bra{\psi^2} &\qquad& Q^2=a\ket{\psi^2}\bra{\phi}-b\ket{\chi}\bra{\psi^1} \\
S_1=c\ket{\psi^2}\bra{\chi}+d\ket{\phi}\bra{\psi^1} &\qquad& S_2=-c\ket{\psi^1}\bra{\chi}+d\ket{\phi}\bra{\psi^2}
 \end{eqnarray}
 From the commutation relation
 \begin{equation}
 \label{ensu21}
\{Q^\al,S_\be\}=L_\al^\be+\delta^\al_{\be}(R+C)
\end{equation}
with $C$ a central element\footnote{Here it was chosen to separate the generator $R$ from $C$ because $C$ is related to the dilatation operator;  in fact the commutation relations of $su(2|1)$ can be expressed only in terms of the $u(1)$ generator $R+C$.}   related to the energy
 \begin{equation}
C=\frac{1}{2}+bc={\bf b\; }
\end{equation}
we obtain the constraint
 \begin{equation}
ad-bc=1.
\end{equation}
 Representations with ${\bf b}$ of opposite sign are conjugate.
The quadratic Casimir of $su(2|1)$ takes the value ${\bf b}^2-1/4$ in the representation mentioned above, and it vanishes for ${\bf b}=\pm 1/2$.
At this value, the four dimensional representation splits into a three dimensional fundamental representation and the trivial representation. As we will see, this happens at $g=0$, where the interaction term  (\ref{btof}) does not exist. 

This representation can be lifted to a four dimensional representation of centrally extended $su(2|2)$.
The commutation relations of $su(2|2)$ are given by \cite{Beisert:2005tm}
  \begin{eqnarray}
\[R^a_b,J^c\]&=&\delta^c_b J^a-\frac{1}{2}\delta^a_b J^c\;,\\
\[L^\al_\be,J_\ga\]&=&\delta^\ga_\be J^\al-\frac{1}{2}\delta^\al_\be J^\ga\;,\\
\{Q^\al_a,S_\be^b\}&=&\de^b_a L_\al^\be+\delta^\al_{\be}R^b_a+\de^b_a\delta^\al_{\be}C
\end{eqnarray}
where the indices $a,b,c$ and $\al,\be, \ga$ take the values $1,2$ and $J^c$ and $J^\gamma$ are any operators carrying the quantum numbers of one of the two $su(2)$ subalgebras.  
This super algebra admits a central extension via
  \begin{eqnarray}
\{ Q^\alpha_a,Q^\beta_b\}=\varepsilon^{\al \be}\varepsilon_{ab} P\\
\{ S_\alpha^a,S_\beta^b\}=\varepsilon_{\al \be}\varepsilon^{ab} K
\end{eqnarray}
To define a consistent action of the centrally extended $su(2|2)$ on the four states of the representation previously considered, we have to work with states with variable length. 
We consider a magnon in a periodic chain of large but otherwise arbitrary length
 \begin{eqnarray}
\ket{\Upsilon}=\sum_n e^{inp}\ket{....ZZ{\buildrel n-1, \over Z}{\buildrel n,\over \Upsilon}{\buildrel n+1,\over Z}...} \quad {\rm with}\\
\nonumber \Upsilon\in\{\phi^1\equiv X,\ \phi^2\equiv Y,\ \psi^1,\ \psi^2\}
 \end{eqnarray}
Upon adding a new site, the one-magnon state transforms as
  \begin{equation}
\ket{\Upsilon Z}\simeq \ket{\Upsilon }\qquad 
\ket{Z\Upsilon }\simeq e^{-ip} \ket{\Upsilon }
 \end{equation}
  The bosonic subalgebra should not modify the length 
 \begin{eqnarray}
 \label{su22su2}
R^a_b&=&\delta_{cb}\ket{\phi^a}\bra{\phi^c}-\frac{1}{2}\delta^a_{b}\ket{\phi^c}\bra{\phi^c}
\\ \nonumber
L_\al^\be&=&\delta_{\ga\be}\ket{\psi^\al}\bra{\psi^\gamma}-\frac{1}{2}\delta^\al_{\be}\ket{\psi^\ga}\bra{\psi^\gamma}
 \end{eqnarray}
while for the action of the fermionic operators there are several possible gradings, the most symmetric being \cite{fabian}
 \begin{eqnarray}
 \label{su22twisted}
Q^\al_a\,\ket{\phi^b}&=&a\, \de^b_a\, \ket{\psi^\al Z^{1/2}}\\ \nonumber
Q^\al_a\,\ket{\psi^\be Z^{-1/2}}&=&b\, \varepsilon^{\al \be}\,\varepsilon_{ab} \, \ket{\phi^b }\\ \nonumber
S_\al^a\,\ket{\phi^b}&=&c\, \varepsilon_{\al \be}\,\varepsilon^{ab}\,  \ket{\psi^\be Z^{-1/2}}\\ \nonumber
S_\al^a\,\ket{\psi^\be  Z^{1/2}}&=&d\, \de^\be_\al\, \ket{\phi^a}\\ \nonumber
P\,\ket{\Upsilon}&=&ab\,\ket{\Upsilon Z}\\ \nonumber
K\,\ket{\Upsilon}&=&cd\,\ket{\Upsilon Z^{-1}}
 \end{eqnarray}
In this notation, the su(2|1) representation displayed in Fig. 5 is realized by $\ket{\phi^1},\ \ket{\phi^2 Z}$, $\ket{\psi^{1,2} Z^{1/2}}$, at fixed length. The extra $su(2|2)$ generators connect copies of this representations at different lengths. The grading (\ref{su22twisted}) makes manifest another $su(2|1)$, with representation spanned by $\ket{\phi^1 Z},\ \ket{\phi^2 }$, $\ket{\psi^{1,2} Z^{1/2}}$.
 \begin{figure}
  \centering
    {%
    \includegraphics[width=0.8\textwidth]{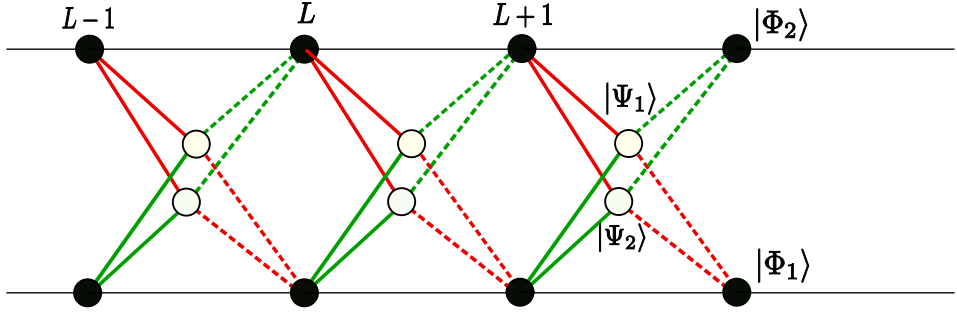}
    \vskip.1cm}
  \caption{Structure of the representation (\ref{su22twisted}) of the centrally extended $su(2|2)$ algebra. The links in red (green) represent action of two $su(2|1)$ subalgebras. The dotted links  disappear at $g=0$, leaving  $su(2|2)$ four dimensional representations at constant length.}
  \label{fig:repsu22}
\end{figure}

The closure of the commutation relation imposes 
 \begin{equation}
ad-bc=1
\end{equation}
and the energy is given again by (\ref{ensu21}). 

For states with several magnons, one has to take the tensor product of the one-magnon representations
above, and impose at the end that 1) the resulting state is invariant by (discrete) translation and 2) the global central charges associated with the central extension $P$ and $K$ vanish. These conditions are equivalent if one sets 
 \begin{equation}
a_kb_k=\alpha (e^{ip_k}-1)\;, \qquad c_kd_k=\beta (e^{-ip_k}-1)
\end{equation}
with the constants $\al$ and $\be$ the same for all the magnons. This relation implies 
 \begin{equation}
 \label{BDSgroup}
a_kb_kc_kd_k={\bf b}_k^2-\frac{1}{4}=4\alpha\beta \sin^2\frac{p_k}{2} \quad {\rm or} \quad C_k=\pm\frac{1}{2}\sqrt{1+16 g^2 \sin^2\frac{p_k}{2}}
\end{equation}
where in the last  formula we identified by physical reasons $\al\be=g^2$. The result is the BDS dispersion relation, which arises here uniquely from supergroup analysis. Another peculiar feature is that, unlike for the usual spin chains, the momentum of the magnon, which is related to the spectral parameter, is now a parameter of the representation. This is specific to the super-spin chains, since representation of super-algebras often involve continuous charges.

The image presented above is valid for large values of $L$. For $L$ finite,
$P$ and $K$ cannot be central elements anymore, since applying $K$ a finite number of times will reduce the length of the chain to zero. Since the centrally extended $su(2|2)$ algebra is a contraction 
of the exceptional superalgebra $d(2,1;\varepsilon)$ for $\varepsilon\to 0$ \cite{Beisert:2005tm}, it is tempting to associate $\varepsilon$ with $1/L$, although this possibility was not explored in the literature. The structure of the magnon representation at finite $L$ is still an open problem.

It is possible to set \cite{Beisert:2005tm}
 \begin{equation}
  \label{BDSparam}
a_k=\gamma_k\;, \quad b_k=-\frac{\alpha }{\gamma_k x_k^+}(x_k^+-x_k^-)\;, \quad c_k=\frac{i g^2 \gamma_k}{\alpha x_k^-}\;, \quad d_k=-\frac{1}{\gamma_k}\beta (x_k^+-x_k^-)\;,
\end{equation}
which allows to retrieve
 \begin{equation}
e^{ip_k}=\frac{x_k^+}{x_k^-}\;, \quad C_k=\frac{1}{2}\(1+\frac{ig^2}{x_k^+}-\frac{ig^2}{x_k^+}\)\;.
\end{equation}
The closure relation $a_kd_k-b_kc_k=1$ corresponds to the constraint
 \begin{equation}
{x_k^+}+\frac{g^2}{x_k^+}-{x_k^-}-\frac{g^2}{x_k^-}=i\;.
\end{equation}
The dispersion relation in (\ref{BDSgroup}) appears with both positive and negative sign, and the two signs correspond to conjugate representations. On the parametrization (\ref{BDSparam}), conjugation corresponds to $x_k^\pm \to g^2/x_k^\pm$ and it reverses the sign of the momentum $p_k$ as well.
This is a sign that the conjugate representations should correspond to antiparticles. This observation is essential to implement the crossing symmetry which will determine the dressing phase. 

The negative branch cannot be seen in perturbative gauge theory. There is no notion of antiparticle in the spin chains, and the correspondence with the Hubbard model in section  \ref{HubBDS} is able only to produce the positive branch of the energy. This is related to the difficulty of obtaining a spin chain description beyond three loops, where the effects of the dressing phase and therefore of the existence of the antiparticles
 will manifest themselves.
In the strong coupling or the string limit, which is described, at least in some sectors, by a relativistic field theory, the antiparticles are unavoidable.  That is the case for example for the $su(2)$ principal chiral model described in section \ref{pcm} where the inversion $x\to 1/x$ interchanges the two symmetries $su(2)_L$ and $su(2)_R$.  It is interesting that the two $su(2)$ sectors cannot be separated in the string limit, as noted by Minahan \cite{Minahan:2005jq}, as they are in the perturbative limit.

The next step is to analyze the scattering of magnons. The two-magnon scattering matrix can be defined along the lines in section \ref{BAHeis}  for the Heisenberg spin chain, as the ratio of the components
of the two-magnon wave function for $x_1\ll x_2$ and $x_2\ll x_1$, where $x_1$ and $x_2$ are the positions of the magnons. Unlike in section \ref{BAHeis}, here the magnons have several flavors corresponding to one of the two centrally extended $su(2|2)$ algebras. The two magnon states live in the tensor product of two representations (\ref{su22twisted}) and the condition that the two-magnon scattering matrix $\CS_{12}$ is invariant under the centrally extended $su(2|2)$ algebra means that 
 \begin{equation}
 \label{symSm}
[J^A_1+J^A_2,\CS_{12}]=0
\end{equation}
where $J^A_{1,2}$ represent any of the generators in the equations (\ref{su22su2}) and (\ref{su22twisted}) in the representations associated to the magnon $1$ and $2$ and characterized by the parameters $x^\pm_1$, $x^\pm_2$.
Beisert \cite{Beisert:2005tm} solved the symmetry constraint (\ref{symSm}) for the $S$ matrix, which depends on the parameters $x^\pm_1$, $x^\pm_2$ as the spectral parameters. The resulting $S$ matrix is uniquely defined, up to a scalar phase factor, and, surprisingly, is related to the Shastry's $R$ matrix \cite{ShastryR} for the Hubbard model \cite{Beisert:2006qh} \cite{Martins:2007hb}. Although this connection to the Hubbard model is seemingly unrelated to that described in section \ref{HubBDS}, the identification of the coupling constants is the same. One of the implications of this connection is that the symmetry which is behind the integrability of the Hubbard model is $su(2|2)$, although in the Hubbard model part of this symmetry is broken, presumably by boundary conditions.

The centrally extended $su(2|2)$ symmetric S-matrix obeys Yang-Baxter equation \cite{Beisert:2005tm}  \cite{Beisert:2006qh} and therefore is compatible with factorized scattering. Diagonalizing it \cite{Beisert:2005tm} through the nested Bethe ansatz procedure reproduces the Beisert-Staudacher equations (\ref{allFNBA}).

The link  of the S-matrix with the Yangian was investigated in \cite{Beisert:2007ds} \cite{deLeeuw:2009hn} \cite{Spill:2008tp}.
The Hubbard model itself possesses a Yangian symmetry, discovered by Uglov and Korepin \cite{Uglov:1993jy}.

\subsection{The dressing phase at all loops}
\label{alldp}

The fact that the $S$ matrix is fixed by symmetry up to a scalar factor is similar to the situation encountered in integrable two-dimensional relativistic field theories, where the scattering matrix of the physical excitation can be obtained postulating a set of properties:
\begin{itemize}
\item symmetry $$[J^A_1+J^A_2,\CS_{12}]=0\;,$$
\item factorized scattering (Yang-Baxter equation) $$\CS_{12}\CS_{13}\CS_{23}=\CS_{23}\CS_{13}\CS_{12}\;,$$
\item unitarity $$\CS_{12}(p_1,p_2)\CS_{21}(p_2,p_1)=1\;,$$
\item crossing symmetry 
\begin{equation}
\label{cross}
\CS_{12}(p_1,p_2)C_1\CS^{t_1}_{12}(-p_1,p_2)C_1^{-1}=1\;,
\end{equation} 
where $C_1\CS^{t_1}_{12}C_1^{-1}$ stands for charge conjugation on the first particle. 
\end{itemize}
The knowledge of the $S$ matrix allows to construct the Hilbert space as a representation of the so-called Zamolodchikov-Faddeev algebra \cite{Zamolodchikov:1978xm} \cite{Faddeev:1980zy}, which is the algebra of the operators creating the physical excitations. 
Arutyunov, Frolov and Zamaklar \cite{Arutyunov:2006yd}
applied this idea to the string sigma model in the light cone gauge.
The symmetry of the excitations is again centrally extended $su(2|2)\times su(2|2)$. Counting the number of physical excitations, one obtains eight bosons and eight fermions, which in the plane wave limit become identical to the ones from equation (\ref{BMNac}). Up to some subtleties related to the definition of the action of the fermionic generators, the symmetry in \cite{Arutyunov:2006yd} is the same as the one
used by Beisert to determine the scattering matrix in the gauge theory. Therefore, the scattering matrix
in the gauge and string theory are the same, up to a twist. The condition of crossing symmetry is the same in the two theories.

\subsubsection{The quantization of the $su(2)$ principal chiral model and the AFS dressing factor}
Let us consider the simple case of the  $su(2)$ principal chiral model,
where the scattering matrix is given by \cite{Zamolodchikov:1992zr}
 \begin{equation}
\CS^{R,L}_{12}(\theta)=S_0(\theta)\frac{\theta+iP_{12}}{\theta-i} \quad {\rm with}\quad 
S_0(\theta)=i
\frac{\Gamma\(\frac{1}{2}-\frac{i\theta}{2}\)\Gamma\(\frac{i\theta}{2}\)}{\Gamma\(\frac{1}{2}+\frac{i\theta}{2}\)\Gamma\(-\frac{i\theta}{2}\)}
\end{equation}
and 
 \begin{equation}
\CS_{12}=\CS_{12}^R\CS_{12}^L
 \end{equation}
 where, again, the indices $R,L$ refer to the two copies of $su(2)$. 
 The crossing relation reads
  \begin{equation}
S_0(\theta+i/2)S_0(\theta-i/2)=\frac{\theta-i/2}{\theta+i/2}\;.
 \end{equation}
The excitations have relativistic dispersion relation, with 
  \begin{equation}
E=m \cosh (\pi \theta)\;,\qquad   p=m\sinh (\pi \theta)\;. 
 \end{equation}
Diagonalizing the scattering matrix via the nested Bethe ansatz, one obtains
 \begin{eqnarray}
 \label{qpcm}
1&=&\prod_\beta \frac{u_j-\theta_\beta-i/2}{u_j-\theta_\beta+i/2}\;\prod_{l\neq j} \frac{u_j-u_l+i}{u_j-u_l-i}\;,\\ \nonumber
e^{-i\CL m\sinh \pi \theta_\al} &=&\prod_{\beta\neq \al}S_0^2(\theta_\al-\theta_\be)\;\prod_j
\frac{\theta_\al-u_j+i/2}{\theta_\al-u_j-i/2}\;\prod_k
\frac{\theta_\al-v_k+i/2}{\theta_\al-v_k-i/2}\;,\\ \nonumber
1&=&\prod_\beta \frac{v_k-\theta_\beta-i/2}{v_k-\theta_\beta+i/2}\;\prod_{n\neq k} \frac{v_k-v_n+i}{v_k-v_n-i}\;,
\ \end{eqnarray}
where $\CL$ is the circumference of the system, $\al, \be=1,\ldots, L$, $j,l=1,\ldots, J_u$ and  $k,n=1,\ldots, J_v$. In the absence of the magnons of type $v$, these equations are similar to the equations for the Hubbard model (\ref{liebwu1}) and (\ref{liebwu2}), except for the value of $S_0(\theta)$

This is the quantum solution of the model presented in section \ref{pcm} and it 
is an interesting question how to obtain the classical limit of the section \ref{pcm} starting from the equations (\ref{qpcm}). This problem was addressed  in reference \cite{Gromov:2006dh}.
The classical limit can be obtained when $m\sim e^{-2\pi g}\to 0$ and $g\sim L\to \infty$.
One rescales the rapidities as $\theta =2g \xi$ and reduces the Bethe equations to the equilibrium equations for a system of 2D Coulomb charges subject to a potential $V(\xi)$
  \begin{equation}
V(\xi)=\CL m \cosh (2\pi g \xi)
 \end{equation}
 When $g\to \infty$, the potential $V(\xi)$ becomes a box with infinite walls situated at $\xi=\pm 1$, such that the $L$ rapidities $\theta_\al/2g=\xi_\al$ will be confined to this interval. 
The algebraic curve which can be obtained from the equations (\ref{qpcm}) in the continuum limit 
has four sheets, unlike the curve in section \ref{pcm}. However, by putting 
  \begin{equation}
 \xi=\frac{1}{2}\(x+\frac{1}{x}\)
 \end{equation}
 the four sheets of the algebraic curve in $\xi$ reduce to two sheets in the variable $x$, while the cut $\xi\in[-1,1]$ transforms into the unit circle $x=1$. The inversion map $x\to 1/x$ maps the $u$ variables into $v$ variables, as expected. 
 
 As already mentioned, the equations (\ref{qpcm}) are similar to the equations for the Hubbard model,
 where $\theta_\al\sim 2g\sin q_\al$, except for the form $S_0(\theta)$. It is therefore interesting to check what happens with the equations if a small number of magnons (of type $u$, for example), are kept and
 if we eliminate the $L$ particles with rapidity $\theta_\al$. This was done in \cite{Gromov:2006cq}, for the leading order in $1/g$. The result obtained was the BDS-like equation (\ref{BDSBAxs}). Since the result obtained from the principal chiral model cannot be trusted beyond the leading order in $1/g$, only the AFS dressing factor (\ref{AFSdress}) can be obtained. A similar result was obtained by Polchinski and Mann \cite{Mann:2005ab}
starting from an $Osp(2m+2|2m)$ sigma model.
 
 \subsubsection{Higher loop corrections to the dressing phase}
 
 The first one loop corrections to the AFS phase were obtained by Beisert and Tseytlin \cite{Beisert:2005cw} by a direct computation in the string sigma model, and the complete one loop answer was obtained by Hern\'andez and L\'opez \cite{Hernandez:2006tk}. They extended the AFS ansatz to
 \begin{equation}
\label{HLdress}
 \sigma^2(x_k,x_l)=\exp{2i\sum_{r=2}^\infty \sum _{s=r+1}^\infty g^{r+s-1}c_{r,s}(g) \Big( q_{r}(x_k)q_{s}(x_l)- q_{s}(x_k)q_{r}(x_l)\Big)}\;,
\end{equation}
where the first two terms in the expansion of $c_{r,s}(g)$ at strong coupling are 
are 
 \begin{equation}
c_{r,s}(g)=\delta_{r+1,s}-\frac{2}{\pi g}\frac{(r-1)(s-1)}{(r+s-2)(s-r)}+\CO(1/g^2)
\end{equation}
if $r+s$ is odd and zero otherwise. The result was confirmed by Freyhult and Kristjansen \cite{Freyhult:2006vr} and Gromov and Vieira
\cite{Gromov:2007aq} \cite{Gromov:2007cd}. A systematic procedure for the semiclassical quantization of the algebraic curves associated with the string sigma model was given in \cite{Gromov:2007aq} \cite{Gromov:2007cd}. It is difficult to extend this procedure  beyond one loop.

Therefore, the only available option to obtain the full expression of the dressing factor is to implement the crossing symmetry. The main difficulty here is that the scattering matrix depends on the rapidities
of the two particle separately and not only on the rapidity difference. This, in turn, can be related to the 
dispersion relation of the magnons, which is not exactly of the relativistic type
  \begin{equation}
  \label{disperrel}
\epsilon(p)=\sqrt{1+16g^2\sin^2\frac{p}{2}}
\end{equation}
It is interesting to know that the dispersion relation (\ref{disperrel}) corresponds to a quantum deformation of the Poincar\'e group in two dimensions \cite{Gomez:2007zr} which is in turn related to the existence of the elliptic parametrization of the rapidity (\ref{defU}). 

While the relativistic dispersion relation can be parametrized in terms of hyperbolic functions
 \begin{equation}
\epsilon(u)=m\cosh \pi u\;, \quad p(u)=m\sinh \pi u\;, \quad \epsilon^2=m^2+p^2
\end{equation}
and the particle-antiparticle transformation corresponds to $u\to u+i$, the dispersion relation  (\ref{disperrel})  can be uniformized by 
  \begin{equation}
\epsilon(s)=\frac{1}{\dn s}\;, \qquad 4g\sin \frac{p(s)}{2} =k\frac{\sn s}{\dn s}\;.
\end{equation}
The change of sign of both the energy and the momentum occur for the translation $s\to s+2i\K'$,
under which we also have
 \begin{equation}
    x^\pm (s+ 2i \K')={1\over    x^\pm (s)}\;.
    \end{equation}
It is this transformation that was used by Janik \cite{Janik:2006dc} to implement the crossing symmetry.
He worked with the $su(2|1)$ representation of the scattering matrix, where the problems of length changing are absent.
On the labels of the $su(2|1)$ representation, this transformation changes the representation labelled by $(1/2,{\bf b})$ to its conjugate labelled by $(1/2,-{\bf b})$. 
In order to implement the crossing relation (\ref{cross}), one has to specify the action of the conjugation matrix $C$ which relates the two conjugated representations.
Once this action is specified, the next step is to solve the equation
\begin{equation}
\label{crossng}
C_1\CS^{st_1}_{12}(s_1+2i\K',s_2)C_1^{-1}=\CS^{-1}_{12}(s_1,s_2)\;,
\end{equation} 
in the tensor product of representations $(1/2,{\bf b}_1)$ and $(1/2,{\bf b}_2)$. The symbol $^{st_1}$ represents supertransposition in the first space and we used the uniformization parameter $s$ for the rapidity.  For generic values of 
${\bf b}_1$ and ${\bf b}_2$, the tensor product decomposes in three irreducible representations, and equating the coefficients of the projectors on these three representations allows to obtain in particular
the condition satisfied by the scalar part of the scattering matrix
\begin{equation}
\label{crossf}
S_0(s_1+2i\K',s_2)S_0(s_1,s_2)=\frac{(g^2/x_1^+-x_2^-)(x_1^+-x_2^+)}
{(g^2/x_1^--x_2^-)(x_1^--x_2^+)}\;,
\end{equation} 
Similarly to the $su(2)$ case, we do not expect that the dressing factor is invariant by shifts of  $4i\K'$ in $s_1$.  

The crossing equation (\ref{crossf}) has many solutions. Among the required properties, it should reproduce the known AFS \cite{Arutyunov:2004vx} and Hern\'andez L\'opez \cite{Hernandez:2006tk}
behavior at strong coupling and the behavior at weak coupling $S_0=1+\CO(g^6)$ at weak coupling.

A solution which satisfies these constraints was constructed by Beisert, Hern\'andez and L\'opez \cite{Beisert:2006ib}. This solution can be written in the form (\ref{HLdress}), 
 \begin{eqnarray}
S^2_0 (x_1,x_2)&\equiv& \frac{x_1^+-x_2^-}{x_1^--x_2^+}\frac{1-g^2/x_1^-x_2^+}{1-g^2/x_1^+x_2^-}\; \sigma^{-2}(x_1,x_2)\;,\\
\nonumber
\sigma(x_1,x_2)&=&\exp{i\sum_{r=2}^\infty \sum _{s=r+1}^\infty g^{r+s-1}c_{r,s}(g) \Big( q_{r}(x_1)q_{s}(x_2)- q_{s}(x_1)q_{r}(x_2)\Big)}\;, 
\end{eqnarray}
with the coefficients $c_{r,s}(g)$, non zero only for $r+s$ odd, expressed as a series in $1/g$
\begin{equation}
\label{crsinf}
c_{r,s}(g)=\sum_{n=0}^\infty c_{r,s}^{(n)}\;g^{-n}
\end{equation}
and the numbers $c_{r,s}^{(n)}$ are determined as
\begin{equation}
\label{cnrs}
c_{r,s}^{(n)}=\frac{\zeta(n)(r-1)(s-1)}{(-2\pi)^n\Gamma(n-1)}\frac{\Gamma\left[\frac{1}{2}(r+s+n-3)\right]
\Gamma\left[\frac{1}{2}(s-r+n-1)\right]}{\Gamma\left[\frac{1}{2}(r+s-n+3)\right]
\Gamma\left[\frac{1}{2}(s-r-n+1)\right]}\delta_{r+s,1{\rm\; mod \;}2}
\end{equation}
For $n=0,1$ the coefficients $c_{r,s}^{(n)}$ are defined as in (\ref{HLdress}). Beisert, Eden and Staudacher  \cite{Beisert:2006ez} gave compelling evidence that the coefficients describing the
dressing function can be expressed by an expansion for small $g$ in the remarkable form
  \begin{equation}
  \label{crszero}
c_{r,s}(g)=\sum_{n=0}^\infty c_{r,s}^{(-n)}\;g^{n}
\end{equation}
where $c_{r,s}^{(-n)}$ can be defined for positive $n$ by analytical continuation of the ones in (\ref{cnrs})
 \begin{eqnarray}
\label{cmnrs}
&\ &c_{r,s}^{(-n)}=\\ \nonumber
&=&\frac{\zeta(1+n)\cos \left[ \frac{\pi n}{2} \right]\; (r-1)(s-1)\Gamma(n+2)\Gamma(n+1)\;\delta_{r+s,1{\rm\; mod \;}2}}
{\Gamma\left[\frac{1}{2}(5+n-r-s)\right]\Gamma\left[\frac{1}{2}(3+n+r-s)\right]\Gamma\left[\frac{1}{2}(3+n-r+s)\right]\Gamma\left[\frac{1}{2}(1+n+r+s)\right]}\;.
\end{eqnarray} 
 As expected, the pertubative expansion contains only even powers of $g$, due to the factor $\cos \pi n/2$. As $\zeta(n)\to 1$ for $n\to\infty$, (\ref{crszero}) is convergent around $g=0$, while the series (\ref{crsinf}) in $1/g$ has zero radius of convergence. This fact has deep consequences and enormously complicates the strong coupling limit. The analysis of the strong coupling limit is the subject of the next section.
 
Being able to express the dressing factor in perturbation in $g$ allowed to check that it is trivial up to three loop order and to compute anomalous dimensions recursively up to the desired order, at least for operators without wrapping corrections. In particular, Beisert, Eden and Staudacher  \cite{Beisert:2006ez}  computed the four loop contribution to the so-called cusp anomalous dimension. Remarkably,  this correction coincided with an impressive four loop computation \cite{Bern:2006ew} of the cusp anomalous dimension from the gluon amplitudes, later confirmed in \cite{Cachazo:2007ad}. The four-loop direct computation of the dilatation operator in the $su(2)$ sector \cite{Beisert:2007hz} confirmed the 
result. 
 
 \subsubsection{The ``magic'' BES formula}
 
 Since the dressing factor is an important ingredient for determining the anomalous dimensions, it is highly desirable to express it in a more compact form than that in (\ref{crsinf}) and (\ref{cnrs}).
 A remarkable representation as a convolution of scattering kernels was given in  \cite{Beisert:2006ez}. 
  Let us define the scattering kernel, as in appendix \ref{appb}, as the derivative of the scattering phase
 \begin{equation}
K(u,u')= \frac{1}{2\pi  }\frac{d}{du} \;\varphi(u,u')\;
\end{equation}
and take its Fourier transform on both variables $u$ and $u'$ with the definition
 \begin{equation}
K(t,t')=\int_{-\infty}^\infty du\int_{-\infty}^\infty du' \;K(u,u')\;e^{i(tu+t'u')}\;.
\end{equation}
An important building block of the Bethe equations is the $su(1|1)$ kernel
 \begin{eqnarray}
 \label{kernm}
K_m(u,u')&=&-2K_{su(1|1)}(u,u')=\\ \nonumber
&=&\frac{i}{\pi }\frac{d}{du}\left[\ln\(1-\frac{g^2}{x^+(u)x^-(u')}\)-\ln\(1-\frac{g^2}{x^-(u)x^+(u')}\)\right]\;.
 \end{eqnarray}
After Fourier transform, due to the structure of the singularities in $u$ and $u'$, we obtain\footnote{Our definitions for the kernels differs from the ones of \cite{Beisert:2006ez}. We try to keep the notation unitary and as close as possible to the different notations in the literature. }
 \begin{equation}
K_m(t,t')=8\pi g^2 (1-\sgn tt')|t| \sum_{n>0}n\frac{J_n(2g|t|)}{2g|t|}\frac{J_n(2g|t'|)}{2g|t'|}e^{-(|t|+|t'|)/2}\;.
 \end{equation}
To avoid complications due to the sign of $t$, we define the Fourier transform only for positive values of $t, t'$. In the next chapter, we are going to explain the significance of this simplification. We define
then
 \begin{equation}
 \label{kernbess}
K_m(t,t')=16\pi g^2 t \sum_{n>0}n\frac{J_n(2gt)}{2gt}\frac{J_n(2gt')}{2gt'}e^{-(t+t')/2}\equiv K_+(t,t')+K_-(t,t')\;,
 \end{equation}
 where $K_+$ and $K_-$ contain the expansion on {\it odd} and {\it even} Bessel functions respectively.
 The ``magic'' formula of BES expresses the dressing kernel as
  \begin{equation}
  \label{kernd}
K_d(t,t')=8g^2 \int_0^\infty\;\frac{dt''}{2\pi}\,K_-(t,t'')\,\frac{1}{1-e^{-t''}}\,K_+(t'',t')
 \end{equation}
 where $K_d(t,t')$ is defined as
  \begin{equation}
  \label{kernddef}
K_d(t,t')=\frac{i}{\pi }\int_{-\infty}^\infty du\int_{-\infty}^\infty\; du' \frac{d}{du}\ln \sigma  (u,u')\;e^{i(tu+t'u')}\;.
 \end{equation}
Inverse Fourier transformed versions of the dressing phase appeared in \cite{Kostov:2007kx} and \cite{Dorey:2007xn}. The Dorey, Hofman and Maldacena formula
\begin{eqnarray}
 \theta (x, y)&=&\chi(x^{+},y^+)+\chi(x^{-},y^-)-\chi(x^{+},y^-)-\chi(x^{-},y^+)\\ \nonumber
  \chi (x, y)&=&-i\oint \frac{dz_1}{2\pi}\oint \frac{dz_2}{2\pi}\frac{1}{(x-z_1)(y-z_2)} \; \ln \Gamma(1+ig(z_1+\frac{1}{z_1}-z_2-
  \frac{1}{z_2}))\;.
\end{eqnarray}
was used to prove 
\cite{Arutyunov:2009kf} \cite{Volin:2009uv} that the dressing factor satisfies the crossing equation (\ref{crossf}) for arbitary values of the coupling constant $g$.

 An attempt to derive the formula from the consistency conditions of the $psu(2,2|4)$ Bethe equations was made by Sakai and Satoh \cite{Sakai:2007rk} \cite{Sakai:2007ie}, although the interpretation of their calculation is still mysterious. Before, a formula with a similar structure was obtained for some special state inside $psu(2,2|4)$ \cite{Rej:2007vm}. Janik and \L ukowski \cite{Janik:2008hs} argued that this pattern is rather common among the 2d integrable field theories.

\section{Strong coupling limit of the asymptotic Bethe An- satz equations}

\subsection{The cusp anomalous dimension}

The most sophisticated checks of the Bethe ansatz equations (\ref{allFNBA}) involve the anomalous dimension for the twist two operator, or the so-called cusp anomalous dimension. In the language of the $\CN=4$ SYM theory, the twist two operator can be defined as 
  \begin{equation}
  \label{twistwop}
\Tr \( Z\; \CD^S \;Z\)+\ldots
 \end{equation}
where $\CD=\CD_0+i\CD_1$ is a light-cone component of the covariant derivative, the dots represent terms with different orderings of the fields, and $S$ is the Lorentz spin. Gross and Wilczek \cite{Gross:1974cs} showed that the corresponding QCD operators has, at one loop in perturbation theory, a dimension which scales logarithmically with $S$. Later  \cite{Korchemsky:1988si} \cite{Korchemsky:1992xv}, arguments were given that the logarithmic scaling should hold at any order in perturbation theory.  Korchemsky
 \cite{Korchemsky:1988si} showed that the anomalous dimension of the twist-two operator is twice the cusp anomalous dimension, which appears  in the divergences associated to the Wilson loops with cusps \cite{Polyakov:1980ca}.
 Therefore, the conformal dimension of the operator (\ref{twistwop}) scales at large $S$ as
   \begin{equation}
\Delta=S+2+f(g)\ln S+\ldots 
\end{equation}
where $f(g)$ will be referred to in the following as the {\it universal scaling function}. The denomination of {\it universal} originates in the fact that this factor does not depend on the value of the twist, at least for finite value of the twist. A twist  $L$ operator  is defined as
  \begin{equation}
  \label{twistLop}
\Tr \(\CD^{k_1} Z\; \CD^{k_2} \;Z\ldots \CD^{k_L} \;Z\)+\ldots\quad {\rm with}\quad k_1+k_2+\ldots k_L=S
 \end{equation}
 The logarithmic scaling persists when the twist scales like the log of the Lorentz spin, $L\sim \ln S$
 \cite{Belitsky:2006en}, although the scaling function  depends now on the  new parameter, see also \cite{Alday:2007mf}
 \begin{equation}
j=\frac{L}{\ln S}\;, \qquad f(g)\to f(g,j)\;.
\end{equation}
The universal scaling function was determined in perturbation in the gauge theory, at three loop order in \cite{KLOV}, who extracted it from a computations for QCD \cite{Moch}, and to the fourth order numerically from the four-gluon amplitude in \cite{Bern:2006ew}, \cite{Cachazo:2007ad},
 \begin{equation}
 \label{cuspert}
f(g)=8g^2-\frac{8}{3}\pi^2g^4+\frac{88}{45}\pi^4 g^6-16\(\frac{73}{630}\pi^6+4\zeta(3)^2\)g^8+\CO(g^{10})
\end{equation}
where we have replaced the numerical value \cite{Bern:2006ew}, \cite{Cachazo:2007ad} by the probable exact value.
From the string point of view, the state (\ref{twistwop}) corresponds to a solution for a folded string spinning in $AdS_5$ \cite{Gubser:2002tv}. The leading term in the strong coupling expansion was computed in \cite{Gubser:2002tv}, while the  one-loop order was computed in \cite{Frolov:2002av} and the two loop order in \cite{Roiban:2007jf,Roiban:2007dq}
 \begin{equation}
  \label{cuspstring}
f(g)=4g-\frac{3\log 2}{\pi}-\frac{\K}{4\pi^2}\frac{1}{g}+\CO\(\frac{1}{g^2}\)\;,
\end{equation}
where $\K$ is the Catalan constant. The two expansions (\ref{cuspert}) and (\ref{cuspstring}) were reproduced 
from the Bethe ansatz equation and they can be performed recursively to arbitrary order in $g$ or $1/g$.  
This result can be obtained by solving an integral equation for the magnon density corresponding to the state
(\ref{twistwop}) with lowest energy, which belongs to the rank one sector $sl(2)$, the non-compact version of $su(2)$.
The resulting integral equation is
\begin{eqnarray}
\label{intsl2}
&\ &\rho(u)+\rho_h(u)=-\frac{L}{2\pi  }\frac{dp}{du}+
\int_{-\infty}^\infty{d u'}\;\CK(u,u')\  \rho(u')\;,
\end{eqnarray}
where $\rho(u)$ and $\rho_h(u)$ are the densities of magnons and holes respectively, normalized such  that
\begin{eqnarray}
\int du\; \rho(u)=S\;, \qquad \int du\; \rho_h(u)=L
\end{eqnarray}
and the kernel $\CK(u,u')$ is composed from three parts
\begin{eqnarray}
\CK(u,u')=K_{sl(2)}(u,u')-K_m(u,u')-K_d(u,u')\;.
\end{eqnarray}
The kernels $K_m(u,u')$ and $K_d(u,u')$ were defined in (\ref{kernm}) and (\ref{kernd}) and they vanish for $g=0$, while 
\begin{eqnarray}
K_{sl(2)}(u,u')=-K_{su(2)}(u,u')=\frac{1}{\pi}\frac{1}{(u-u')^2+1}\;.
\end{eqnarray}
In the lowest energy state, $L-2$ holes are closely packed around $u=0$ and two are situated at around $u=S/\sqrt{2}$, see {\it e.g.} \cite{Freyhult:2007pz,Bombardelli:2008ah}. 
The equation (\ref{intsl2}) was solved, for $L/S$ finite and $g=0$, in 
\cite{Beisert:2003ea} where it was shown that the density $\rho(\tilde u)$ with $\tilde u=u/S$ has supports on two intervals on the real axis symmetric with respect to the origin, $[-b,-a]$ and $[a,b]$. When $L/S\to 0$, the gap around the origin closes, $a\to 0$. The way the gap closes depends on the way $L/S$ goes to 0 \cite{Belitsky:2006en}. When $L$ is finite and $g=0$, the equation (\ref{intsl2}) becomes singular. Korchemsky \cite{Korchemsky:1995be} solved this equation in the rescaled  variable $\tilde u=u/S$ 
\begin{eqnarray}
\rho_0(\tilde u)=\frac{1}{\pi}\ln \frac{1+\sqrt{1-4\tilde u^2}}{1-\sqrt{1-4\tilde u^2}}\;.
\end{eqnarray}
Let us note that the density is normalized as $\int d\tilde u\; \rho_0(\tilde u)=1$, which expresses the fact that the total umber of magnons is $S$.
At $\tilde u\to 0$ the Korchemsky  density has a logarithmic divergency $\rho_0(\tilde u)\sim -\frac{1}{\pi}\ln(2\tilde u^2)$. This divergency can be removed if one solves
the equation in the original variable $u$  around $u=0$ \cite{Korchemsky:1995be,IK,Bombardelli:2008ah} to get
\begin{eqnarray}
\rho_0 (u)\simeq -\frac{1}{\pi}\[\psi(1/2-iu)+\psi(1/2+iu)+\ln\frac{2}{S^2}
\]\;.
\end{eqnarray}
At scales $1\lesssim u\ll S$, the leading contribution in$S$ in $\rho_0(u)$  is the constant part $\rho_0=\frac{2}{\pi}\ln S$.
Eden and Staudacher \cite{Eden:2006rx}  proposed to treat the case $g\neq 0$ by separating the one-loop contribution from the rest and writing an equation for the density for the fluctuation $\sigma(u)$ defined by
\begin{eqnarray}
\label{ESfluct}
\rho(u)=[\sigma_0-\sigma(u)]\,\ln S+\CO(S^0)
\end{eqnarray}
with $\sigma_0=\frac{2}{\pi}$. This separation works provided we take the limit $S\to \infty$ and neglect $1/S$ corrections. Then $\sigma(u)\to 0$ when $u$ goes to infinity.
The equation satisfied by $\sigma(u)$ is called the BES equation \cite{Beisert:2006ez}
\begin{eqnarray}
\label{intBES}
\sigma(u)= \int_{-\infty}^\infty{d u'}\;\CK(u,u')\  (\sigma(u')-\sigma_0)
\end{eqnarray}
and the universal scaling function is, up to a constant, the normalization of the density $\sigma(u)$
\begin{eqnarray}
f(g)=2 \int_{-\infty}^\infty du\; \sigma(u)
\end{eqnarray}
This expression, due to Lipatov,  can be obtained using the expression for the energy  and the integral equation (\ref{intBES}). The fact that $\sigma (u)$ has a non-zero norm may seem paradoxical, since it is the leading term in the expansion in $1/S$ of difference of two densities $\rho_0(u)$ and $\rho(u)$ which are normalized to $S$. 
The reason is that the expansion in $1/S$ of the density is not uniform in $u$.
Another remark is that a shift of the density $\sigma(u)$ by a constant leaves the equation (\ref{intBES}) unchanged.
Therefore, the right solution is obtained from the condition $\sigma(u)\to 0$ when $u\to \infty$, which is necessary in order to have a finite normalization.

The BES equation can be solved relatively easily in perturbation in $g$ by a recursive procedure, since the expansion of the kernel $K(u,u')$ around  $g$ is well-behaved. 
The first four orders in $g^2$  obtained in \cite{Beisert:2006ez} coincide with the ones obtained in perturbative gauge theory (\ref{cuspert}). The expansion 
in $1/g$ was much more difficult to obtain, due to the fact that $1/g=0$ is an accumulation point of singularities.
The limit of the kernel $K(u,u')$ is not uniform in $u$ and one has to deal differently with different regions in the rapidity space $u$ \cite{Kostov:2007kx,Maldacena:2006rv} .
The leading term in  (\ref{cuspstring}) was first obtained numerically \cite{Benna:2006nd}, and then analytically in \cite{Kotikov:2006ts,Alday:2007qf,Kostov:2007kx,Beccaria:2007tk}. The next term was obtained by Casteill and Kristjansen \cite{Casteill:2007ct} and Belitsky  \cite{Belitsky:2007kf}, and the third term and a recursive procedure for the next terms 
by Basso, Korchemsky and Kotanski \cite{Basso:2007wd} and in \cite{Kostov:2008ax}.

The finite-twist operators in ${\cal N}=4$ SYM were extensively studied, including the wrapping corrections at five loops for the twist-three operator \cite{Beccaria:2009eq} and logarithmic corrections to the scaling with the Lorenz spin \cite{Fioravanti:2009ei}. The finite twist operators obey an interesting property called reciprocity \cite{Basso:2006nk} \cite{Beccaria:2010tb}, which was observed previously in QCD  \cite{Moch:2004sf}. Reciprocity was studied from the point of view of Bethe ansatz in \cite{Beccaria:2008fi}.

\subsection{The generalized scaling function}

Essentially the same methods which were used to compute the universal scaling function can be used to analyze the 
generalized scaling function $f(g,j)$. Freyhult, Rej and Staudacher \cite{Freyhult:2007pz} proposed use parametrization similar to  
(\ref{ESfluct}) but now for the ensemble of magnon and hole densities
\begin{eqnarray}
\label{FRSfluct}
\rho(u)+\rho_h(u)=[\sigma_0-\sigma(u)]\,\ln S+\CO(S^0)
\end{eqnarray}
The support of the hole density is the interval $[-a,a]$ and the support of the magnon density is the complement
$\IR \backslash [-a,a]$.  The density fluctuation is now subject to the FRS equation \cite{Freyhult:2007pz}
\begin{eqnarray}
\label{intFRS}
\sigma(u)=\frac{j}{2\pi}\frac{dp}{du}+ \int_{\IR \backslash [-a,a]}{d u'}\;\CK(u,u')\  (\sigma(u')-\sigma_0)
\end{eqnarray}
The relation between $j$ and $a$ can be obtained via the normalization condition 
\begin{eqnarray}
\label{FRSnorm}
j=\int_{-a}^a du \;\(\sigma_0-\sigma(u)\)
\end{eqnarray}
and the generalized scaling function is given by
\begin{eqnarray}
\label{FRSgensc}
f(g,j)=j+2\int_{-\infty}^\infty du\;  \sigma(u)
\end{eqnarray}
At small $g$ and $j$, the equation (\ref{intBES}) can be solved perturbatively in both parameters. 
The equation (\ref{intBES}) was rederived and analyzed perturbatively in $j$ at any $g$ up to eight order in $j$ in \cite{Fioravanti:2008bh}. When $a\ll 2g$, the hole density can be approximated with the density
of excitations in the $O(6)$ sigma model \cite{Basso:2008tx}. In this regime
\begin{eqnarray}
j\sim m=k g^{1/4} e^{-\pi g}(1+\CO(1/g))\;, \quad {\rm with}\quad k=\frac{(8\pi)^{1/4}}
{\Gamma(5/4)}
\end{eqnarray}
The relation to the $O(6)$  sigma model was
previously identified in \cite{Alday:2007mf}.
A different regime corresponds to $a\gtrsim 2g$, where the relevant parameter is 
\begin{eqnarray}
\ell =\frac{j}{4g}
\end{eqnarray}
In this regime, the universal scaling function was shown to behave \cite{Roiban:2007ju} at strong coupling, for small $\ell$ and up to two loop order
\begin{eqnarray}
f(g,j)&=&f(g)-j\\ \nonumber
&+&\ell^2\[{2g}+\frac{1}{\pi}\(\frac{3}{2}-2\ln \ell\)+\frac{1}{4\pi^2 g}\(q_{02}-6\ln \ell+8\ln^2\ell\)+\CO\(\frac{1}{g^2}\)\]+\CO(\ell^4)
\end{eqnarray}
which means that the $\ell=0$ point is singular, therefore the limits $j\to 0$ and $g\to \infty$ do not commute. This structure can be predicted from the connection to the $O(6)$ sigma model  \cite{Alday:2007mf}.
The coefficient $q_{02}$ was predicted from an early string computation \cite{Roiban:2007ju} to be
\begin{eqnarray}
q_{02}^{\rm string}=-2\K-\frac{3}{2}\ln 2 +\frac{7}{4}
\end{eqnarray}
while the computation \cite{Gromov:2008en} using the Bethe Ansatz equations, confirmed by
 \cite{Bajnok:2008it,Volin:2008kd} predicts 
\begin{eqnarray}
\label{BAgsf}
q_{02}^{\rm BA}=-\frac{3}{2}\ln 2 +\frac{11}{4}\;.
\end{eqnarray}
Recently, a revised string computation \cite{Giombi:2010fa} reproduced the Bethe ansatz result (\ref{BAgsf}).

Finally, it would be interesting to analyze the transition between the two previous regimes, especially 
the regime when $a-2g\sim 1$, which should occur for values $j\sim g^{3/4}$. 

Other computations in the strong coupling limit concern the correction \cite{Freyhult:2009my} \cite{Fioravanti:2009xt} to the logarithmic scaling, the so-called virtuality $B_L(g)$, defined as
\begin{eqnarray}
\Delta-S-L=f(g)(\ln S+\gamma_E-(L-2)\ln 2)+B_L(g)+\CO(1/S)
\end{eqnarray}
where $\gamma_E$ is the Euler-Mascheroni constant.

\subsection{The strong coupling limit}
\label{sec:strg}

Several approaches were taken to approach the strong coupling limit of the BES equation (\ref{intBES}) or other similar integral equations. One of them proposed in \cite{Benna:2006nd} and which was successfully used for numerical computations, uses the representation
of the Fourier transforms of the kernels in terms of Bessel functions  (\ref{kernbess}), which are used as
building blocks of the magnon densities. This method was used in  \cite{Benna:2006nd,Alday:2007qf,Basso:2007wd}.
Other authors used the rapidity representation, starting either from the BES equation, like in \cite{Kotikov:2006ts,Kostov:2007kx,Kostov:2008ax}, or directly from the Beisert-Staudacher equations for large twist $L$ and then taking the small $L$ limit, like in \cite{Casteill:2007ct,Belitsky:2007kf}.

Here, the method of \cite{Kostov:2008ax} will be presented. It allows to obtain the original result of  \cite{Basso:2007wd} by using standard analyticity assumptions with a transparent physical interpretation, and in the same time
it may be helpful to make contact with other problems at strong coupling, like to quantization of the algebraic curve
(\ref{algcur}). 

In order to avoid clumsy formulas, we are going to use notations which are better adapted for the strong coupling limit
\begin{eqnarray}
\la{newu} \e\equiv\frac 1{4g}, \qquad u=\frac{u_{{\rm old}}}{2g}\,
 , \end{eqnarray}
 as well as the variable $x(u)$, related to $u$ by
 \begin{eqnarray}
 u(x)\equiv \frac 12\(x+\frac 1x\),\quad x(u)=u\(1+\sqrt{1-\frac
 1{u^2}}\)\;. \end{eqnarray}
Note the branch cut of $x(u)$ for $u\in [-1,1]$.  In the intermediate
steps we will also use the notations
 \begin{eqnarray}
  x^{\pm }(u)
\equiv x(u \pm i\e) \, .
  \end{eqnarray}
    The kernels will be rescaled by $4g^2$ and the densities by $2g$.

\subsubsection{The resolvent}

Instead of solving the BES equation for the density, we are going to write the equation satisfied by the resolvent
\begin{eqnarray}
R_{\textrm{phys}}(u)=\int_{-\infty}^\infty  du' \;\frac
{\s(v)}{u-u'}\, .
\end{eqnarray}
The resolvent is analytic in the complex plane except on the support of the density. If the density
is normalizable, the resolvent behaves at $u\to \infty$ as
\begin{eqnarray}
\label{infbeh}
R_{\textrm{phys}}(u)\sim \frac{1}{u}\int_{-\infty}^\infty du' \;
\s(u') .
\end{eqnarray}
In the case we are going to consider, the support  of the density is the real axis, so that 
\begin{eqnarray}
\la{density} \s(u)= \frac 1{2\pi
i}\[R_{\textrm{phys}}(u-i0)-R_{\textrm{phys}}(u+i0)\]
\end{eqnarray}
The resolvent is closely related to the Fourier transform of the density, $\sigma(t)$, defined as
\begin{equation}
 \sigma(t)=\int_{-\infty}^\infty du\; e^{itu}\sigma(u)\;.
\end{equation}
 Assuming that $u$ is in the upper half plane (UHP), we have
  \begin{eqnarray}
 R_{\textrm{phys}}(u)&=&-i\int _{-\infty}^\infty du' \int_0^\infty
dt\; e^{it(u-u')}\sigma(u') \\ &=& -i\int_0^\infty dt\;
e^{itu}\sigma(-t)\, .
\end{eqnarray}
 Since the density we are working with is symmetric $\sigma(u)=\sigma(-u)$, the resolvent is antisymmetric $R_{\textrm{phys}}(u)=-R_{\textrm{phys}}(-u)$, so we can retrieve its value in the lower half plane (LHP) once we know that in the UHP. In the following we are going to call resolvent the function $R(u)=R_{\textrm{phys}}(u)$ in the UHP and which is analytically continued to the LHP. Therefore, $R(u)$ is analytic in the UHP, but it may have singularities in the LHP and 
\begin{eqnarray}
\la{densityp} \s(u)= -\frac 1{2\pi
i}\[R(u+i0)+R(-u+i0)\]
\end{eqnarray}

\subsubsection{The linearized BES equation}
\label{sec:linBES}

Our starting point is the BES equation written in Fourier transformed form
\begin{eqnarray}
\label{fourierB}
\[1-K_{sl(2)}(t)\]\sigma(t)=-\int_0^\infty \frac{dt'
}{2\pi}\(K_m(t,t')+K_d(t,t')\)(\sigma(t')-\sigma_0(t'))\;, 
\end{eqnarray}
where $\sigma_0(t)=8\e \delta(t)$ and $K_{sl(2)}(t)=e^{-2\e t} $.
In abstract notations it can be written as
\begin{eqnarray}
\label{abstractBES}
-S^{-1}\sigma=\(K_++K_-+2K_-S K_+\)(\sigma-\sigma_0)\;, 
\end{eqnarray}
where we used 
\begin{eqnarray}
&S^{-1}\equiv{1-K_{sl(2)}}\;,&\\ \nonumber
& K_m=K_++K_-\;, \qquad
 K_d=2K_-\,S\, K_+\;,&
 \end{eqnarray}
with $K_+$ and $K_-$  defined in  (\ref{kernbess})  and the dressing kernel $K_d$ is
given by the ``magic formula'' (\ref{kernd}).
The relation (\ref{abstractBES}) can be rewritten as
\begin{eqnarray}
\label{abstractBESp}
-2\sigma=\[(1+2SK_+)(1+2SK_-)-1\](\sigma-\sigma_0)\;, 
\end{eqnarray}
This  equation can be transformed into a system of equations where the kernels $K_\pm$ appear linearly,
at the expense of introducing an extra density $\tau$ such that
\begin{eqnarray}
\label{decoup}
\tau+\sigma&=&-2S K_+(\sigma-\sigma_0)\, \\
\tau-\sigma&=&-2S K_-(\tau+\sigma_0)\;. \nonumber
\end{eqnarray}
This procedure was used in different forms by Kotikov and Lipatov \cite{Kotikov:2006ts}, Eden
\cite{Eden} and Basso, Korchemsky and Kotanski \cite{Basso:2007wd} and is essential to obtain
equations which can be exploited at strong coupling. Going back to the rapidity representation, these equations
can be written as 
\begin{eqnarray}
 H(u+2i\e)-H(u)+R(u+2i\e)-R(u)&=&8i\e \frac{d}{du}\frac{1}{x^+}+2\int du'\; K_+(u,u') R(u')\nonumber
\\ \label{holBES}
H(u+2i\e)-H(u)-R(u+2i\e)+R(u)&=&2\int du'\; K_-(u,u') H(u')\;
\end{eqnarray}
where $H(u)$ is an auxiliary resolvent associated to the density $\tau(u)$. These equations are derived for $\Im u>0$ and then analytically continued for any $u$. A set of similar, equivalent equations hold for $\Im u<0$. 
Since $R(u)$ and $H(u)$ are analytic in the UHP, any term in $K_\pm(u,u')$ which is analytic in $u'$ in the UHP
will give a zero contribution to the integrals in the r.h.s.  We therefore can set without loss of generality
\begin{eqnarray}
K_-(u,u')&=&-\frac{1}{2\pi i}
\frac{d}{du}\Big[\ln\(1-\frac{1}{x^+y^-}\)+\ln\(1+\frac{1}{x^+y^-}\)\Big]\no
\\
\label{kernelspm}
K_+(u,u')&=&-\frac{1}{2\pi i}
\frac{d}{du}\Big[\ln\(1-\frac{1}{x^+y^-}\)-\ln\(1+\frac{1}{x^+y^-}\)\Big]\;
. 
\end{eqnarray}
where we used the notation
$$u\pm i\e=\frac{1}{2}\(x^\pm+\frac{1}{x^\pm}\)\;, \qquad u'\pm i\e=\frac{1}{2}\(y^\pm+\frac{1}{y^\pm}\)$$
The key observation is that the kernels $K_\pm(u,u')$ depend on the coupling constant $\e=1/4g$ only through
the shifts $u,u'\pm i\e$, and the shifts can be absorbed by a change of variable and the shift of the integration contour. The equations can be rewritten in terms of the kernels
\begin{eqnarray}
 \K_\pm(u,u')&=&\lim_{\e\to 0}\;K_\pm(u,u')\;,\\ \nonumber
  \K_\pm(u,u')&=&\frac{1}{2\pi i}\,\frac{2}{1-x^2}\(\frac{1}{y-\frac{1}{x}}\mp\frac{1}{y+\frac{1}{x}}\)
\end{eqnarray}
which depend on the  variables $x,\ y$ with a cut on $[-1,1]$ with $x(u+i0)=1/x(u-i0)$.
Upon introducing 
\begin{eqnarray}
 R_\pm(u)
   =\frac{1}{2} [R(u)\pm H(u)]  
\end{eqnarray}
as well as the related functions $r_\pm(u)$
\begin{eqnarray}
  r_\pm(u) = R_\pm(u-i\e)-R_\pm (u+i\e)\, .
 \label{defrpm}
\end{eqnarray}
we arrive at the equations
\begin{eqnarray}\label{eq:theequationpm} 
 r_+(u)& = &\frac
 {8i\e}{x^2-1}-\int_{\IR-i0} du' \; \K_+(u,u')\, [R_+(u'+i\e)+R_-(u'+i\e)]\no\\
  r_-(u)&=&\int_{\IR-i0} du'\; \K_-(u,u')\,
  [R_+(u'+i\e)-R_-(u'+i\e)]\, .  
\end{eqnarray}
The r.h.s. of these equations have a single branch cut of order two for $u\in [-1,1]$ and they have definite symmetry
properties under $u\to -u$. The same properties are shared by the l.h.s., so we deduce that $r_+(u)$ and $r_-(u)$
are symmetric, respectively antisymmetric functions with a single branch cut on  $[-1,1]$. At infinity they behave as
\begin{eqnarray}
r_+(u)\propto \frac 1{u^2}\; ,\; \qquad r_-(u)\propto \frac
  1{u^3}\; \; \qquad (u\rightarrow\infty) 
 \label{infrpm}
\end{eqnarray}
  and at $u\to\pm 1$ they have singularities at most of the type
\begin{eqnarray}
r_\pm(u)\propto  \frac {1}{\sqrt{u^2-1}} \propto \frac{1}{x^2-1} \;.
 \label{onerpm}
\end{eqnarray}
  %
The relation (\ref{defrpm}) can be inverted to obtain 
\begin{eqnarray}
 R_\pm(u)=\sum _{n=0}^\infty r_\pm (u+(2n+1)i\e)\, .
 \label{invdefrpm}
\end{eqnarray}
  Therefore, the resolvents $ R_\pm(u)$ have a set of equidistant branch cuts in the LHP, separated by a distance $2i\e$.
  In the strong coupling limit, the distance between cuts  vanishes and we are left with a single cut on the real axis plus non-perturbative terms. The contour of integration in (\ref{eq:theequationpm}) gets pinched between two branch cuts. It is therefore better to define the action of the kernels $\K_\pm$ in a manner which can be exploited at strong coupling. 

\subsubsection{The functional equations}
  
On functions $F(u')$ analytic in the UHP with the real axis included (as $R_\pm(u'+i\e)$ are), we can deform the integration contours for $u'$ defining the action of the kernels $\K_\pm(u,u')$ such that the new contour encircles the branch cut $u'\in [-1,1]$, and then transform the contour integral into a line integral using $y(u'+i0)=1/y(u'-i0)$,
\begin{eqnarray}
 \label{eq:kernewaction} \K_\pm F (u)&=&\frac
  2{1-x^2}\int_{-1+i0}^{1+i0}\ \frac {du'}{2\pi i}\(\frac
  {-yx}{y-x}\pm\frac {yx}{y+x}-\frac 1{y-\frac 1x}\mp\frac 1{y+\frac
  1x}\) F(u') \nonumber \\&=& \int_{-1+i0}^{1+i0}\ \frac {du'}{2\pi i}\;\frac
  {y-\frac 1y}{x-\frac 1x}\(\frac 1{u'-u}\mp\frac 1{u'+u}\) F(u')\, .  \end{eqnarray}
  Finally, the second term in the last line can be transformed, using $u'\to -u'$ , so that we get
\begin{eqnarray}
\label{eq:fdefinition}
  \K_\pm F(u)
  &=&\int_{-1}^{1}\frac {du'}{2\pi   }\; \sqrt{\frac {  1- u'^2} { u^2-1
  }}\ \frac { F(u'+i0)\pm F(-u'+i0)}{u'-u } \, . 
 \end{eqnarray}
In this form, the kernel can be applied to functions with a branch cut on the real axis, as the resolvents $H(u)$ and $R(u)$ become in the limit $\e=1/4g\to 0$. The kernel in equation (\ref{eq:fdefinition}) is very similar to a Cauchy kernel. If one defines the function $\tilde F_\pm(u)$ by
\begin{eqnarray}
 \tilde F_\pm(u)\, =\,
  \begin{cases}
     \ \ \ F(u)  & \text{  if  $\Im u>0 $  }, \\
      \pm F(-u) & \text{ if  $\Im u<0$}\, ,
\end{cases}
\end{eqnarray}
then the kernels $\K_\pm$ can be written as Cauchy integrals
\begin{eqnarray} 
\K_\pm F(u)=\oint \frac {du'}{2\pi i} \ \tilde F_\pm(u') \, 
\sqrt{\frac {
  u'^2-1} { u^2-1 }}\, \frac 1{u'-u}\;.
  \end{eqnarray}
This definition is useful when $\tilde F_\pm(u)$ is analytic outside the $[-1,1]$, in particular when $\tilde F_\pm(u)=F(u)$,
which means that $F(u)$ is symmetric/antisymmetric. On the space $\CL_\pm$ of functions $f_\pm$ which are { i)} symmetric/antisymmetric,
ii) analytic everywhere except a branch cut on $[-1,1]$ and iii) vanish faster than $1/u$ at infinity the kernels $\K_\pm$
act as
\begin{eqnarray} 
\K_\pm f_\pm=f_\pm\;.
  \end{eqnarray}
Since $\K_\pm f(u)$ belongs to the space $\CL_\pm$, for any function $f(u)$ we deduce that the kernels $\K_\pm$ are idempotents
\begin{eqnarray} 
\K_\pm ^2=\K_\pm \;.
  \end{eqnarray}
  Finally, we notice that for the constant function $f(u)=1$, which does not belong to the spaces  $\CL_\pm$, we have
\begin{eqnarray} 
K_+\;1=\frac{2}{1-x^2}\;, \qquad K_-\; 1=0 \;.
  \end{eqnarray}  
  At this point, we can express the equations (\ref{eq:theequationpm}) as an equation for the zero modes of the kernels $\K_\pm$. By redefining
\begin{eqnarray} \label{defGpm}
  \Gamma_{+} (u)+\Gamma_{-} (u)&\equiv&R_+(u-i\e)+R_-(u+i\e)+4i\e\no\\
  \Gamma_{+} (u)-\Gamma_{-} (u)&\equiv&R_-(u-i\e)-R_+(u+i\e)+4i\e\, .
  \end{eqnarray}
we arrive at the functional form of the BES equation
\begin{eqnarray}\label{KGG}
\K_+( \Gamma_{+}  + \Gamma_{-}  )
=0,
\\
\K_-(\Gamma_{+}  -\Gamma_{-}  )
=0\, .
  \end{eqnarray}
 As it can be easily deduced from the expression (\ref{eq:fdefinition}), the zero mode condition can be expressed as
 a functional equation on the cut. The two equations  (\ref{KGG}) are equivalent to the functional equation
\begin{eqnarray}\label{eq:funcexactongammas}
  \Gamma_{+}  (u+i0)+\Gamma_{-}  (-u+i0)=0,\;\; u\in [-1,1]\, .
 \end{eqnarray}
Note that this functional equation is valid at {\it any} value of $g$, and it is equivalent with the original BES equation,
provided we specify the class of functions to which $\Gamma_\pm(u)$  belong, namely functions with a collection of 
equidistant square root cuts  
\begin{eqnarray}\la{Gmr}
     \Gamma_{-} (u)
     = \hf r_+(u) - \hf r_-(u) + \sum_{n=1}^\infty r_+(u+2ni\e)
      \, ,
     \\
	  \la{Gpr} \Gamma_{+} (u) =4i\e + \hf r_+(u) +\hf r_-(u) +
\sum_{n=1}^\infty r_-(u+2ni\e) \, .   \end{eqnarray}

\subsubsection{The perturbative solution at strong coupling} 

At strong coupling, the distance between the square root cuts vanishes and the functions  $\Gamma_\pm(u)$
reduce to a functions with a single cut on $[-1,1]$ and a non-perturbative contribution. Let us consider the combinations
\begin{eqnarray}
\label{Gperiodic}
\Gamma_{\pm}  (u) \mp  \Gamma_{\pm}  (-u)
        &=&\mp  \sum  _{n\in\MZ} r_\mp( u+ 2in \e)
         \,
 \end{eqnarray}
The functions in the r.h.s are periodic in $u$ with period $2i\e$, therefore they are purely non-perturbative. In the perturbation expansion in $\e$ they can be put to zero, so that the functions $\Gamma_\pm(u)$ are now symmetric/antisymmetric
\begin{eqnarray}
\label{symgamma}
\Gamma_{\pm}  (u) =\pm  \Gamma_{\pm}  (-u)\;.
        \end{eqnarray}
Introducing the functions
\begin{eqnarray}
G_{\pm}  (u) = \Gamma_{+}  (u)\pm  i\Gamma_{-}  (u)\;.
        \end{eqnarray}
  the symmetry property becomes (\ref{symgamma}) becomes
  \begin{eqnarray}
G_+  (u) = G_-(-u)\;.
        \end{eqnarray}
        while the functional equation (\ref{eq:funcexactongammas}) becomes
      \begin{eqnarray}
G_\pm  (u+i0) = \mp iG_\mp(-u+i0)=\mp iG_\pm(u-i0)\;.
\label{moneq}
        \end{eqnarray}   
This equation can be considered as a monodromy equation for $G_\pm(u)$ around the cut. The monodromy is of order four, so we deduce that $G_\pm(u)$ have a branch cut of order four between $[-1,1]$. 
Introducing the uniformisation variable  $s$  via
 \begin{eqnarray}
e^{2s}=\frac{u+1}{u-1}
        \end{eqnarray}
        we can write the general solution of the monodromy equations (\ref{moneq}) as
\begin{eqnarray}
\label{solGs} G_{\pm}  (s)&=& 4i\e \sum_{n\in \MZ} c_{ n} (\e) \, e^{\pm
(2n+1/2)s} \no\\
& =&  4i\e    \sum_{n\in  \MZ}   c_{ n}  (\e) \,
 \({u+1\over u-1}\)^{\pm n  \pm {1\over 4}}
 \, , \end{eqnarray}
where the coefficients $c_{ n} (\e)$ are to be fixed. Basso, Korchemsky and Kotanski \cite{Basso:2007wd} supposed 
the coefficients $c_{ n} (\e)$ can be expanded in a series in integer powers in  $\e$. We suppose the same thing 
here, although this is not generally true for generic states, see for example the $su(1|1)$ sector analyzed in \cite{Beccaria:2008dv}.
From the behavior at infinity of the two functions 
 \begin{eqnarray}
\label{expGpminf} G_{\pm} (u) = \sum _{n\ge 0} {W^\pm_n\over u^{n }} 
 \end{eqnarray}
we can extract the universal scaling function, which is encoded into the physical resolvent $R_{phys}(u)$, according to equation (\ref{infbeh}).
We obtain the conditions
 \begin{eqnarray}
 \label{univcn}
 1=  \sum_{n\in  \MZ}   c_{ n}  (\e) \;, \qquad f(g) = {1\over \e} \sum_{n\in\MZ} (4n+1)c_n(\e)\;.
 \end{eqnarray}

The coefficients $c_{ n} (\e)$  can be determined, at least recursively, by analyzing the structure of the general solution
(\ref{solGs}) in the vicinity of the points $u=\pm 1$. We introduce the rescaled variables
 \begin{eqnarray}
  z= {u-1\over 2\e}, \qquad \bar z =- {u+1 \over 2\e}\, .  
 \end{eqnarray}
 The variable $z$ coincides, up to a shift by $ 2g$, with the original
 (before rescaling by $2g$) rapidity in the BES equations. 
 In the variable $z$, the functions $G_\pm$ still have a set of distinct cuts separated by $i$. When $g\to \infty$, these
 cuts become semi-infinite, extended to from $z=0$ to $z=-\infty$, and they can be expressed in terms of the 
 functions $r_\pm(z)$ which have, as discussed in section (\ref{sec:linBES}), the following expansion
 \begin{eqnarray}
 \label{smallz}
r_\pm (z) = \sum _{n\ge 0} b_n^\pm(\e) \, z^{n-1/2} + \sum
_{n\ge 0} d^\pm_n(\e) \, z^{n} \, \, \qquad (\, |z|<1\, ). 
\end{eqnarray}
For a more symmetric presentation, one can introduce the notations
 \begin{eqnarray}
 g_\pm(z)=r_+(z)\pm ir_-(z)\;, 
 \end{eqnarray}
 The functions $ g_\pm(z)$ are related to $G_\pm(z)$ via
  \begin{eqnarray}\label{ggpmGGpm} g_{\pm} = {1\pm i \over D^2\mp i}\, (D^2-1) \, G_{\pm}  \, ,
\end{eqnarray} 
 with $D$  the shift operator $Df(z)=f(z+i/2)$.

The compatibility of the expansion (\ref{smallz}), and the similar expansion around $\bar z=0 $, with (\ref{solGs}) requires that
 \begin{eqnarray}
    \la{DSL}
    c_n (\e) =\e^{|n| }\, { \a_{n}(\e) },
    \qquad
    \a_n(\e) =   \sum_{p=0}^\infty \a_{n,p}\, \e^p
 \qquad (n\in \MZ)\,  .
\end{eqnarray}
 In particular, this implies that at the leading order in $\e$ 
 $$c_0=1+\CO(\e)\;, \quad {\rm and}\quad  f(g)=\frac{1}{\e}+\CO(\e^0)$$
 with the corresponding resolvent
 \begin{eqnarray}
  R(u)_{|_{\e\to0}}
 & =& {-4i\e} \( 1-{ 1\over  \sqrt{1-{1/ x^2}}}
 + i{1/x\over \sqrt{1-{1/ x^2}}}   \) \, 
\end{eqnarray}
which is the solution which was obtained in \cite{Alday:2007qf} \cite{Kostov:2007kx}.

 In order to determine the higher orders in $\e$, we have to implement the compatibility condition between the 
 (large) $u$ expansion (\ref{solGs}) and the (small) $z$ expansion (\ref {smallz}). We first write
 \begin{eqnarray}
\label{solGr}
  G_{\pm}  (z)
  =  2 i\e  \sum_{n\in\MZ}     \e^{|n|}  \, {\a_{ n} (\e)  }   \,
 \({ 1+\e z\over  \e z}\)^{\pm n  \pm {1\over 4}}\, 
 .
\end{eqnarray}
and then take the inverse Laplace transform defined by
 \begin{eqnarray}
\label{invlaplace}
 \tilde f(\ell) ={1\over 2\pi i}
\int_{i \MR + 0}  d  z \, e^{z\ell}\,  f(z)\,
.
\end{eqnarray}
  The explicit expression for the
inverse Laplace transform of (\ref{solGr}) is a series of confluent
hypergeometric functions of the first kind
 \begin{eqnarray}\label{soll} 
\tilde G_\pm(\ell) &=&   \pm 4
    i     \sum_{n\in \MZ }    \e^{|n|}  \a_{ n}  (\e) \,
    \( n +\frac{1}{4} \)
    \, _1F_1\left(1\mp  \frac{1}{4}  \mp n;2;- {\ell/\epsilon }\right)
\\ 
 &\simeq&  4 i \sum _{n\in\MZ} \e^{|n|} \a_n(\e) { (\e/\ell)^{1\mp {1\over 4}
\mp n} \over \G( \pm {1\over 4}\pm n)} \, _2F_0\left( 1\mp \frac{1}{4} \mp
n,\mp \frac{1}{4} \mp n ;;\ {{\e/ \ell}}\right) \;, \nonumber
 \end{eqnarray}
where in the second line we have used the transformation properties of the hypergeometric function
and we have neglected terms of the order $e^{-\ell/\e}$. This is an expansion in quarter powers of $\ell$,
while $\tilde g_\pm(\ell)$ will contain half-integer powers of $\ell$.  
Under the inverse Laplace transform, the relation (\ref{ggpmGGpm}) becomes 
 \begin{eqnarray}
\tilde g_{\pm}(\ell)= {\sqrt{2} \sin({\ell\over 2})\over \sin({\ell\over
2} \pm {\pi\over 4} )} \, \tilde G_{\pm} (\ell) \, . 
 \la{FEell}
\end{eqnarray}
  Now we represent, as in \cite{Basso:2007wd}, the ratio of the sine
functions in (\ref{FEell}) as
 \begin{eqnarray} { \sin({\ell\over 2})\over \sin({\ell\over 2} \pm {\pi\over 4} )}=
{S_\pm(\ell)\over T_\pm(\ell)}\, , 
\end{eqnarray}
where $S$ and $T$  represent ratios of Gamma functions:
 \begin{eqnarray} S_\pm(\ell) =\pm {\G({1\over 2}+ {\ell \over 2\pi} \mp {1\over
4})\over \G({\ell\over 2\pi} )}\, , \qquad T_\pm(\ell)&=& {\G(1-{\ell
\over 2\pi})\over \G( {1\over 2}-{\ell\over 2\pi} \pm {1\over 4})} \,
,\end{eqnarray}
which expand
 at $\ell=+\infty$ and at $\ell=0$ respectively as
 \begin{eqnarray} S_\pm(\ell) &=& \pm \( \ell /2 \pi \)^{{1\over 2}\mp {1\over 4}}
\( 1+ \sum_{n=1}^\infty S ^\pm_n \ell^{ -n} \) \, , \la{Sexp} \\
  T_\pm (\ell)&=&\frac{1}{\Gamma \left(\frac{1}{2} \pm
  \frac{1}{4}\right)} \(1+\sum_{n=1}^\infty T^\pm_n \ell^n\) \, .
  \label{Texp} 
 \end{eqnarray}
 If we rewrite the equation (\ref{FEell}) as
 \begin{eqnarray}\label{FEM} { \tilde G_{\pm} (\ell)\over T_\pm(\ell) } &=&{1 \over
\sqrt{2}} \ { \tilde g_{\pm}(\ell)\over S_\pm(\ell) } \la{Fuec}\, ,
\end{eqnarray}
then the l.h.s. is analytic everywhere except the negative real axis,
while the r.h.s. is analytic everywhere except the positive real axis.
As a consequence, neither of the sides has poles and the only
singularities can be branch points at $\ell=0$ and $\ell=\infty$.
This means, in particular, that the expansion of the r.h.s. as a power
series at $\ell =\infty$ coincides with the expansion of the l.h.s. at
$\ell =0$. Both sides expand in the same fractional powers, so that if we multiply both sides of equation
 (\ref{FEM}) by
by $(\ell/\e)^{1\mp 1/4}$ we obtain
\begin{eqnarray}
 \label{seriesGT} {\tilde G_{\pm} (\ell)\over\tilde T_\pm(\ell)} \(
 {\ell / \e}\)^{ 1 \mp {1\over 4}}\ = \sum_{n\in \MZ} \ C_{n}^\pm
 (\e)\ \ell^{-n }\, , 
\end{eqnarray}
where the coefficients $C_n^\pm(\e)$ should be understood as formal
series in $\e$,
\begin{eqnarray}\label{TaylC}
C_n^\pm(\e)= \sum_{p=0}^\infty C_{n,p}^\pm \, \e^p\, .
\end{eqnarray}
From (\ref{smallz}), (\ref{Sexp}) and the relation (\ref{FEM}) we
 deduce that the coefficients in front of the non-negative powers of
 $\ell$ vanish,
 \begin{eqnarray} \label{Constraint}
   C_n^\pm (\e) = 0  \quad  {\rm for} \quad n=-1,-2, \dots\, \ .
 \end{eqnarray}  
Solving the contraints (\ref{univcn}) and (\ref{Constraint}) order by
order in $\e$ one can evaluate recursively the Taylor coefficients
$\a^\pm_n$ of the series (\ref{DSL}).  The recurrence procedure is
possible because at each order in $\e$ the sum on the r.h.s. of
(\ref{seriesGT}) contains only a finite number of negative powers of
$\ell$.

In the leading order in $\e$ the series expansion of $\tilde G_\pm$ at
$\ell=0$, given by 
 \begin{eqnarray}\label{leadingNFSR} \tilde G_\pm(\ell) =2i (\e/\ell)^{1\mp{1\over
4}}\( \sum_{n= 0}^\infty {\a_{\pm n,0}\over \G( n\pm {1\over 4} )} \,
\ell^{ n } + {\cal O} ( \e)\) \,,  \end{eqnarray}  
  contains only non-negative
powers in $\ell$.  Therefore the sum on the r.h.s. of (\ref{seriesGT})
contains only the term with $n=0$, and we have
\begin{eqnarray}
\label{leadingNFSRbis} 
(\ell/\e)^{1\mp{1\over 4}} \tilde G_\pm(\ell)
=2i \sum_{n= 0}^\infty {\a_{\pm n,0}\over \G( n\pm {1\over 4} )} \,
\ell^{ n } \, = T_\pm(\ell) \, C_{0,0}^\pm \, .  
 \end{eqnarray}
From the constraint (\ref{univcn}), which in the leading order gives
$\a_{0,0}=1$, we evaluate
 \begin{eqnarray} C^+_{0,0}= {2i} {\G({3\over 4})\over \G({1\over 4})}\, , \ \
  C^-_{0,0}= {2i } {\G({1\over 4})\over \G(-{1\over 4})} \, .   \end{eqnarray}
For  the other coefficients  we find 
\begin{eqnarray} 
\a_{\pm n,0} &=& {\G(n \pm {1\over 4})\over \G(\pm {1\over 4})} \
T^\pm _n \, , 
 \end{eqnarray}
 where $T^\pm _n$ are the coefficients in the expansion (\ref{Texp}).
Continuing this procedure order by order in $\e$, we reproduce the result  of \cite{Basso:2007wd}
for the  universal scaling function,
\begin{eqnarray} 
f(\e)&=&
{1\over\e}     + 4   \sum_{n=1}^\infty \e^{|n|-1} n\,  \a_n(\e) \\ \nonumber
&=&{1\over\e}  -\frac{3\log 2}{\pi }  -\frac{K}{\pi ^2} \e + \dots
\ . \end{eqnarray}

The method presented here can be extended to the computation of other quantities, in particular it can be adapted to the computation of the generalized scaling function, as it was done by Volin in the regime $j\sim g$ \cite{Volin:2008kd}. It was also used by Volin
\cite{Volin:2009wr} to compute analytically the mass gap for the $O(N)$ sigma model.
   
\section{Finite size corrections and the thermodynamical Bethe ansatz}   

In the current state of the development of the subject, there is compelling evidence that the dilatation operator of the 
$\CN=4$ SYM theory is integrable and the results for its spectrum are compatible with the AdS/CFT correspondence. The above picture is coherent, although the problem of computing conformal dimensions for operators of finite length is not yet solved. The Beisert-Staudacher equations  (\ref{allFNBA}) are correct
for lengths of the chain $L$ large. The finite size corrections are associated to the so-called {\it wrapping interactions} because they involve spin interactions that wrap around the circumference of the spin chain. Those terms appear in perturbation theory as terms of the order $g^{2L}$, hence they are exponentially suppressed with the size of the chain. A first 
glimpse of how they should appear in a spin chain is given by the Inozemtsev model. There, the Bethe ansatz disappear at finite $L$ and the two-magnon energy has a non-additive part which also vanishes exponentially with the system size
\cite{Inozemtsev:2002vb}. From an (integrable) field theory point of view, the finite size corrections should come from the virtual particles propagating around the cylinder \cite{Ambjorn:2005wa}. A method to compute these terms for a 
generic two dimensional field theory was given by L\"uscher
 \cite{Luscher:1983rk, Luscher:1986pf}.  The L\"uscher method was implemented for AdS/CFT by Janik and \L ukowsky
  \cite{Janik:2007wt} and it was pursued in a vast number of circumstances, culminating with the computation of the wrapping correction at four  \cite{Bajnok:2008bm} and five \cite{Bajnok:2009vm}  loops for the Konishi operator . The result  was shown to agree with the diagrammatics computation in $\CN=4$ SYM  \cite{Fiamberti:2007rj,Velizhanin:2008jd}. The same method was used to show  \cite{Bajnok:2008qj} that the wrapping corrections are compatible with the analyticity requirements imposed by the BFKL equation on the anomalous dimension of the twist two operator \cite{Kotikov:2002ab}  and which were not met by the asymptotic Bethe ansatz \cite{Kotikov:2007cy}.

A method to compute the finite-size corrections  for integrable field theories and integrable models in general is the so-called Thermodynamic Bethe Ansatz (TBA). Thermodynamic functions for integrable systems where first computed by Yang and Yang \cite{YangYang} for the Bose gas with delta interaction and soon after by Gaudin for the XXZ model \cite{Gaudin:1971gt}
and Takahashi \cite{Takahashi1971} for the XXX model. The distributions of roots were found to obey certain functional equations, known in recent years under the name of Y systems.
Similar relation were found around 1990 among the transfer matrices for various integrable models \cite{Pearce:1991ty,Bazhanov:1989yk,Kuniba:1993cn}. 
These relations, also known as fusion relations, can be encoded in terms of the Hirota equation. For models based on Lie supergroups of type $su(M|N)$ Hirota equation was first obtained by Tsuboi \cite{Tsuboi97}. A connection between these seemingly different equations, at the origin of the
method for  computing finite size corrections using the thermodynamic Bethe ansatz  \cite{Zamolodchikov:1989cf}, is the interpretation of the finite size $L$ of an integrable system as the inverse temperature in the system with the space and time interchanged. The system of functional equations associated to the thermodynamical Bethe ansatz contains data from a Dynkin diagram \cite{Ravanini:1992fi}. 
An equivalent formulation of TBA in  terms of a non-linear integral equation was given by Destri and De Vega \cite{Destri:1994bv}.
The TBA approach was initially used to compute the finite size corrections to the vacuum energy, but later it has been understood how to extend it to excited states \cite{Dorey:1996re,Fioravanti:1996rz}.  

 In a relativistically invariant system, the rotated system, called {\it mirror} theory in \cite{Arutyunov:2007tc}, is equivalent to the initial one. In the context of AdS/CFT, because of the choice of the light cone gauge, the excitations do not have a relativistically invariant dispersion relation, but the BDS dispersion relation (\ref{BDSen}). The systematic study of the mirror model, including the bound states, was initiated in \cite{Arutyunov:2007tc} and continued in \cite{Arutyunov:2008zt,Arutyunov:2009zu,Arutyunov:2009ur}. 

The TBA method was recently applied to the AdS/CFT system \cite{Gromov:2009tv}, 
after working out in detail the simpler $su(2)$ principal chiral model \cite{Gromov:2008gj}, with which it bears some similarity. In particular, this method allowed to compute the four loop wrapping correction to the Konishi operator \cite{Gromov:2009tv} and the five loop numerically \cite{Arutyunov:2010gb} and analytically At strong coupling, a numerical analysis of the Y-system was done \cite{Gromov:2009zb}  which confirmed the leading $\lambda^{1/4}$ and permitted to fix the first few coefficients in the strong coupling expansion. Compared with the prediction from the string theory \cite{Roiban:2009aa}, the result differs in the coefficient of the $\lambda^{-1/4}$ term. In \cite{Gromov:2009tq,Gromov:2010vb} the finite size corrections for operators with $L\sim M\sim g$ were analyzed and the corresponding results were compared with the predictions from the quasi-classical analysis of the algebraic curve.

In  \cite{Bombardelli:2009ns,Gromov:2009bc} the Y-system was derived from the Beisert-Staudacher equations for the mirror model. The Y-system is a system of finite-difference equations obeyed by the densities of magnons/holes 
at finite  temperature. These equations can be restated as Hirota equations \cite{Gromov:2008gj,Gromov:2009tv},
which are usually obeyed by the transfer matrices associated to particular representations of $su(n)$  \cite{Krichever:1996qd}  or $su(n|m)$ \cite{Kazakov:2007fy} algebras. 

In the following, I will explain how the formalism of the Hirota equation and Y-system  can be defined  for the $su(2)$ principal chiral model, which is a simpler version of the  Y-system which appears in the AdS/CFT 
problem. Then, I will present the AdS/CFT Y-system, which was used to compute the dimension of the Konishi operator at weak and strong coupling. The presentation follows mainly the references \cite{Gromov:2008gj,Gromov:2009tv}.

\subsection{Y system for the principal chiral field}
\def\m{{\rm m}}

We are starting with the Bethe equations for the principal chiral field in a large volume. 
 \begin{eqnarray}
 \label{baepcf}
1&=&\prod_\beta \frac{u_j-\theta_\beta-i/2}{u_j-\theta_\beta+i/2}\;\prod_{l\neq j} \frac{u_j-u_l+i}{u_j-u_l-i}\;,\\ \nonumber
e^{-i\m \CL \sinh \pi \theta_\al} &=&\prod_{\beta\neq \al}S_0^2(\theta_\al-\theta_\be)\;\prod_j
\frac{\theta_\al-u_j+i/2}{\theta_\al-u_j-i/2}\;\prod_k
\frac{\theta_\al-v_k+i/2}{\theta_\al-v_k-i/2}\;,\\ \nonumber
1&=&\prod_\beta \frac{v_k-\theta_\beta-i/2}{v_k-\theta_\beta+i/2}\;\prod_{n\neq k} \frac{v_k-v_n+i}{v_k-v_n-i}\;,
\ \end{eqnarray}
with the scattering matrix $S_0(\theta)$ satisfying the crossing equation
\begin{equation}
\label{croseq}
S_0(\theta+i/2)S_0(\theta-i/2)=\frac{\theta-i/2}{\theta+i/2}\;.
\end{equation}
The energy of a state is given by
\begin{equation}
\label{eneq}
E=\sum_\al m\cosh \pi \theta_\al\;.
\end{equation}
In order to introduce the thermodynamic Bethe ansatz, one has to identify the string solutions which are supported by these equations. It is easy to see that the strings  corresponding to (\ref{baepcf}) are given by
\begin{eqnarray}
u_{j,k}^{(n)}=u_j^{(n)}+i(n+1)/2-ik,\qquad k=1,\ldots,n\\ \nonumber
v_{j,k}^{(m)}=v_j^{(m)}+i(m+1)/2-ik,\qquad k=1,\ldots,m
\end{eqnarray}
For convenience, we are going to use labels $n>0$ for $u$ strings, $n<0$ for $v$ strings and $n=0$ for the $\theta$ solutions, which do not form strings. When going to the continuum limit, the equations (\ref{baepcf}) become equations for the density of magnon strings $\rho_n$ and string holes $\bar \rho_n$
\begin{equation}
 \rho_n+\bar \rho_n =\m\CL \cosh \pi \theta\; \delta_{n,0} +\sum_{m=-\infty}^\infty K_{n,m} \; \rho_m
\end{equation}
The kernel $K_{n,m}$ acts by convolution and it vanishes for $nm<0$. For $nm>0$, $K_{n,m}=K_{-n,-m}$  and its Fourier transform has a relatively simple expression 
\begin{equation}
(1-\hat K)^{-1}=1-\hat s\;I\;, \quad {\rm where} \quad \hat s(t)=\frac{1}{2 \cosh t/2}
\end{equation}
and $I$ is the incidence matrix of an $A_\infty$ graph, $I_{n,m}=\delta_{n,m+1}+\delta_{n,m-1}$. In the rapidity space, the operator $s$ can be expressed via the shift $D$ by $s=1/(D+D^{-1})$, where $Df(\theta)=f(\theta+i/2)$. The remaining kernels are given by
\begin{equation}
\hat K_{0,0}=\frac{e^{-|t|/2}}{2\cosh t/2}\;, \qquad \hat K_{0,n}=-\hat K_{n,0}=e^{-|t|/2}\;.
\end{equation}
The thermodynamical functions can be obtained by minimizing the free energy density $f=F/\CL$
\begin{equation}
f=\int d\theta \(\m \rho_0\cosh \pi \theta-T\sum_n\[\rho_n\ln \(1+\frac{\bar \rho_n}{\rho_n}\)+\bar \rho_n\ln \(1+\frac{ \rho_n}{\bar\rho_n}\)\]\)\;.
\end{equation}
Taking into account that the variation of densities of strings and holes are related by $\delta \bar \rho_n+\delta \rho_n=\sum_{m=-\infty}^\infty K_{n,m}\delta \rho_m$, we obtain at equilibrium
\begin{equation}
\frac{\m}{T}\cosh \pi \theta\; \delta_{m,0} =\ln\(1+\frac{\bar \rho_m}{\rho_m}\)+\sum_{n=-\infty}^\infty(K_{n,m}-\delta_{n,m})\ln \(1+\frac{ \rho_n}{\bar \rho_n}\)
\end{equation}
By denoting $Y_n=\frac{\bar \rho_n}{\rho_n}$ for $n\neq 0$ and $Y_0=\frac{\rho_0}{\bar \rho_0}$, and considering the cases the $m=0$, $m>0$, $m<0$ separately, see \cite{Gromov:2008gj} for details, one can write the above equation as
\begin{equation}
\frac{\m}{T}\cosh \pi \theta\; \delta_{m,0} =\sum_{n=-\infty}^\infty(I_{n,m}s-\delta_{n,m})\ln\(1+Y_n\)+\ln \(1+Y_m^{-1}\)\;.
\end{equation}
Multiplying the previous equation by the operator $s^{-1}=D+D^{-1}$, for which the l.h.s. is a zero mode, and taking into account that $I_{m,n}$ is an incidence matrix, we obtain
\begin{equation}
\label{Ypcf}
\ln \[Y_n(\theta+i/2) Y_n(\theta-i/2)\]=\ln \[(1+Y_{n+1})(1+Y_{n-1})\]\;.
\end{equation}
This is the so-called Y-system for the principal chiral field. It is interesting to note that the information about the potential term, which contains the size of the system, has disappeared from the equation and it enters to the solution only via boundary conditions.
The equations (\ref{baepcf}) can be retrieved if one imposes the boundary condition at large $\theta$
\begin{equation}
\label{Ypcfbc}
Y_n\sim e^{-\m \CL \cosh \pi \theta\;\delta_{n,0}}
\end{equation}
The Y functions of the principal chiral field are defined on a one-dimensional lattice. More generally, the Y system is defined on a two dimensional lattice, $Y_{a,s}$
\begin{equation}
\label{Ysys}
Y_{a,s}(\theta+i/2) Y_{a,s}(\theta-i/2)=\frac{(1+Y_{a,s+1})(1+Y_{a,s-1})}{(1+1/Y_{a+1,s})(1+1/Y_{a-1,s})}
\end{equation}
with some specific boundary conditions. 
The Y system can be reformulated as  the Hirota equation via the transformation
\begin{equation}
\label{YHs}
Y_{a,s}=\frac{T_{a,s+1}T_{a,s-1}}{T_{a+1,s}T_{a-1,s}}\;,
\end{equation}
where $T_{a,s}$ is defined up to a gauge transformation $T_{a,s}\to g_1(\theta+i(a+s)/2)g_2(\theta+i(a-s)/2)g_3(\theta-i(a+s)/2)g_4(\theta-i(a-s)/2)T_{a,s}$.
The equation obeyed by $T$ is
 \begin{equation}
\label{Hirota}
T^+_{a,s}T^-_{a,s}=T_{a,s+1}T_{a,s-1}+T_{a+1,s}T_{a-1,s}\;,
\end{equation}
where $T^\pm_{a,s}=T_{a,s}(\theta \pm i/2)$.
This type of equation is obeyed by transfer matrices of spin chain with auxiliary spaces corresponding to representations with rectangular Young tableaux of size $(a,s)$ \cite{Bazhanov:1989yk,Pearce:1991ty,Kuniba:1993cn,Tsuboi97}.
On the border of the domain, the Hirota equation becomes the discrete d'Alembert equation in 1+1 dimensions, with solution 
({\it e.g.} on the boundary $a=0$)
 \begin{equation}
T_{0,s}(\theta)=g_1(\theta+is/2)g_2(\theta-is/2)\;.
\end{equation}
Coming back to the Hirota system for the principal chiral field, the boundary conditions can be reduced, by a gauge transformation,
to
 \begin{equation}
T_{0,s}(\theta)=\Phi(\theta+is/2)\;, \qquad T_{2,s}(\theta)=\bar \Phi(\theta-is/2)\;.
\end{equation}
If one imposes a reality condition $\[Y_s(u)\]^*=Y_s(u^*)$, taking into account that 
 \begin{equation}
1+Y_s(\theta)=\frac{T_{1,s}(\theta+i/2)T_{1,s}(\theta-i/2)}{\Phi(\theta+is/2)\bar \Phi(\theta-is/2)}\;.
\end{equation}
we conclude that $\Phi$ and $\bar \Phi$ are complex conjugate functions\footnote{For $su(n)$ one can take as symmetry condition $T_{a,s}(\theta)=T^*_{n-a,s}(\theta)$.} , $\bar \Phi(\theta)=\Phi^*(\theta)$.
The Hirota equation, quadratic in $T$, can be seen as the compatibility condition for a system of equations linear in $T$.
In the simple $su(2)$ case, this system reads, in the notation $T_s\equiv T_{1,s}$
  \begin{eqnarray}
&\ &T_{s+1}(\theta)Q(\theta+is/2)-T_{s}(\theta-i/2)Q(\theta+i(s+2)/2)=\Phi(\theta+is/2)\bar Q(\theta-i(s+2)/2)\nonumber \\ 
&\ &T_{s-1}(\theta)\bar Q(\theta-i(s+2)/2)-T_{s}(\theta-i/2)\bar Q(\theta-is/2)=-\bar \Phi(\theta-is/2)Q(\theta-is/2)\;. \nonumber \\
&\ & \label{Hlin}
\end{eqnarray}
The solution of the Hirota equation can be represented in a determinant form
 \begin{equation}
T_s(\theta)=h(\theta+is/2)\left|\begin{array}{cc}Q(\theta+i(s+1)/2) & R(\theta+i(s+1)/2) \\ \bar Q(\theta-i(s+1)/2)& \bar R(\theta-i(s+1)/2)\end{array}\right|\;,
\end{equation}
where $h(\theta)$ is a periodic function, 
$h(\theta+i)=h(\theta)$ and $Q,\bar Q$ and $R, \bar R$ are two linearly independent solutions of the linear system (\ref{Hlin}), related by the Wronskian relation
\begin{equation}
\Phi(\theta)=h(\theta+i/2)\left|\begin{array}{cc}R(\theta) & Q(\theta) \\ R(\theta+i)& Q(\theta+i) \end{array}\right|\;.
\end{equation} 
Specialized to the value $s=0$, the equations (\ref{Hlin}) take the form of the Baxter equations
  \begin{eqnarray}
&\ &T_{1}(\theta)\,Q(\theta)=T_{0}(\theta-i/2)\,Q(\theta+i)+\Phi(\theta)\,\bar Q(\theta-i)\nonumber \\ \nonumber
&\ &T_{-1}(\theta)\,\bar Q(\theta-i)=T_{0}(\theta-i/2)\,\bar Q(\theta)-\bar \Phi(\theta)\,Q(\theta)\;.
\end{eqnarray}
The asymptotic Bethe ansatz equations of the principal chiral model (\ref{baepcf}) can be obtained from the Hirota equations supplemented with the boundary condition (\ref{Ypcfbc}). We have from the definition (\ref{YHs})
 \begin{equation}
 \label{Ybound}
Y_0(\theta)=\frac{T_{1}(\theta)T_{-1}(\theta)}{\Phi(\theta)\bar \Phi(\theta)}\sim e^{-\m \CL \cosh \pi \theta}\to 0 \quad {\rm when} \quad \CL\to \infty\;,
\end{equation}
therefore one of $T_{-1}(\theta)$ or $T_{1}(\theta)$ is exponentially small. Since the two objects play symmetric roles, one deduces that the exchange between the cases $T_{-1}(\theta)\to 0$ and $T_{1}(\theta)\to 0$ is done by a gauge transformation which also depends exponentially in $\CL$. Following \cite{Gromov:2008gj}, we denote the transfer matrix in the two gauges $T^u_s(\theta)$
and $T^v_s(\theta)$, such that
 \begin{equation}
T_{-1}^u(\theta)\to 0\;, \qquad T_{1}^v(\theta)\to 0\;.
\end{equation}
The gauge transformation preserving the reality condition of $T_k$ are given by
  \begin{eqnarray}
T_s^u(\theta)= g(\theta+is/2)\bar g(\theta-is/2)\;T_s^v(\theta)\\ \nonumber
\Phi^u(\theta)=  g(\theta+i/2) g(\theta-i/2)\;\Phi^v(\theta)\\ \nonumber
\bar \Phi ^u(\theta)= \bar g(\theta+i/2) \bar g(\theta-i/2)\; \bar \Phi^v(\theta)
\end{eqnarray}
so that
 \begin{equation}
T_{-1}^u(\theta)=g(\theta-i/2)\bar g(\theta+i/2) T_{-1}^v(\theta)\;,
\end{equation}
where $g(\theta)$ is some function to be determined.
Let us choose the first of the two gauges.  From the Hirota equation, we deduce that
\begin{equation}
\label{Hbound}
T_{0}^u(\theta+i/2) T_{0}^u(\theta-i/2)=\Phi^u(\theta)\bar \Phi^u(\theta)
\end{equation}
A solution of this equation is given by $T_0^u(\theta+i/2)=\Phi^u(\theta)$.
Supposing that $T_0^u$ and $Q^u$ are polynomials, $T_0^u=\prod_\al(\theta-\theta_\al)\equiv \phi(\theta)$,  and 
$Q^u=\prod_j(\theta-u_j)$ we see that the solution of the 
 equation (\ref{Hlin}) coincides with the first line of the Bethe ansatz  equations (\ref{baepcf}). 
 The last line of (\ref{baepcf}) can be obtained by using the other gauge and $T_0^u=T_0^v =\phi(\theta)$ and $\bar Q^v=\prod_j(\theta-i-v_j)$.
 We can solve the equivalent of the equation (\ref{Hbound}) by $T_0^v(\theta-i/2)=\Phi^v(\theta)$. 
 this implies that the gauge transformation  $g(\theta)$ obeys
 \begin{equation}
\label{gugeeq}
g(\theta+i/2)g(\theta-i/2)=\frac{\Phi^u(\theta)}{\Phi^v(\theta)}=\frac{\phi(\theta+i/2)}{\phi(\theta-i/2)}\;.
\end{equation}
This equation can be solved in terms of the scattering matrix $S(\theta)=\prod_\al S_0(\theta-\theta_\al)$ which obeys the crossing condition
 \begin{equation}
\label{crosscond}
S(\theta+i/2)S(\theta-i/2)=\frac{\phi(\theta-i/2)}{\phi(\theta+i/2)}\;.
\end{equation}
The solution of the difference equation (\ref{gugeeq}) is defined up to a zero mode, satisfying $g_0(\theta+i/2)g_0(\theta-i/2)=1$. To fix the zero mode we use the boundary condition (\ref{Ybound}), which implies $|g_0(\theta-i/2)|^2=e^{-\m \CL \cosh \pi \theta}$, and 
$|g_0(\theta)|^2=1$. We conclude that
 \begin{equation}
\label{gugesol}
g(\theta)=S^{-1}(\theta)\,e^{-\frac{i}{2}\m \CL \sinh \pi \theta}\;, \qquad \bar g(\theta)=S(\theta)\,e^{\frac{i}{2}\m \CL \sinh \pi \theta} \;.
\end{equation}
The explicit form of the gauge transformation allows to derive the central equation in (\ref{baepcf}). It suffices to consider the condition 
 \begin{equation}
\label{bacentr}
-1=Y_0(\theta_\be+i/2)=\frac{T^u_{1}(\theta)T^v_{-1}(\theta)}{\Phi^u(\theta)\bar \Phi^u(\theta)}g(\theta+i/2)\bar g(\theta-i/2)|_{\theta=\theta_\be+i/2} \;.
\end{equation}
and use the equations (\ref{Hlin}) for $s=0$, the first in the $u$ gauge and the second in the $v$ gauge. 

{\it Vacuum solution at finite $\CL$.} The simplest  solution of the Hirota equations for $s\geq 0$ is $T^u_{s-1}=s$ and  $T^v_{s+1}=s$ for $s\leq 0$, which corresponds to the vacuum with strictly infinite $\CL$.  However, $T^u_{-1}$ should not be zero, but $T^u_{-1}=T_{-1}^v |g(\theta-i/2)|^2\sim e^{-\m \CL \cosh \pi \theta}$.  Since in  the vacuum $T^u_{s}=T^v_{-s}$, we are going to consider only the $u$ gauge and drop the gauge index in the following. An improved solution is
  \begin{eqnarray}
  \label{solvac}
T_{s-1}(\theta)&=& s+G(\theta-is/2)-G(\theta+is/2)\;, \\ \nonumber
\Phi(\theta)&=& 1+G(\theta+i0)-G(\theta+i)\;,\\
\bar \Phi(\theta)&=& 1+G(\theta-i)-G(\theta-i0)\;, \nonumber
\end{eqnarray}
where $G(\theta)$ is the resolvent associated with the "density"  $T_{-1}(\theta)$ 
 \begin{equation}
 \label{soljump}
G(\theta)=\frac{1}{2\pi i}\int_{-\infty}^\infty \frac{d\theta'}{\theta-\theta'}\,T_{-1}(\theta')\;
\end{equation}
and therefore it is analytic everywhere except of a cut on the real axis.
In particular, we have 
\begin{equation}
\label{gugesol1}
T_0(\theta)=1+G(\theta-i/2)-G(\theta+i/2)=\Phi(\theta-i/2)=\bar \Phi(\theta+i/2) \;.
\end{equation}
The function $T_0(\theta)$ is analytic between the cuts at $\theta=\pm i/2$ and the functions $\Phi$ and $\bar \Phi$ are, up to a shift, the determination of $T_0$ beyond these cuts. The ansatz (\ref{soljump}) is motivated by the fact that jump in $T_0$ should be equal 
to $T_{-1}$, {\it cf.} the first equation (\ref{Hlin}) at $s=-1$ and $Q=1$,
\begin{equation}
\label{Hjump}
T_{-1}(\theta)=T_{0}(\theta+i/2)-\Phi(\theta) \;.
\end{equation}
The solution (\ref{solvac}) is determined by an unique function, $f(\theta)=T_{-1}(\theta)$. In turn, this function is constrained by the
gluing of the two wings in the Hirota equation, realized via the function $Y_0$
  \begin{eqnarray}
  Y_0(\theta)&=&\frac{T_{1}(\theta)T_{-1}(\theta)}{\Phi(\theta)\bar \Phi(\theta)}\;,\\
  Y_0(\theta+i/2) Y_0(\theta-i/2)&=&(1+Y_1(\theta))^2=\(\frac{T_1(\theta+i/2)T_1(\theta-i/2)}{\Phi(\theta+i/2)\bar \Phi(\theta-i/2)}\)^2
  \end{eqnarray}
Eliminating $Y_0$ from the two above equations, one obtains formally
\begin{equation}
\label{eqtone}
T_{-1}(\theta)=T_1(\theta)\frac{\Phi(\theta)\bar \Phi(\theta)}{(\Phi(\theta+i/2)\bar \Phi(\theta-i/2))^{\frac{2}{D+D^{-1}}}}  e^{-\m \CL \cosh \pi \theta}\;.
\end{equation}
This equation can be solved numerically for arbitrary length, taking as initial values $T_{1}(\theta)=2$, $\Phi(\theta)=\bar \Phi(\theta)=1$ and computing iteratively $T_{-1}(\theta)$ until it converges to its actual value. The results for the vacuum energy
as function of $\m\CL$ are given in \cite{Gromov:2008gj}.

{\it Solution at finite $\CL$ in the $u(1)$ sector.} In this sector, there are excitations of the type $\theta_\al$, but no $u_j$ or $v_j$.
The linear Hirota equation is again of the form (\ref{Hjump}), but now $T_0(\theta)$ has zeroes on the real axis, $T_0(\theta_\al)=0$.
The infinite length solution is provided this time by $T_{s-1}=P(\theta+is/2)-P(\theta-is/2)$, where $P(\theta)$ is a polynomial satisfying $$P(\theta+i/2)-P(\theta-i/2)=\phi(\theta)=\prod_\al(\theta-\theta_\al)\;.$$ 
Let us now define the function
\begin{equation}
G(\theta)=\frac{\phi(\theta-i/2)}{2\pi i}\int_{-\infty}^\infty \frac{d\theta'}{\theta-\theta'}\,\frac{T_{-1}(\theta')}{\phi(\theta'-i/2)}
\end{equation}
with the property that on the real axis $G(\theta-i0)- G(\theta+i0)=T_{-1}(\theta)$ and $G(\theta)+\bar G(\theta)$ is analytic. 
A solution of the Hirota equation can be generated from
  \begin{eqnarray}
  \label{solvacu}
T_{0}(\theta)&=& \phi(\theta)+\bar G(\theta-i/2)+G(\theta+i/2)\;, \\ \nonumber
\Phi(\theta)&=& \phi(\theta+i/2)+\bar G(\theta+i0)+G(\theta+i)\;,\\
\bar \Phi(\theta)&=& \phi(\theta-i/2)+\bar G(\theta-i)+ G(\theta-i0)\;. \nonumber
\end{eqnarray}
The solution can be put in a determinant from via
  \begin{eqnarray}
  \label{solvacuk}
T_{k}(\theta)&=&R(\theta+i(k+1)/2)-R(\theta-i(k+1)/2)\;,\\
R(\theta)&=&P(\theta)+\frac{1}{D-D^{-1}}\[G(\theta+i/2)+\bar G(\theta-i/2)\]\;.
\end{eqnarray}
The function $R(\theta)$ is a solution to the system (\ref{Hlin}) linearly independent from $Q(\theta)$ and the transition between the two solutions is done by the gauge function $g(\theta)$
\begin{equation}
1=Q^v(\theta)=g(\theta-i/2)R^u(\theta).
\end{equation}
The equation determining $T_{-1}$ recursively is the same as (\ref{eqtone}), and it can be considered as an equation for $g(\theta)$.
The generic case when both $Q^u(\theta)$ and $Q^v(\theta)$ are non-trivial can be treated similarly, by relating the jump of the function $T_0(\theta)$ at $\theta\pm i/2$ to the gauge function $g(\theta)$ and deriving a close equation for this function.

\subsection{Y-system for the AdS/CFT Bethe ansatz}

The Beisert-Staudacher equations (\ref{allFNBA}) resemble to the asymptotic Bethe ansatz equations for the principal chiral model 
(\ref{baepcf}), in the sense that the momentum carrying node is the central node and that the two wings are symmetric and they correspond to an inhomogeneous spin chain with $su(2|2)$, respectively with $su(2)$ symmetry. It is therefore tempting to conjecture
that the Y-system (and therefore the T-system) corresponding to the Beisert-Staudacher equations is that of two $su(2|2)$ chains, glued together via the central node.
This hypothesis was confirmed by analyzing the bound states of the Beisert-Staudacher equations and writing the corresponding Y-system \cite{Bombardelli:2009ns,Gromov:2009bc}. Although the resemblance with the principal chiral field is important, there are a few points which are slightly different and which render the analysis of the equations considerably more complicated. The kernels depend on the Jukovski  transform $x$ of the spectral parameter, $u/g=x+1/x$. The existence of this parameter can be traced back to the existence of the central extension of the $su(2|2)$ algebra \cite{Beisert:2005tm}.
The so-called mirror transformation \cite{Arutyunov:2007tc}, which transforms
$(p,E)\to(iE,ip)$ and which corresponds in the principal chiral model to the translation $u\to u+i/2$ is, for the AdS/CFT system
realized by $(x^+,x^-)\to (x^+,1/x^-)$. In terms of the elliptic parametrization (\ref{ellipar}), this transformation is equivalent to the translation of the elliptic argument $s\to s+{\rm K}-i{\rm K'}$, which exchanges the real and imaginary parts of the momentum, $(p,\; \beta)\to (\tilde p=-i\beta, \; \tilde \beta=ip)$. In the Jukovski plane, parametrized by the variable $x$, the mirror transformation sends the real line into the unit circle and conversely.  The mirror system is not equivalent to the original one, since the corresponding dispersion relations are different. Let us remind that the particle-antiparticle transformation appearing in the crossing equation is the square of the mirror transformation\footnote{The transformation $s\to s+2{\rm K}$ has no effect on $(x^+,x^-)$. } $s\to s+2{\rm K}-2i{\rm K'}$ and it takes $(x^+,x^-)\to (1/x^+,1/x^-)$.
 
\subsubsection{Hirota equation and the Backlund transform for supersymmetric spin chains}

Let us first review the Hirota equation for a spin chain with $GL(K|M)$ symmetry, following  \cite{Kazakov:2007fy,Zabrodin:2007rq} \cite{Hegedus:2009ky}. The case relevant for  AdS/CFT is $K=M=2$.
The domain where $T^{K,M}_{a,s}(u)\neq 0$ is non-zero is the ''fat hook`` with $0\leq a\leq K$ or 
$0\leq s\leq M$ and $a,s$ non-negative. On the boundary  $a=0$ the Hirota equation becomes the discrete d'Alembert equation, with the solution
\begin{equation}
T^{K,M}_{0,s}(u)=g_-(u-is/2)g_+(u+is/2)\;.
\end{equation}
We can make use of the gauge freedom in order to fix the boundary condition as 
\begin{equation}
\label{bcsusp}
T^{K,M}_{0,s}(u)=Q_{K,M}(u-is/2)\;, \qquad T^{K,M}_{a,0}(u)=Q_{K,M}(u+ia/2)\;.
\end{equation}
For spin chains, $Q_{K,M}(u)$ is a polynomial.
The linear auxiliary problem, which is a higher rank analogue of (\ref{Hlin}), relates the transfer matrices whose rank differ by one unit,  $({k,m})\to ({k-1,m})$ or $({k,m})\to ({k,m-1})$. For more details on this relation, called Backlund transformation, see \cite{Kazakov:2007fy,Zabrodin:2007rq}. The Backlund transformation is related to the nested Bethe ansatz, where the diagonalization of the transfer matrix is done by a succession of operations which reduce the rank of the algebra by one unit up to rank one. 
For Lie superalgebras, the reduction to rank one can be done in different ways, depending on the order in  which of the two elementary Backlund transformation are taken. Each path connecting the point $(K,M)$ to $(0,0)$ by going either to left or downwards corresponds to a different Dynkin diagram of the Lie superalgebra and is associated to a different form of the nested Bethe ansatz equations. 

\begin{figure}
  \centering
    {%
    \includegraphics[width=0.4\textwidth]{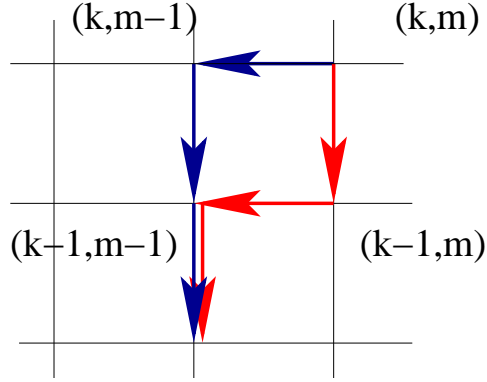}}
  \caption{Two different paths for successive Backlund transformations. The blue and red pieces of path correspond to the two pieces of Dynkin diagrams represented in fig. \ref{fig:dualDyn}.}
  \label{fig:zerocurv}
\end{figure}
The Backlund transformations can be formulated in a rather intuitive way in terms of generating functions.
\begin{eqnarray}
\label{gentrans}
W_{k,m}(u)&=&\sum_{s\geq 0}\frac{T^{k,m}_{1,s}(u+i(s-1)/2)}{Q_{k,m}(u)}D^{2s}\;,\\
W^{-1}_{k,m}(u)&=&\sum_{a\geq 0}(-1)^a D^{2a} \frac{T^{k,m}_{a,1}(u-i(a+1)/2)}{Q_{k,m}(u-i)}
\end{eqnarray}
where again $D\equiv e^{i\partial_u/2}$.
By setting
\begin{eqnarray}
X_{k,m}(u)&=&\frac{Q_{k,m}(u+i)Q_{k-1,m}(u-i)}{Q_{k,m}(u)Q_{k-1,m}(u)}\\ \nonumber
Y_{k,m}(u)&=&\frac{Q_{k,m-1}(u+i)Q_{k,m}(u-i)}{Q_{k,m-1}(u)Q_{k,m}(u)}\;
\end{eqnarray}
one can prove the following recursion relations which encode the two kinds of Backlund transformation
\begin{eqnarray}
W_{k-1,m}(u)&=&(1-X_{k,m}(u)D^2)\;W_{k,m}(u)\\ \nonumber
W_{k,m+1}(u)&=&(1-Y_{k,m+1}(u)D^2)\;W_{k,m}(u)\;.
\end{eqnarray}
Going from the point $(k,m)$ to $(k-1,m-1)$ on two different ways as in figure \ref{fig:zerocurv}
we obtain
\begin{eqnarray}
\label{prezerocurv}
W_{k-1,m-1}&=&(1-X_{k,m-1}D^2)(1-Y_{k,m}D^2)^{-1}W_{k,m}\\
&=&(1-Y_{k-1,m}D^2)^{-1}(1-X_{k,m}D^2)W_{k,m}\;,
\end{eqnarray}
which implies the zero-curvature condition
\begin{equation}
\label{zerocurv}
(1-Y_{k-1,m}D^2)(1-X_{k,m-1}D^2)=(1-X_{k,m}D^2)(1-Y_{k,m}D^2)\;.
\end{equation}
This relation
gives for the term linear in $D^2$
\begin{equation}
Y_{k-1,m}(u)+X_{k,m-1}(u)=X_{k,m}(u)+Y_{k,m}(u)\;.
\end{equation}
 which translates into relations for the $Q$ polynomials which are equivalent to the Bethe ansatz equations.
The zero curvature condition (\ref{zerocurv}) insures that the Bethe ansatz equations corresponding to different paths (or different Dynkin diagrams) are equivalent,
and it is related to the duality transformations described in section (\ref{fulloneloop}).
From (\ref{gentrans}) we can obtain the transfer matrix for the fundamental representation 
\begin{equation}
\frac{T_{1,1}^{K,M}(u)}{Q_{K,M}(u)}=\sum_{{\rm path}} (Y_{k,m}- X_{k,m})
\end{equation}
\begin{figure}
  \centering
    {%
    \includegraphics[width=0.6\textwidth]{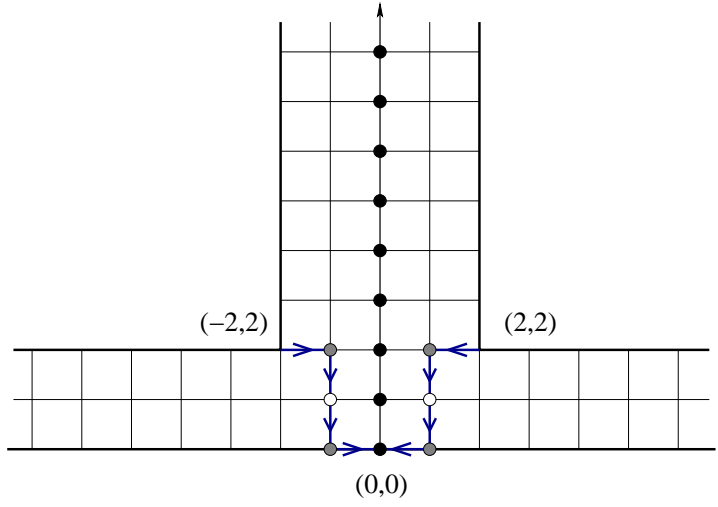}}
  \caption{The T hook corresponding to the Hirota equation for $psu(2,2|4)$. The two wings correspond to two spin chains with symmetry $su(2|2)$, glued together on the black nodes. The blue (gray) path connecting the points $(2,2)$ and $(-2,2)$ to $(0,0)$
  represent a chain of Backlund transformations associated to the second Dynkin diagram in fig. \ref{twdynkin}. }
  \label{thook}
\end{figure}
where $Y_{k,m}$ is associated to a vertical step and  $X_{k,m}$ to a horizontal step starting at $(k,m)$. For example, for the $su(2|2)$ spin chain with Dynkin diagram from figure \ref{su22} , we have
\begin{equation}
W_{2,2}=(1-Y_{2,2}D^2)(1-X_{2,1}D^2)^{-1}(1-X_{1,1}D^2)^{-1}(1-Y_{0,1}D^2)W_{0,0}
\end{equation}
so that 
\begin{figure}
  \centering
    {%
    \includegraphics[width=0.22\textwidth]{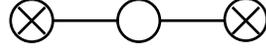}}
  \caption{A Dynkin diagram for a spin chain with $su(2|2)$ symmetry.  }
  \label{su22}
\end{figure}
\begin{eqnarray}
\label{adscftfdt}
T_{1,1}^{2,2}&=&Q_{2,2}(X_{1,1}+X_{2,1}-Y_{2,2}-Y_{0,1})\\ \nonumber
&=&Q_{2,2}\(\frac{Q^{++}_{1,1}Q^{--}_{0,1}}{Q_{1,1}Q_{0,1}}+\frac{Q^{++}_{2,1}Q^{--}_{1,1}}{Q_{2,1}Q_{1,1}}-
\frac{Q^{++}_{2,1}Q^{--}_{2,2}}{Q_{2,1}Q_{2,2}}-\frac{Q^{++}_{0,0}Q^{--}_{0,1}}{Q_{0,0}Q_{0,1}}\)\;.
\end{eqnarray}
where and indices $\pm\pm$ represent a shift by $\pm i$. The Bethe ansatz equations are obtained as the conditions of regularity for $T_{1,1}^{2,2}(u)$ at the zeroes of the polynomials $Q_{k,m}(u)$. With the identification $Q_{0,0}=1$, $Q^{-}_{0,1}=Q_1$,  $Q_{1,1}=Q_2$, $Q_{2,1}^+=Q_3$ and $Q_{2,2}=Q_4$, $Q_n=\prod_{j=1}^{K_n}(u-u_{n,j})$ they are equivalent with the $su(2|2)$ wings of the one-loop
system {\it i.e.} the first three or last three equations in (\ref{FNBA}).
It is interesting to note that for $K=M$, if $T^{K,K}_{a,s}(u)$ is a solution of the Hirota equation with boundary condition (\ref{bcsusp}) defined by $Q_{K,K}(u)=\prod_j (u-u_{K,K;j})$,
then $T^{K,K*}_{s,a}(u)$ is also a solution with similar boundary condition defined by $Q^*_{K,K}(u)=\prod_j (u-u^*_{K,K;j})$.

\subsubsection{Hirota equation and the AdS/CFT integrable system}

The results from the previous section are valid at one loop in gauge theory. The all-loop dependence of the Bethe ansatz can be incorporated formally by redefining in equation (\ref{adscftfdt}) $Q_{2,2}(u)=R^{-(+)}(u)$,  $Q^{--}_{2,2}(u)=R^{-(-)}(u)$, $Q_{0,0}=B^{+(-)}(u)$, $Q^{++}_{0,0}(u)=B^{+(+)}(u)$ with
\begin{equation}
R_n^{(\pm)}(u)=\prod_{j=1}^{K_n}\frac{x(u)-x^{\mp}(u_{n,j})}{(x^{\mp}(u_{n,j}))^{1/2}}\;, \qquad B_n^{(\pm)}(u)=\prod_{j=1}^{K_n}\frac{\frac{1}{x(u)}-x^{\mp}(u_{n,j})}{(x^{\mp}(u_{n,j}))^{1/2}}\;.
\end{equation}
We have $Q_n^{\pm}(u)=(-g)^{K_n}R_n^{(\pm)}(u)B_n^{(\pm)}(u)$ and 
\begin{equation}
T_{1,1}(x)=
R^{-(+)}\(\frac{Q^{++}_2Q^{-}_1}{Q_2Q^+_1}+\frac{Q^{+}_3Q^{--}_2}{Q_3^-Q_2}-
\frac{Q^{+}_{3}R^{-(-)}}{Q^-_{3}R^{-(+)}}-\frac{B^{+(+)}Q^{-}_{1}}{B^{+(-)}Q^+_{1}}\)\;.
\end{equation}
Let us note that $T_{1,1}$ enjoys the interesting transformation property
\begin{equation}
\label{tconj}
T^*_{1,1}(1/x)=\frac{B^{+(-)}}{R^{-(+)}}\frac{Q_1^+Q_3^-}{Q_1^-Q_3^+}T_{1,1}(x)\;
\end{equation}
The Y-system for the full Beisert-Staudacher system can be obtained by gluing the T-functions for two $su(2|2)$ spin chains, $T^l$ and $T^r$ via a gluing condition similar to (\ref{Ybound}),
\begin{equation}
\label{igreczeo}
Y_{a,0}(x)=\(\frac{x(u-ia/2)}{x(u+ia/2)}\)^L\frac{f(u-ia/2)}{f(u+ia/2)}\;T^l_{a,-1} \;T^r_{a,1}\;
\end{equation}
where the first two factors are zero modes, {\it i.e.} solution of the equation $\frac{f_a^+f_a^-}{f_{a+1}f_{a-1}}=1$. The dependence on $L$ is fixed by requiring a behavior of $Y_{a,0}(u)$ at large $u$ compatible with the asymptotic Bethe ansatz. The function $f(u)$ can be determined from the requirement that the asymptotic Bethe Ansatz is reproduced from the equation $Y_{1,0}(u_{4,k})=-1$.

The energy of a state is given by
\begin{equation}
\label{enconj}
E-E_0=\sum_{j=1}^{K_4}\epsilon_1(u_{4,j})+\sum_{a=1}^\infty\int_{-\infty}^\infty\frac{du}{2\pi i}\,\frac{\partial \tilde \epsilon_a}{\partial u}\,\ln (1+\tilde Y_{a,0}(u))\;,
\end{equation}
where $E_0$ is the classical dimension, $ \epsilon_a=2ig(1/x^{a+}-1/x^{a-})$ and the tilde indicates that the corresponding quantities are evaluated in the mirror dynamics.

There are four different presentations of the Beisert-Staudacher equations corresponding to the values $\eta_1, \eta_2=\mp 1$ in \cite{Beisert:2005fw}. These different presentations can be traced back to the choice $T^{r,l}=T_{a,1}(x)\ {\rm or }\ T^*_{1,a}(x)$ respectively in (\ref{igreczeo}) and they lead to different values for the function $f^-/f^+$. The choice $\eta_1=\eta_2=1$ corresponds to a central node of $su(2)$ type, $\eta_1=\eta_2=-1$ corresponds to $sl(2)$ and $\eta_1\eta_2=-1$ corresponds to $su(1|1)$. 
The asymptotic Bethe ansatz equations with $sl(2)$ central node 
\begin{equation}
\frac{f^-}{f^+}=\(\frac{S(x)}{R^{+(+)}}\)^2\;\frac{Q^{++}}{Q^{--}}\;\frac{B_{1l}^+\;B_{3l}^-}{B_{1l}^-\;B_{3l}^+}\;\frac{B_{1r}^+\;B_{3r}^-}{B_{1r}^-\;B_{3r}^+}\;,
\end{equation}
where $S^2(x)=\prod_{j=1}^{K_4}\sigma^2(x,x_{4,j})$
and we used the following convention for the indices of the nodes of the nested Bethe ansatz $(1,2,3,4,5,6,7)\to (1r,2r,3r,\ ,3l,2l,1l)$.

The symmetry condition\footnote{The complex conjugation in $Y^*(x)$ is understood to send $x^\pm \to x^\mp$ and of course it assumes $g$ is real.} $Y_{1,0}(1/x)=Y^*_{1,0}(x)$ translates into the equation for the scattering factor
\begin{equation}
\frac{S(x)}{S^{*}(1/x)}=\frac{R^{-(-)}B^{+(-)}}{B^{+(+)}R^{-(+)}}=\frac{R^{-(-)}B^{-(+)}}{B^{+(+)}R^{+(-)}}\;,
\end{equation}
which is a consequence of the crossing equation for the elementary dressing factor $\sigma(x,y)$
 (\ref{crossf})
 \begin{equation}
\sigma(x,y)\sigma(1/x,y)=\frac{y^-}{y^+}\frac{x^--y^+}{x^+-y^+}\frac{1/x^--y^-}{1/x^+-y^-}\;.
\end{equation}
if we take into account that the dressing factor is a pure phase for particles (or anti-particles) with physical dynamics, 
$S^*(x)=S^{-1}(x)$ and if we use the momentum condition $\prod_{j=1}^{K_4} x_{4,j}^+/x_{4,j}^-=1$.

\subsubsection{Four loop Konishi operator and the Y-system}

The Konishi operator is the simplest operator which is not protected by supersymmetry, ${\rm Tr} \Phi_i^2$. It belongs to a super-multiplet and we find it more convenient to work with a representative situated within the $su(2)$ sector, ${\rm Tr} \, ZXZX$.
The length of the spin chain is $L=4$ and the magnon number is $M=K_4=2$ corresponding to two roots on the central node, $u_{1,4}=-u_{2,4}=1/2\sqrt{3}+\CO(g^2)$. The Y functions associated to the central node are given by
\begin{equation}
\label{igreczeosu}
Y_{a,0}(x)=\(\frac{x^{a-}}{x^{a+}}\)^L\frac{f^{a-}}{f^{a+}}\;T^{l*}_{-1,a} \;T^{r*}_{1,a}\;
\end{equation}
where the index $^{a\pm}$ denotes a shift of the argument by $\pm ia/2$. 
The Beisert-Staudacher equations in the $su(2)$
presentation are obtained if 
\begin{equation}
\frac{f^-}{f^+}=S^2(x)\frac{Q^{++}}{Q^{--}}\(\frac{R^{+(-)}}{R^{+(+)}}\)^2\frac{B_{1l}^-\;B_{3l}^+}{B_{1l}^+\;B_{3l}^-}\;\frac{B_{1r}^-\;B_{3r}^+}{B_{1r}^+\;B_{3r}^-}\;.
\end{equation}
where we have made a gauge transform $T_{1,a}\to T_{1,a}/R^{a-(+)}$ such that $T_{0,a}=1$.
Because we are in the $su(2)$ sector, the $B, R,Q$ polynomials associated to roots other than the central root are equal to $1$. In order to compute the finite length corrections to the energy, the $Y$ functions should be computed in the mirror dynamics for the main argument, so we have to transform $(x^{a+},x^{a-})\to (x^{a+},1/x^{a-})$. For small $g$ we have 
\begin{equation}
\tilde Y_{a,0}(x)\sim \(\frac{g^2}{u^{a+}u^{a-}}\)^L\sim g^{2L}
\end{equation}
which is in agreement with the fact that the wrapping corrections are of order $g^{2L}$ for a chain of length $L$. In particular, we notice that for $L=4$ at fourth loop order the main correction comes from this factor and it is sufficient to consider the other quantities at leading order.
Using (\ref{gentrans}) we obtain for the transfer matrices
\begin{equation}
T^*_{1,a}=\[(a+1)-a\frac{R^{a+(+)}}{R^{a+(-)}}-a\frac{B^{a-(-)}}{B^{a-(+)}}+(a-1)\frac{R^{a+(+)}}{R^{a+(-)}}\frac{B^{a-(-)}}{B^{a-(+)}}\]
\end{equation}
where at leading order in the mirror dynamics
\begin{equation}
\frac{R^{a+(+)}}{R^{a+(-)}}=\frac{Q_0^{(a+1)+}}{Q_0^{(a-1)+}}\;, \qquad \frac{B^{a-(-)}}{B^{a-(+)}}=\frac{Q_0^{(a+1)-}}{Q_0^{(a-1)-}}\;,\end{equation}
with $Q_0^{a\pm} =\prod_{j=1,2} (u-u_{j,4}^0\pm ia/2)$. After simplification we have 
\begin{equation}
T^*_{1,a}=\frac{ 24a (-4 + 3 a^2 + 12 u^2)}{144\; Q_0^{(a-1)-}\,Q_0^{(a-1)+}}\;.
\end{equation}
The scalar factor gives, again at leading order in $g$
\begin{equation}
\frac{f^-}{f^+}=\frac{Q_0^2}{Q_0^{--}Q_0^{++}}\;, \qquad \frac{f^{a-}}{f^{a+}}=\frac{Q_0^{(a-1)-}Q_0^{(a-1)+}}{Q_0^{(a+1)-}Q_0^{(a+1)+}}\;.
\end{equation}
Putting all the pieces together we obtain
\begin{equation}
\label{igreczeosu1}
\tilde Y_{a,0}(x)=g^8 \( \frac{24a (-4 + 3 a^2 + 12 u^2)}{(u^2+a^2/4)^2}\)^2\frac{1} {y_a y_{-a} }    +\CO(g^{10})\;
\end{equation}
where $y_a=144\, Q^{(a-1)-}\;Q^{(a-1)+}=16 - 48 a + 60 a^2 - 36 a^3 + 9 a^4 + 48 u^2 - 144 a u^2 + 
 72 a^2 u^2 + 144 u^4$. Inserting this result in the expression of the energy (\ref{enconj}) and taking the integrals, for example using the residue method, we obtain the wrapping correction
 $\delta E_{{\rm Konishi}}=g^8(648+864\zeta(3)-1440\zeta(5))+\CO(g^{10})$. 
 
At five loop order, the Bethe roots acquire wrapping corrections, which will manifest themselves in the individual magnon energies, that is in the first term in r.h.s. of equation (\ref{enconj}). The second term will receive corrections from the dressing phase, since in the mirror dynamics, $\tilde S^2(x)=1+\CO(g^2)$. The difficult part of the computation is to solve the Bethe Ansatz equations  $Y_{1,0}(u_{4,k})=-1$ beyond the asymptotic regime. This was done numerically in \cite{Arutyunov:2010gb} and analytically in \cite{Balog:2010xa} and the results agree with the computation of \cite{Bajnok:2009vm} using L\"uscher techniques.

The success of the Y system in reproducing various  finite size corrections is encouraging and it is likely that these equations are exact at any size.
It is important to understand the finite size corrections from the spin chain point of view, with a formulation of the dilatation operator as a lattice Hamiltonian
which would be valid for any value of the coupling constant and for any length. In this direction, it would be interesting to clarify the connections with the Hubbard model and how to treat the length-changing operations for finite chains, which were elegantly circumvented by Beisert \cite{Beisert:2005tm} in the case of infinite chains.

\section{Recent developments and open problems}

Another important question which awaits to be solved is that of the origin, and the formal proof, of integrability
in the $\CN=4$ gauge theory. It is commonly believed that the all-loop integrability of the spectral problem discussed above is due to an interplay between planarity and supersymmetry. The discovery \cite{Aharony:2008ug} of another duality between the 
$\CN=6$ supersymmetric Chern-Simons theory in three dimensions and strings in the $AdS_4\times CP^3$ background which is presumably also integrable at all loops \cite{Minahan:2008hf} \cite{Gromov:2008qe} \cite{Arutyunov:2008if} \cite{Stefanski:2008ik}, reinforced the opinion that at the core of the AdS/CFT correspondence stays the same tightly constrained symmetry which includes supersymmetry. 

Most of the effort concerning integrability in $\CN=4$ gauge theory was concentrated on the problem of computing the spectrum of conformal dimensions, but there are clear indications that the integrability of the dilatation operator is related to other special properties of the theory. It was discovered \cite{Anastasiou:2003kj}  that the four-gluon amplitudes possess a recursive structure where the one-loop contribution is the main building block for the higher loop amplitudes. This structure led to the conjecture by Bern, Dixon and Smirnov (BDS) \cite{Bern:2005iz} that the all loop amplitude can be obtained by exponentiating the one-loop result. The BDS conjecture \cite{Bern:2005iz} was verified at strong coupling \cite{Alday:2007hr} for four gluon amplitudes. For six and more gluons, the situation is more subtle and the BDS conjecture does not hold anymore \cite{Bern:2008ap,Drummond:2008aq}. 

The integrals entering the gluon amplitudes have unexpected conformal properties \cite{Drummond:2006rz} when expressed in the variables $\tilde x_i$ defined by
$p_i=\tilde x_i-\tilde x_{i+1}$ where $p_i$ is the four-momentum of the $i^{th}$ gluon. This {\it dual} conformal symmetry, which 
was shown \cite{Drummond:2008vq} to extend to a dual {\it super}-conformal theory prompted the discovery \cite{Drummond:2007aua}
  of a duality between the multi-gluon amplitudes and the Wilson loops formed by light-like segments with cusps, and where $\tilde x_i$ are interpreted as the positions of the cusps. The (super) conformal symmetry of these Wilson loops is broken by the cusps, and the anomalous Ward identities associated with this symmetry were used \cite{Drummond:2007au} to compute the  
 the dependence of the Wilson loops on the positions of the cusps $x_i$. 
 
 At strong coupling, the dual conformal symmetry and the duality with the Wilson loops with cusps arose via a T-duality \cite{Alday:2007hr}, which was extended to a fermionic T-duality in \cite{Berkovits:2008ic}. The action of this duality on the integrable structure of the $AdS_5\times S^5$ superstring was analyzed in \cite{Beisert:2008iq} and reviewed in \cite{Beisert:2009cs}. 
 
 In a recent paper \cite{Drummond:2009fd} it was argued that the generators of the super-conformal symmetry and the ones of the
 dual super-conformal symmetry constitute the generators of a Yangian. This would imply the existence of an integrable structure at the level of the amplitudes, and not only of the conformal dimensions. Such an integrable structure was identified by Lipatov \cite{Lipatov:2009nt} in a certain high energy approximation, where the amplitudes obey BFKL-like \cite{Fadin:1975cb,Balitsky:1978ic} equations. The corresponding integrable Hamiltonian is very similar to the one discovered in the high energy limit of large $N$ QCD \cite{Lipatov:1993yb,Faddeev:1994zg}.
 It is conceivable that this Hamiltonian could be generalized to a integrable Hamiltonian which would describe the amplitudes in the $\CN=4$ gauge theory for arbitrary values of the momenta. 
 
 Very recently, the problem of computing the strong coupling limit of gluon amplitudes with arbitrary number of legs was mapped to
 the problem of minimization of the free energy of a Hitchin system
 \cite{Alday:2009dv}. In \cite{Alday:2010vh} the problem was formulated in terms of a Y system for the Hitchin equation.

\bigskip

\leftline{\bf Acknowledgments}

\noindent I would like to thank the members of the jury of the habilitation thesis, V. Kazakov, G. Korchemsky, J.-M. Maillet, J. Minahan, A. Tseytlin and K. Zarembo, as well as S. Alexandrov, N. Gromov, R. Janik, I.Kostov, I. Shenderovich, M. Staudacher, D. Volin for discussions and for useful remarks on the manuscript. I would like to thank KITP Santa Barbara, where part of the manuscript was written, for hospitality.
This work has been partially supported by the  ANR program INT-AdS/CFT (contract
ANR36ADSCSTZ). 

\appendix

\section{Dynkin diagrams}
\label{appa}

The commutation relations of a Lie (super) algebra ${\mathfrak{g}}$ can be encoded into a Dynkin diagram. 
One starts form a commuting ensemble of generators, the Cartan generators, satisfying
\begin{equation}
[H_i,H_j]=0\;, \quad i,j=1,\ldots, {\rm rank}({\mathfrak{g}}).
\end{equation}
The commutations relations of the Cartan generators with the raising/lowering operators
can be put on the form
\begin{equation}
[H_i,J_r]=r_i\;J_r
\end{equation}
where $r$ is now an index labeling  the raising/lowering operator (the root index) and  $r_i$ 
is a real coefficient (Dynkin label). The roots can be seen as vectors in a linear space of dimension
rank$({\mathfrak{g}})$ with basis vectors $\e_i$, $r=r_i \e_i$, endowed with a scalar product. The non-zero roots can be divided into
positive and negative roots with respect to an arbitrary hyperplane passing through the origin
and not containing any of the roots. Among the positive roots, one can choose a set of {\it simple}
roots as the minimum set generating the other positive roots by linear combinations with positive coefficients. 
Finally, the Cartan matrix encodes the scalar products of the simple roots when the square of the largest root
is normalized to two
 \begin{equation}
A_{ij}=r_i\cdot r_j\;.
\end{equation}

For $su(N)$ algebras, the simplest way to construct the root system is to introduce an extra Cartan
element associated to the identity and extend the algebra to $u(N)$. The basis of the root system
can be defined using the vectors $\e_i$, $i=1,\ldots,N$ with the scalar product
\begin{equation}
\e_i\cdot \e_j=\delta_{ij}
\end{equation}
The positive roots are given by
\begin{equation}
\e_i-\e_j\;,\quad  i>j
\end{equation}
while the simple roots are
\begin{equation}
r_i=\e_{i+1}-\e_i\;,\quad  i=1,\ldots,N-1\;.
\end{equation}

For the Lie algebras, all the possible choices of the set of simple roots are equivalent. The Cartan matrices and the Dynkin diagrams corresponding to the different choices of simple roots are the same. This 
is not the case for the graded (or super) Lie algebras. The simplest case is that of the $sl(N,M)$ algebras. The root space of this super algebra can be constructed using $\e_i$, $i=1,\ldots,N$ and
$\de_k$, $k=1,\ldots,M$ with scalar products
\begin{equation}
\e_i\cdot \e_j=\delta_{ij}\;, \quad \de_k\cdot \de_l=-\delta_{kl}\quad {\rm and} \quad \e_i \cdot \de_k=0\;.
\end{equation}
such that positive roots are given by 
\begin{equation}
\e_{i}-\e_j\;,\quad \de_k-\e_j\;, \quad \de_{k}-\de_l\;, \quad {\rm for} \quad {i>j}\;, \quad  {k>l}
\end{equation}
The positive roots are not equivalent anymore, in particular there are roots of zero norm, which correspond to the odd or fermionic generators.
Therefore the choice of the simple roots is not unique and the Cartan matrix and the Dynkin diagram are not unique. 
A particular system of simple roots is the one with the minimum number of odd roots. This is called the {\it distinguished} root system
\begin{equation}
r_i=\e_{i+1}-\e_i\;,\quad \de_1-\e_N\;, \quad \de_{k+1}-\de_k\;.
\end{equation}
and the corresponding Dynkin diagram has only one gray (or sermonic) node which separates (N ? 1) + (M ? 1) white (or bosonic) node, as for example in Fig. 9.
Other, inequivalent Dynkin diagrams can be obtained by permuting $\de_k$ with $\e_i$. This operation corresponds to a Weyl reflection with respect to an odd root.
\begin{figure}
\label{distinguished}
  \centering
    \reflectbox{%
    \includegraphics[width=0.7\textwidth]{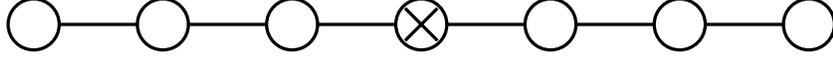}}
  \caption{The distinguished Dynkin diagram for $psu(2,2|4)$. }
\end{figure}

\section{Bethe ansatz equations as integral equations}
\label{appb}

Let us consider the thermodynamical limit of the Bethe ansatz equation for a state with the maximum number of magnons, like the antiferromagnetic state in a $su(2)$ spin chain or the fermionic state in the
$su(1|1)$ sector. The discrete equations for the $M$ magnons are 
\begin{equation}
e^{ip_kL}=\prod_{l=1}^M e^{i\varphi(u_k,u_l)}\, ,
\end{equation}
or by taking the logarithm of both sides we obtain
\begin{equation}
\label{discreteba}
\frac{p_k}{2\pi }=\frac{k}{L} +\frac{1}{2\pi L}\sum_{l=1}^M \varphi(u_k,u_l)\, .
\end{equation}
where we have taken as the mode numbers of magnons $n_k=k$.
When the length of the chain is very large, we can take the continuum
limit.  We introduce the variable $t=k/L$ and the density of
rapidities $\rho(u)=-dt/du$.  The derivative with respect to $u$ of
the equation (\ref{discreteba}) gives
\begin{eqnarray}
\label{intgauge}
&\ &\rho(u)=-\frac{1}{2\pi  }\frac{dp}{du}+
\int_{-\infty}^\infty{d u'}\;K(u,u')\  \rho(u')\;,
\end{eqnarray}
where the kernel is the derivative of the scattering phase 
\begin{equation}
\label{kergen}
K(u,u')= +\frac{1}{2\pi  }\frac{d}{du} \;\varphi(u,u')\;.
\end{equation}
Solving the integral equation (\ref{intgauge}) we obtain the magnon density for the state with the maximal number of magnon, let us call it the AF state in analogy with the $su(2)$ case.

At this point, we want to introduce vacancies in the magnon distribution, vacancies which will be called {holes}. Let us introduce a single hole at $k_h$, such that the magnon mode numbers are now
$n_k=k+\theta(k-k_k)$, and call the new magnon density, in the presence of the hole, $\rho_m(u)$.
The new magnon density obeys
\begin{eqnarray}
\label{intgaugeh}
&\ &\rho_m(u)+\frac{1}{L}\delta(u-u_h)=-\frac{1}{2\pi  }\frac{dp}{du}+
\int_{-\infty}^\infty{d u'}\;K(u,u')\  \rho_m(u')\;.
\end{eqnarray}
If we denote $\rho_h(u)=\frac{1}{L}\delta(u-u_h)$ and
parametrize
\begin{eqnarray}
\label{fluctdens}
 \rho_m(u)+\rho_h(u)
=\rho(u)+\frac{1}{L}K_h(u;u_h)
\end{eqnarray}
the perturbation $K_h(u;u_h)$ obeys the integral equation
\begin{eqnarray}
\label{kmkh}
K_h(u;u_h)=-K(u,u_h)+
\int_{-\infty}^\infty{d u'}\;K(u,u')\  K_h(u';u_h)\;.
\end{eqnarray}
Moreover, for an arbitrary number of holes with density $\rho_h(u)=\frac{1}{L}\sum_n\delta(u-u_{h,n})$
the equation (\ref{fluctdens}) becomes
\begin{eqnarray}
\label{intgaugehh}
&\ &\rho_m(u)+\rho_h(u)=\rho(u)+
\int_{-\infty}^\infty{d u'}\;K_h(u;u')\  \rho_h(u')\;.
\end{eqnarray}
which ressembles to the integral equation (\ref{intgaugeh}), except that it is an integral equation for the hole density. This allows to interpret $K_h(u;u_h)$ as a hole scattering kernel (and to replace the semicolon by a comma).
The equation (\ref{kmkh}) allows to express the hole kernel from the magnon kernel as
\begin{equation}
\label{kh}
K_h=-\frac{K}{1-K}\;.
\end{equation}
In the $su(2)$ case the magnon kernel is a difference kernel
\begin{equation}
K(u,u')=K(u-u')=-\frac{1}{\pi}\frac{1}{(u-u')^2+1}\; \quad {\rm or}\quad K(t)=-e^{-|t|}
\end{equation}
where $K(t)=\int du\; e^{iut}K(u)$ is the Fourier transform of the magnon kernel. From (\ref{kh}) we deduce
that 
\begin{equation}
\label{kht}
K_h(t)=\frac{e^{-|t|}}{1+e^{-|t|}}\quad {\rm or} \quad K_h(u)=\frac{1}{2\pi} \int_0^\infty dt \frac{e^{-t/2}}{\cosh t/2}\cos tu\;.
\end{equation}
giving for the hole (spinon) scattering phase 
\begin{equation}
\label{phih}
\varphi_h(u)=\int_0^\infty \frac{dt}{t} \frac{e^{-t/2}}{\cosh t/2}\sin tu
\end{equation}
The expression of the hole momentum can be also read off  (\ref{intgaugehh})
\begin{eqnarray}
\label{peh}
p_h(u)&=&2\pi \int^u du'\; \rho(u')\;{\buildrel XXX \over=}\; \int^u \frac{\pi du'}{\cosh \pi u'}=\arctan \sinh \pi u+p_0\\
\epsilon_h(u)&=&-\epsilon_m(u)+\int du'\, \epsilon_m(u')\,K_h(u',u)\;{\buildrel XXX \over=}\;-\frac{\pi}{\cosh \pi u}
\end{eqnarray}
where the magnon energy for the ferromagnetic $XXX$ model is taken to be $\epsilon_m(u)=\frac{1}{u^2+1/4}$.
It is interesting that by this procedure we have obtained the crossing-symmetric phase factor $S_0(\theta)$ in the $su(2)$ principal chiral model \cite{Zamolodchikov:1992zr}. The reason is
presumably that the low energy excitations around the antiferromagnetic state, the spinons with $u\gg 1$, are described by a relativistic field theory with $su(2)$ symmetry, therefore their scattering matrix should obey crossing
symmetry. 

Finally let us specify the scattering data for the BDS chain
\begin{equation}
K^{BDS}(t)=-e^{-|t|/2g}\ \ {\buildrel g\to \infty \over \longrightarrow}\  \ -1\;, \qquad  K_h^{BDS}(t)=\frac{e^{-|t|/2g}}{1+e^{-|t|/2g}}\ \ {\buildrel g\to \infty \over \longrightarrow}\ \ \frac{1}{2}\;.
\end{equation}
In the large $g$ limit, one retrieves the scattering data for the Haldane-Shastry spin chain. If one defines the statistical parameter $\lambda=1-K$, then one obtains as expected
\begin{equation}
\lambda_{\rm magnon}=2\;, \qquad \lambda_{\rm spinon}=\frac{1}{2}\;.
\end{equation}

The generalization to higher rank cases is straightforward and we will exemplify it on $su(3)$.
The equations for the two type of roots are
\begin{eqnarray}
\label{magmat}
\rho^1_m+\rho^1_h&=&s+K^{11}\rho^1_m+K^{12}\rho^2_m\\\nonumber
\rho^2_m+\rho^2_h&=&K^{21}\rho^1_m+K^{22}\rho^2_m
\end{eqnarray}
with $s$ the source term and the kernel matrix in the Fourier representation
\begin{equation}
K(t)=\left(\begin{array}{cc}-e^{-t} & e^{-t/2} \\e^{-t/2} & -e^{-t}\end{array}\right)
\end{equation}
In order to obtain the hole scattering kernel, we have to invert  the equation (\ref{magmat}), such that
\begin{equation}
(1-K)^{-1}=1-K_h\;, \quad {\rm or} \quad 1-K_h(t)=\frac{1}{e^{-2t}+e^{-t}+1}\left(\begin{array}{cc}1+e^{-t} & e^{-t/2} \\e^{-t/2} & 1+e^{-t} \end{array}\right)
\end{equation}
In the case of purely statistical interaction, these matrices are given by
\begin{equation}
1-K=\left(\begin{array}{cc}2 & -1 \\-1 & 2\end{array}\right)\;, \qquad 1-K_h=\frac{1}{3}\left(\begin{array}{cc}2 & 1 \\1 & 2\end{array}\right)\;.
\end{equation} 
We deduce that the scattering kernel for the momentum-carrying holes is given by
\begin{equation}
K_h^{11}(t)=\frac{e^{-2t}}{e^{-2t}+e^{-t}+1}\;, \qquad K_h^{11}(u)=\int_0^\infty dt \frac{e^{-t}}{1+2\cosh t}\cosh ut
\end{equation}


\end{document}